\documentclass[a4paper,fleqn,useAMS,usenatbib]{mnras}

\usepackage{graphicx,amsmath,amssymb,multirow,units,lpic}
\usepackage{lscape}
\usepackage[autolanguage]{numprint}
\usepackage{xcolor}
\usepackage{caption}
\usepackage[T1]{fontenc}
\usepackage{ae,aecompl}

\definecolor{purple}{rgb}{0.75,0.0,0.75}

\newcommand{\msun}{\mbox{\,$\rm M_{\odot}$\,}}        
\newcommand{\lsun}{\mbox{\,$\rm L_{\odot}$\,}}        
\newcommand{\lsub}{\mbox{\,$ L_{850}$\,}} 
\newcommand{\lcoa}{\mbox{\,$ L^{\prime}_{\rm{CO}}$\,}}

\newcommand{\lci}{\mbox{\,$ L^{\prime}_{\rm CI}$\,}}
\newcommand{\Md}{\mbox{\,$M_{\rm d}$\,}}
\newcommand{\Ms}{\mbox{\,$M_{\ast}$\,}}
\newcommand{\Mh}{\mbox{\,$\rm M_{H2}$\,}}
\newcommand{\Mmol}{\mbox{\,$\rm M_{mol}$\,}}

\newcommand{\Lir}{\mbox{\,$\rm L_{IR}$\,}}
\newcommand{\COa}{\mbox{\,$\rm{^{12}CO(1-0)}$\,}}

\newcommand{\CI}{\mbox{\,[C{\sc i}]\,}}
\newcommand{\CIfull}{\mbox{\,[C{\sc i}]($^3$P$_1$-$^3$P$_0$)\,}}
\newcommand{\Xci}{\mbox{\,$\rm{X_{CI}}$\,}}
\newcommand{\kd}{\mbox{\,$\rm{\kappa_{d}}$\,}}

\newcommand{\asec}{\ensuremath{^{\prime\prime}}}

\newcommand{\kms}{\mbox{\,$\rm{km\,s^{-1}}$\,}}
\newcommand{\td}{\mbox{\,$\rm{T_d}$\,}}

\newcommand{\mwtd}{\mbox{\,$\rm{T_{mw}}$\,}}

\newcommand{\aco}{\mbox{\,$\alpha_{\rm{CO}}$\,}}

\newcommand{\aci}{\mbox{\,$\alpha_{\rm{CI}}$\,}}
\newcommand{\asub}{\mbox{\,$\alpha_{\rm{850}}$\,}}

\newcommand{\aunit}{\mbox{\,$\rm{M_{\odot}\,(K\,kms^{-1}\,pc^2)^{-1}}\,$}}
\newcommand{\gdr}{\mbox{\,$\delta_{\rm{GDR}}$\,}}
\newcommand{\fg}{\mbox{\,$f_{\rm{g}}$\,}}

\newcommand{\mic}{$\mu $m\,}

\title[Self consistent molecular gas estimates via dust continuum, CO and \CI lines.]{Dust continuum, CO and \CI 1-0 lines: Self-consistent
$\rm H_2$ mass estimates and the possibility of globally CO-`dark' galaxies at z=0.35}

\author[L. Dunne et al.]{
L. Dunne,$^{1,2}$\thanks{E-mail:DunneL6@cardiff.ac.uk} 
S. J. Maddox,$^{1}$
C. Vlahakis,$^{3}$
H. L. Gomez$^{1}$
\\
$^{1}$School of Physics \&\ Astronomy, Cardiff University, Queens Buildings, The Parade, Cardiff, CF24 3AA, UK \\
$^{2}$SUPA, Institute for Astronomy, University of Edinbugh, Royal Observatory, Blackford Hill, Edinbugh EH9 3HJ, UK\\
$^{3}$National Radio Astronomy Observatory, 520 Edgemont Road, Charlottesville, VA 22903-2475, USA\\
}

\date{\vspace*{-5em}\today}

\pubyear{2020}

\begin{document}
\label{firstpage}
\pagerange{\pageref{firstpage}--\pageref{lastpage}} 
\maketitle

\begin{abstract}
We present ALMA observations of a small but statistically complete
sample of twelve 250\mic selected galaxies at $z=0.35$ designed to
measure their dust submillimeter continuum emission as well as their
\COa and atomic carbon \CIfull spectral lines. This is the first
sample of galaxies with global measures of all three $H_2$-mass
tracers and which show star formation rates (4--26 \msun yr${-1}$) and
infra-red luminosities ($1-6\times10^{11}\,\lsun$) typical of star
forming galaxies in their era.  We find a surprising diversity of
morphology and kinematic structure; one-third of the sample have
evidence for interaction with nearby smaller galaxies, several sources
have disjoint dust and gas morphology. Moreover two
  galaxies have very high \lci/\lcoa ratios for their {\it global}
  molecular gas reservoirs; if confirmed, such extreme intensity
  ratios in a sample of dust selected, massive star forming galaxies
  presents a challenge to our understanding of ISM. Finally, we use
the emission of the three molecular gas tracers, to determine the
carbon abundance, \Xci, and CO--$\rm{H_2}$ conversion \aco in our
sample, using a weak prior that the gas-to-dust ratio is similar to
that of the Milky Way for these massive and metal rich
galaxies. Using a likelihood method which
  simultaneously uses all three gas tracer measurements, we find mean
  values and errors on the mean of $\langle\aco\rangle
  =3.0\pm0.5\,\rm{\msun\,(K\,kms^{-1}\,pc^2)^{-1}}$ and
  $\langle\Xci\rangle=1.6\pm0.1\times 10^{-5}$ (or
  $\aci=18.8\,\aunit$) and $\gdr=128\pm16$ (or
  $\asub=5.9\times10^{12}\,\rm{W\,Hz^{-1}\,\msun^{-1}}$), where our
starting assumption is that these metal rich galaxies have an average
gas-to-dust ratio similar to that of the Milky Way centered on
$\gdr=135$.
\end{abstract}


\section{Introduction}

The cold neutral gas (i.e. Cold Neutral Medium (CNM) H{\sc i} and H$_2$ gas)
 in galaxies is the main fuel for star formation and thus a driver
 of galaxy evolution across cosmic time, with the CNM being the
 phase where the $\rm H_I\rightarrow H_2$ phase transition takes
 place (\citealp[e.g.][]{PPP2002} and references therein).
 
 Measuring the cold neutral gas content for large representative
 galaxy samples is thus a key requirement for understanding how
 galaxies form from gas clouds in dark matter haloes to the
 agglomerations of stars we see in the local Universe.  There are,
 however, considerable difficulties involved in observing the cold gas
 phase.  Neutral hydrogen (H{\sc i}) produces 21-cm line radiation,
 but its very low $\rm L_{HI}/M_{HI}$ ratio means that large arrays of
 radio telescopes such as the JVLA operating at the long cm range are
 required to observe it at any redshift beyond the very local Universe
 ($z>0.1$). This state of affairs will change dramatically once the
 so-called Square Kilometer Array (SKA) comes into operation.

Cold   $\rm{H_2}$   gas,   with    densities   starting   from   $\sim
(200-500)\,\rm cm^{-3}$ is  the more intimate  fuel for star  formation in
galaxies but unlike the case for H{\sc i}, the  lowest  transition available
for  $\rm{H_2}$ mass tracing is S(0): $\rm J_u-J_l = 2-0$, with $\rm E_{20}/k_B\sim 510\,K$
which will not be excited appreciably for the typical temperature range of
most of the molecular  gas in Giant Molecular Clouds  ($\sim (15-60)\,K$). Indeed
 S(0) is mostly excited  in shocks,  containing only a few percent of the
total H$_2$ gas  mass.\footnote{Even if the S(0) line  could somehow be excited
for the bulk  of the mass of a typical  GMC, its rest-frame wavelength
of $\sim  28\,\mu$m  precludes routine  observations from  the ground
over most  of the cosmic  volume as the  atmosphere is opaque  at such
wavelengths.}

This is the reason why the emission of tracer elements mixed with the
$\rm H_2 $ gas are always used to determine its mass.  These are low-J
rotational lines of CO (the next most abundant molecule in the Universe
after $\rm H_2$ itself), its most abundant isotopologue $^{13}$CO
(though often very faint), and dust continuum emission (which, for
star-forming ISM, is mostly mixed with $\rm H_2$ rather than H{\sc
  i}).  Thus, unlike the case of H{\sc i}, where an H{\sc i} line is
used to trace its mass, for molecular gas, the relative abundance (R)
of the species used to trace $\rm H_2$ will always insert a first,
common, uncertainty in the computed $\rm M(H_2) $ mass.  This
uncertainty enters non-linearly in the case of the typically optically
thick low-J CO lines J=1--0, 2--1, where a (FUV/CR)-regulated $\rm
R_{co}=[CO/H_2]$ abundance enters the mass estimates via the
non-linear dependence of the CO-$\rm H_2$ (cloud volume)-filling factor on
$\rm R_{co}$. The uncertainty is linear only in the cases of optically thin tracers such as low-J $^{13}$CO
lines ($\rm R_{13CO}=[^{13}CO/H_2]$) or dust continuum emission ($\rm R_{dust}=M(dust)/M(H_2)$).

The low-J CO lines (J=1--0, 2--1), with their low critical densities
($\rm n_{10,21}\sim (410\,, 2.7\times 10^3)\,cm^{-3}$,
\citealt{Jansen1995}), and small $\rm E_{10}/k_B\sim 5.5\,K$, $\rm
E_{21}/k_B\sim 11\,K$, are very easily excited even in the coldest and
most diffuse regions of molecular clouds, and thus remain luminous
throughout their volumes.  Their typically high optical depths ($\rm
\tau _{10}\sim 5-10$), and the resulting radiative trapping, make them
even easier to excite by lowering the corresponding critical densities
to: $\rm n^{(eff)}_{crit}\sim [(1-e^{-\tau})/\tau]\, n_{crit}$, but
are also responsible for a much less straightforward link between those
luminosities and the underlying $\rm H_2$ gas mass, than would have
been the case should these CO lines be optically thin. This is what
gives rise to the vagaries of the so called $\rm
\alpha_{CO}=M(H_2)/L_{CO}$ factor for the CO(1--0) line, (see \citealt{Bolatto2013} for a good review).

There are quite a few effects in the use of CO lines as $\rm{H_2}$
tracers which become more severe at high redshifts, namely:

\begin{enumerate}

\item{The conversion between \COa luminosity (\lcoa) and molecular gas
  mass is  sensitive to  metallicity in a  non-linear fashion  that is
  hard to calibrate, especially at  the low ($Z\la 0.2$) metallicities
  expected in the outer regions of  local spirals, and even more so at
  high redshifts  where present-epoch metallicity gradients  are being
  built by cosmic chemical evolution.}

\item{\COa  cannot be  observed  at high  redshifts  ($\rm z\ga  0.4$)
  without having to  resort to the JVLA (the only  array with adequate
  sensitivity).   Any accessible  higher-J CO  lines (from  J=3--2 and
  higher) trace  only the warmer and  denser H$_2$ gas, which  may not
  necessarily  be  the bulk  of  H$_2$  gas  mass even  in  vigorously
  star-forming galaxies.}

\item{The underlying assumption of CO line luminosities emanating from
  macro-turbulent clouds (where gas  cells are radiatively decoupled in
  CO line emission)  which are self-gravitating may not  hold in some
  extreme environments where gas motions may be unbound or responding
  to significant concomitant stellar mass inside the clouds \citep{Solomon1997,Downes1998}.}
 
\item{At very high redshifts ($\rm z\ga 4$), the CO J=1--0, 2--1 lines
  lose their contrast against a brightening CMB, significantly reducing
  the visibility of \COa in the high-z Universe \citep{daCunha2013}, but even
  more importantly the velocity fields and scale-lengths outlined by these lines
  are rendered nearly invisible inducing biases in the assessment of e.g. the
  dynamical mass enclosed \citep{Zhang2016}.}
\end{enumerate}

The latter is rather disconcerting given that, despite early claims of
dusty distant starbursts containing mostly warm and dense gas (which
could be traced using only high-J CO lines \citealp{tacconi06,
  tacconi08}), subsequent JVLA CO(1--0) line imaging of such galaxies
found this not to be true and, in fact, their H$_2$ gas was found to be in a
cold/diffuse state where only only CO(1--0), (and 2-1) lines would be
excited \citep{Ivison2011}.  At $\rm z\ga4 $ large regions of such
low-J CO line emission can rendered nearly invisible by the CMB effect
described by \citep{Zhang2016}.


Furthermore, multi-line studies of large (U)LIRG samples challenged the
long-held view  of $\rm \alpha_{CO}$  $\sim 1/5$ the Galactic value in  these merger
systems \citep{PPP12xco}. It is worth recalling  that this long-held  view was  based on the  study of
four local (U)LIRGs \citep{Downes1998}, and on the premise that
the total mass of dynamically unsettled merger systems can be computed
via the virial theorem.  Unfortunately  the remedy is rather expensive
in telescope time,  demanding the observations of both  low-J CO lines
as  well as  observations of  rotational lines  of heavy  rotor
molecules such as HCN which are much fainter.


The emergent complications of using an $\rm \alpha_{CO}$ factor in the
distant Universe (or a simple bimodal application of it for disks and
mergers) has been recently confounded by the discovery that Cosmic Rays
(CRs) can be effective destroyers of the CO molecule, while leaving
$\rm H_2$ intact, especially in starbursts \citep{Bisbas2017}. 

The combined effects of CRs, metallicity and UV fields have led to
several theoretical and observational studies highlighting the
possibility, and detection of, CO-`dark' molecular gas
\citep{PPP2002,PlanckCOdark,Pelupessy2009,Wolfire2010,Pineda2013,Bisbas2015,Remy2017,Liszt2018}.
This gives more prominence to alternative $\rm H_2$ gas mass tracers
such as dust continuum and the two \CI lines.

Since the first statistical sub-mm survey of 100 local Far-Infrared
(FIR) bright galaxies (SLUGS:\citealp{Dunne2000,Dunne2001}), it has
been clear that sub-mm derived dust masses ($\propto L_{850\mu m}$) and CO
based molecular gas masses ($\propto L_{\rm CO}$) are tightly
correlated. The recent availability of large sub-mm selected samples
from {\em Herschel} have seen an increase in the exploitation of dust
as a measure of molecular gas content
\citep{Magdis2012,Rowlands2014,Scoville2014,Scoville2016} and several
large studies have demonstrated the potential for dust emission to be
a good alternative tracer of molecular gas through comparisons of
850\mic emission and CO measurements
\citep{Scoville2014,Scoville2016,Hughes2017,Orellana2017}. So far, however, there
is no independent verification that dust or CO are tracing the total
molecular component effectively.


The third tracer of molecular hydrogen is the atomic carbon line
\CIfull (hereafter \CI(1--0)), with the \CI(1--0) line being optically thin
in most extragalactic situations. Its promise as an alternative
extragalactic H$_2$ mass tracer was first comprehensively described by
\citet{PPP2004,PPP&Greve2004} (though see the early pioneering work by
\citealp{Keene1985,Keene1997}). Early models of the chemical
evolution of dense molecular clouds predicted that atomic carbon would be found
only in a thin layer between the molecular CO form of C and the
ionised C{\sc ii} (\citealt{Langer1976, Tielens1985}: the standard PDR
view). Subsequent observations of \CI and CO in Galactic clouds were
at odds with this model, measuring much higher $\rm{[C^0/CO]}$ abundances than
the standard PDR view predicted and also revealing that the \CI emission was
distributed widely throughout the cloud \citep{Keene1985}. A review of
the models explored at that time to explain these high Carbon abundances
was presented by \citet{Keene1985}. One avenue of thought was that
balance between the competing dynamical processes of accretion and
evaporation of carbon atoms onto dust grain mantles would determine
the Carbon abundance \citep{Boland1982, Williams1984} although there was
disagreement about the main processes driving the evaporation
(turbulence carrying cells to the cloud surface, or sputtering by
shocks within the cloud). Another suggestion \citep{Prasad1983} was
that UV-photons could be produced deep within clouds by the
interaction of cosmic rays with hydrogen molecules. This photon flux
would dissociate CO and produce \CI throughout the cloud. The idea of
a cosmic ray regulated chemistry has been explored further in recent
numerical work by \citet{Bisbas2017, Clark2019} and could explain the
existence of some clouds in the Galactic Center Region which have
unusually high $\rm{[C^0/CO]}$ abundances \citep{Tanaka2011} as well as the central
regions of starburst galaxies and active nuclei \citep{Israel2002,
  Israel2020, Izumi2020}. Clouds in the process of forming are also expected to
have higher Carbon abundances because of the timescale taken to reach
equilibrium in the conversion of C$\rightarrow$ CO
\citep{Suzuki1992}.

Atomic carbon has a simple partition function of 3 levels (unlike CO's
rotational ladder), while the abundance ($\rm \Xci= [C^0/H_2]$)
uncertainty is one common to all $\rm H_2$ tracer species
(indeed \Xci may actually be the easiest of the three to model in
terms of the prevailing ISM conditions, if it is indeed mostly
CR-controlled \citep{Bisbas2015}). Moreover, while much fainter than
C{\sc ii}, the \CI(1--0) line traces {\it only} molecular hydrogen,
unlike C{\sc ii} which also traces the H{\sc i} or H{\sc ii} gas
components (which can be substantial for `young' and forming high-z
systems) \citep{PPP2004,Liszt2011,PerezB2015,Clark2019}.

The two \CI transitions are, however, highly absorbed by the
atmosphere, making large scale observational work in the local
Universe challenging, even from the driest sites on Earth like the
Atacama plateau where ALMA is located. As a result there has
been little observational work surveying this line in extra-galactic
sources, and most have been restricted to a few nearby starbursts,
local (U)LIRGs and high-z SMG and QSO
\citep{Israel2006,Weiss2005,Walter2011,Zhang2014,krips16}. {\em
  Herschel} expanded the observations of the \CI lines in nearby
galaxies with the FTS spectrometer
\citep{Kamenetzky2014,Lu2017,Jiao2017,Jiao2019,Wilson2017,Crocker2019}
but its limited sensitivity meant that the targets were mostly still
(U)LIRGs or only the central regions of very nearby galaxies. The
recent development of the Atacama Large Millimetre Array (ALMA) and
upgrades to NOEMA have allowed an expansion of studies focused on
using \CI as a tracer of molecular gas at high redshifts
\citep{AZ13,Bothwell2017,Valentino2018,Valentino2020,Bourne2019}, but
what is lacking is a well defined sample of non-extreme star forming
galaxies with global observations of all three main gas tracers:-- \COa,
\CI(1--0) and sub-mm observations of cold dust.

This paper describes the observations, data reduction and results for
a first dedicated study of the dust, \COa and \CI(1--0) emission
in a small but homogeneously selected sample of galaxies in the
relatively local Universe at $z=0.35$. This is possible thanks to the
{\em Herschel}-ATLAS \citep{Eales2010}: the first unbiased survey of
the dust content of local galaxies, covering 660 sq. deg and sensitive
to the cold dust component which dominates the mass of dust in
galaxies. The large areal coverage lends itself to producing samples
of (cold ISM)-selected sources at a variety of redshifts with which to
revisit the relationships of dust, CO and \CI emission with the aim to
cross-calibrate each tracer of molecular gas mass in massive
metal-rich galaxies. A companion paper (Dunne et al. {\em in prep}) will
combine this sample with numerous others from the literature to
provide the first joint analysis of the calibration of $\rm H_2$ gas
mass tracers using all three methods in a self-consistent way.

In Section \ref{Sample} we describe the sample, observations and data
reduction. In Section~\ref{resS} we discuss the results and
comparisons of the three gas tracers. In Section~\ref{gasS} we present
our method for deriving the gas calibration factors for this
sample. We also find the very surprising result that there is an
overdensity of a factor 4--6 in the number of high redshift dusty
galaxies found in these fields compared to blank field surveys. This
is discussed in a companion paper, \citep{Dunne2020smg}. Throughout we
use a cosmology with $\Omega_m = 0.308$ and $H_o = 67.9\,
\rm{km\,s^{-1}\,Mpc^{-1}}$ \citep{planckcosmo}.

\section{Sample and Data}
\label{Sample}

\begin{table*}
\caption{\label{sampleT} Properties of galaxies in our sample.}  \centering
\begin{tabular}{lcccccccccc}
\hline
H-ATLAS IAU & SDP & GAMA  & R.A & Dec. & z & $S_{250}$ & $\rm{r_{pet}}$ & Spectral & Kinematic  & SMG \\  
            & name & ID   & (hh:mm:ss)    & (dd:mm:ss)     &   & (mJy)    & & type &   class    &   \\
\hline
J$090506.2+020700$ & 163  & 347099 & 09:05:06.1 & 02:07:02.2 & 0.34507 & 107.6 & 18.8 & Sy2 & Simple &  N \\
J$090030.0+012200$ & 1160 & 301774 & 09:00:30.1 & 01:22:00.2 & 0.35309 & 48.4  & 19.15 & LIN & Simple & Y\\
J$085849.3+012742$ & 2173 & 376723 & 08:58:49.4 & 01:27:41.0 & 0.35506 & 46.2  & 18.71 & SF & Simple & Y\\
J$091435.3-000936$ & 3132 & 575168 & 09:14:35.3 & $-$00:09:35.6 & 0.35859 & 40.6 & 19.02 & SF & Simple & N\\
J$090450.0-001200$ & 3366 & 574555 & 09:04:50.1 & $-$00:12:03.0 & 0.35401 & 40.3 & 18.93 & Sy2 & Disturbed & Y\\
J$090707.7+000003$ & 4104 & 210168 & 09:07:07.9 & 00:00:02.1 & 0.35032 & 46.2 & 19.38 & LIN & Disturbed & Y\\
J$090845.3+025322$ & 5323 & 518630 & 09:08:45.3 & 02:53:20.0 & 0.35282 & 28.6 & 18.98 & SF-c & Disturbed &   N\\
J$090658.6+020242$ & 5347 & 382441 & 09:06:58.4 & 02:02:44.7 & 0.34744 & 32.7 & 19.01 & -- & Disturbed &  Y\\
J$090444.9+002042$ & 5526 & 600545 & 09:04:44.9 & 00:20:48.2 & 0.34174 & 31.2 & 19.23 & SF & Simple & N\\
J$090844.8-002119$ & 6216 & 204249 & 09:08:44.8 & $-$00:21:18.0 & 0.35215 & 36.2 & 18.75 & SF-c & Simple & N \\
J$090402.3+010800$ & 6418 & 372500 & 09:04:02.2 & 01:07:58.2 & 0.34665 & 31.6 & 18.96 & LIN-c & Disturbed & Y\\
J$090849.4+022557$ & 6451 & 387660 & 09:08:49.5 & 02:25:56.9 & 0.35259 & 33.7 & 19.08 & SF-c & Simple & N\\
\hline
\end{tabular}
\flushleft{\small{\bf Notes:} Positions and redshifts refer to the
  optical cross-identifications from \citet{Bourne2016}. $z$ is the
  spectroscopic redshift from GAMA \citep{Hopkins2013, Baldry2018} 250\mic flux is
  from the H-ATLAS DR1 release \citep{Valiante2016}. $\rm{r_{pet}}$ is
  the SDSS r-band petrosian magnitude from GAMA. Spectral type uses
  the Blue diagnostic from \citet{Lamareille2010} since we only have
  O[{\sc ii}], H$\beta$ and O[{\sc iii}] lines in the GAMA spectral
  range. SF-c are in the SF region but composite SF/LIN are also found
  here. LIN-c are in the LINER region but composite LIN/SF are also
  found here. Kinematic class describes whether the CO and \CI have similar
  kinematics and are aligned with the optical redshift (simple), or
  are disturbed with multiple kinematic components or large
  differences in the CO and \CI properties. SMG indicates whether a
  background high-redshift SMG is detected at $>5\sigma$ in the Band 7
  field.}
\end{table*}

The sample was selected at 250$\mu$m from the Herschel-ATLAS Science
Demonstration Phase (SDP) equatorial field at R.A. 09h. H-ATLAS was
the widest area extragalactic survey carried out with the {\em
  Herschel} Space Observatory \citep{Pilbratt2010}, imaging 600
deg$^2$ in five bands centered on 100, 160, 250, 350 and 500$\mu$m,
using the PACS \citep{Poglitsch2010} and SPIRE instruments
\citep{Griffin2010}. One of the primary aims of the {\em
  Herschel}-ATLAS was to obtain the first unbiased survey of the local
Universe at sub-mm wavelengths, and as a result was designed to
overlap with existing large optical and infrared surveys. The {\em
  Herschel} observations consist of two scans in parallel mode
reaching a 4$\sigma$ point source sensitivity of 28 mJy beam$^{-1}$ at
250\mic. The angular resolution is approximately 9\asec, 13\asec,
18\asec, 25\asec\ and 35\asec. While the original sample for the
proposal was selected from the SDP public release catalogue described
in \citet{Rigby2011} to have $S_{250}>5\sigma$ and a reliable optical
identification with spectroscopic redshift from \citet{DJBSmith2011},
we update the {\em Herschel} photometry and optical parameters in this paper
to those from the H-ATLAS DR1 release
\citep{Valiante2016,Bourne2016}. We use the SPIRE MADX
\citep{Maddox2020} matched filter photometry from the DR1 release, as
these are all point sources and this is the most likely estimate of
their flux. In the case of PACS, the LAMBDAR algorithm of
\citet{Wright2016} produces, in our opinion, a more robust measure of
the PACS fluxes and errors as instead of using a top hat aperture, it
convolves the optical r-band aperture with the PACS PSF and so
measures flux in a PSF-weighted aperture. Where there was a
significant difference between the two catalogues, we returned to the
original H-ATLAS PACS maps and remeasured our own photometry. The four
sources this applies to are marked with $^M$ in
Table~\ref{photomT}. Spectroscopic redshifts and UV-22\mic photometry
are provided by the Galaxy and Mass Assembly (GAMA) survey
\citep{Driver2011,Liske2015,Wright2016}.

In order to fulfil the requirements of the ALMA Cycle 1 call where
only Band 7 and Band 3 were available, all the sources had to be
within 12 deg of each other on the sky and had to be observed with no
more than five tunings. This meant that only sources in a very limited
redshift range around $z=0.34-0.36$ could be selected in order to be
able observe \COa and \CIfull. We selected {\em all} the H-ATLAS SDP
sources within this redshift range, making this sample of twelve
representative of sources from a blind 250\mic selected sample at
$z\sim 0.3-0.4$. Details of the sample are given in
Table~\ref{sampleT}.

\subsection{Properties of the 250\mic selected galaxies}
\begin{figure}
\includegraphics[width=0.49\textwidth]{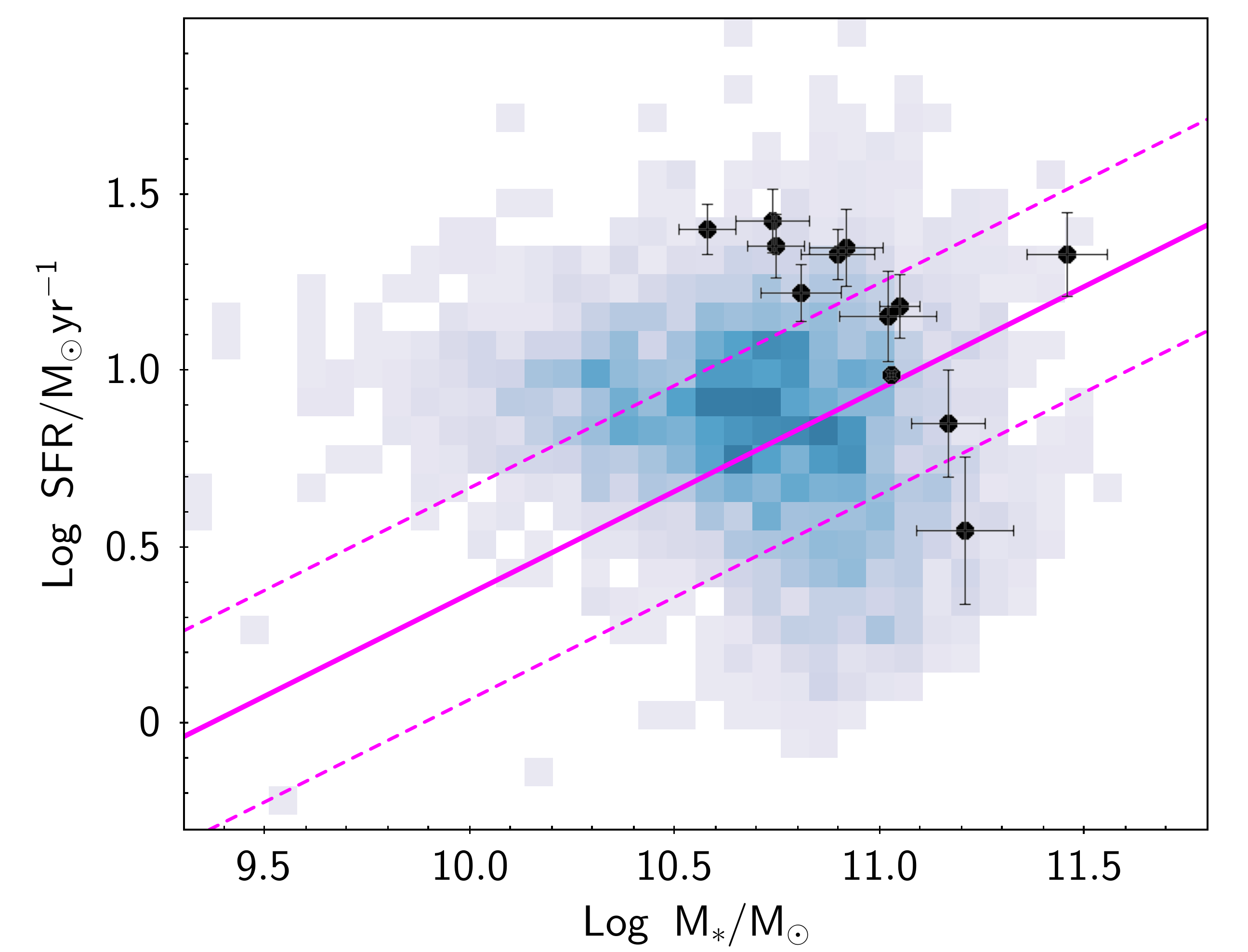}
\caption{\label{MSplotF} SFR versus stellar mass for the $z=0.35$
  sources with the best fit Main Sequence (solid) and $\pm0.3$ dex
  intrinsic scatter (dashed lines) from \citet{Speagle2014}. The
  coloured region represents the density of galaxies in the range
  $0.33<z<0.36$ with log $\rm{sSFR>-11.0}$ and
  $\rm{\sigma_{M_{\ast}}<0.15,\,\sigma_{SFR}<0.2}$ from GAMA
  \citep{Driver2016}. This also shows the effect of the 250\mic
  selection at this redshift, where we sample the leading edge of the
  distribution in SFR-\Ms for the optical selection $r<19.8$ from
  GAMA. Most of the 250\mic sample is consistent with the expected MS
  at this redshift, however, three of the lowest mass galaxies lie
  above the relation (SDP.3132, SDP.6216 and SDP.6451).}
\end{figure}

Earlier studies have shown that the 250\mic population is evolving
rapidly over the $0<z<0.5$ interval \citep{Dye2010,Dunne2011} and so
we note that this sample probes an era where the galaxies were more IR
luminous and gas and dust rich than those at $z=0$. The sample has a
narrow range of $\rm{L_{IR}} = 1.2\times10^{11}-6\times10^{11}\,\lsun$
making them far more `typical' of galaxies at this redshift than
previous very luminous IR samples \citep[e.g.][]{Combes2011}. At these
IR luminosities the sample should be dominated by galaxies which are
forming stars under normal conditions, rather than displaying the
properties of extreme star-bursts \citep{Sargent2012}. The
star-formation rates are between 4--26 $\rm{M_{\odot}\,yr^{-1}}$ and
stellar masses are in the range
$\Ms=4\times10^{10}-3\times10^{11}\,\msun$. A comparison with
optically selected galaxies in the same redshift range from the GAMA
survey \citep{Driver2016, Baldry2018} is shown in
Figure~\ref{MSplotF}. The magenta solid and dashed lines show the
`main sequence' fit from \citet{Speagle2014} with $\pm0.3$ dex
intrinsic scatter. Most of the sample lies within the 0.3 dex scatter
of the galaxy star forming main sequence at $z=0.35$ (when allowing
for the measurement errors), and three sources are above the main
sequence by a factor 4--6 (SDP.3132, SDP.6216, SDP.6451). It is also
apparent that the 250\mic selection picks out the leading edge of the
optical cloud of galaxies, i.e. only the most massive or highly star
forming galaxies at this redshift make it above the {\em Herschel\/}
flux limit.

Three colour optical/near-IR images of our sample from VST KIlo Degree
Survey \citep{deJong2017} and VISTA VIKING, \citep{Edge2013} are shown
in Figure~\ref{163F}--\ref{6451F} and the optical spectra from GAMA
\citep{Baldry2018} are shown in Figure~\ref{A163F}--\ref{A6418F}. The
galaxies have a range of morphologies and colours, from compact to
disk-like, with many displaying signs of disturbance such as stellar
caustics and asymmetric morphology. The optical spectra predominantly
show an older stellar population with superimposed emission lines
indicating ongoing star formation, similar to that seen in stacked
spectra in the much larger H-ATLAS sample analysed by
\citet{Eales2018}. One source (SDP.6216) has a clear starburst
signature in the optical spectrum, and it also has the highest offset
from the main sequence in Figure~\ref{MSplotF}.

Metallicity is a property which affects the emission of all three gas
tracers, but its most complicated effects are on CO because of a
combination of non-linear FUV photon shielding effects (Pak et
al. 1998; Bolatto et al. 1999; Wolfire et al. 2010).\footnote{The
  recently explored effect of CRs only compound these difficulties
  (Bisbas et al. 2015; Bisbas, van Dishoeck et al. 2017).} The effects
on \CI and dust emission are more straight-forward and the abundances
of these tracers are roughly proportional to metallicity. We cannot
directly estimate metallicities for our galaxies as their optical
spectra do not extend as far into the red as H$\alpha$ at this
redshift, i.e. there are no strong line-based methods available to
measure metallicity directly (without H$\alpha$ to correct for
reddening we would not trust methods which use only the ratios of
H$\beta$, O[{\sc iii}] and O[{\sc ii}] in this dust selected
sample). Fortunately, since we have a sample of massive galaxies with
log $\Ms/\msun>10.5$, we can use the fact that the mass metallicity
relation flattens at this mass in this redshift range, hence we expect
these galaxies to have metallicities between 0.8--1.2 $Z_{\odot}$
using a variety of metrics measured at this redshift
\citep{ValeAsari2009,LaraLopez2010,Bothwell2016,Torrey2019}. In this
metallicity range we do not expect any of the molecular gas tracers to
be adversely impacted because of metallicity and thus we do not
consider metallicity-dependence within our sample any further.

\subsection{ALMA observations and data reduction}
\label{COdata}

\begin{figure*}
\includegraphics[trim=3.5cm 0cm 4cm 1.8cm, clip=true, width=0.49\textwidth]{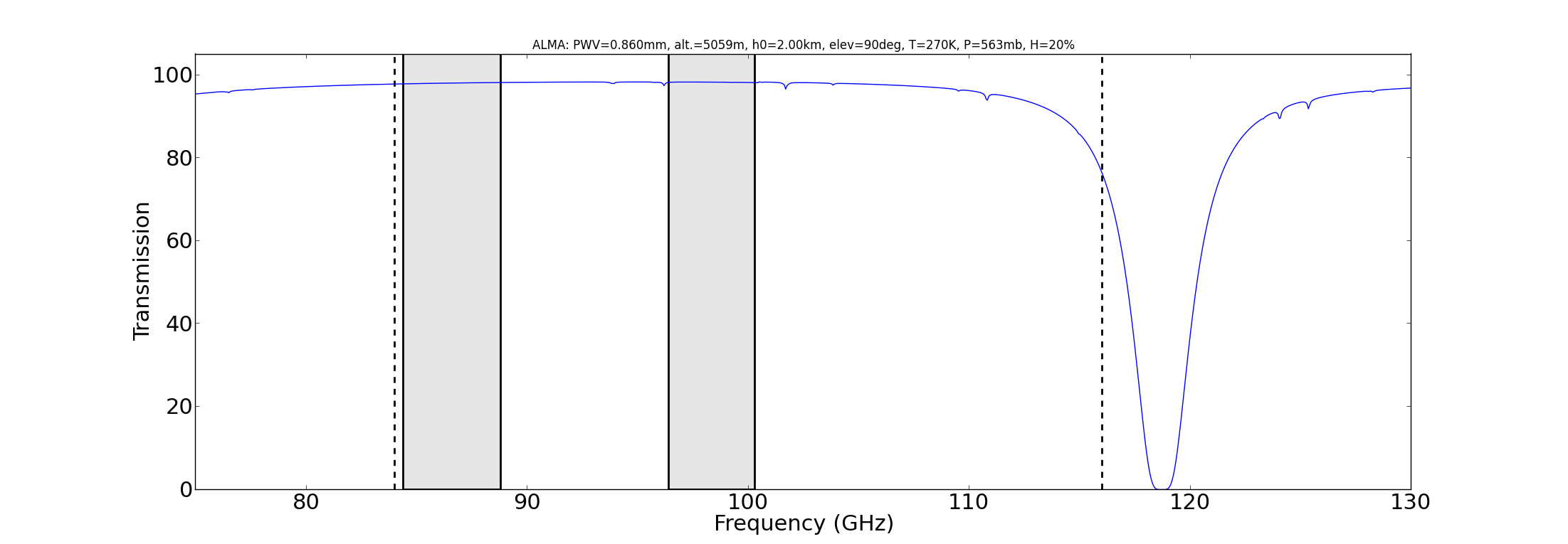}
\includegraphics[trim=3.5cm 0cm 4cm 1.8cm, clip=true, width=0.49\textwidth]{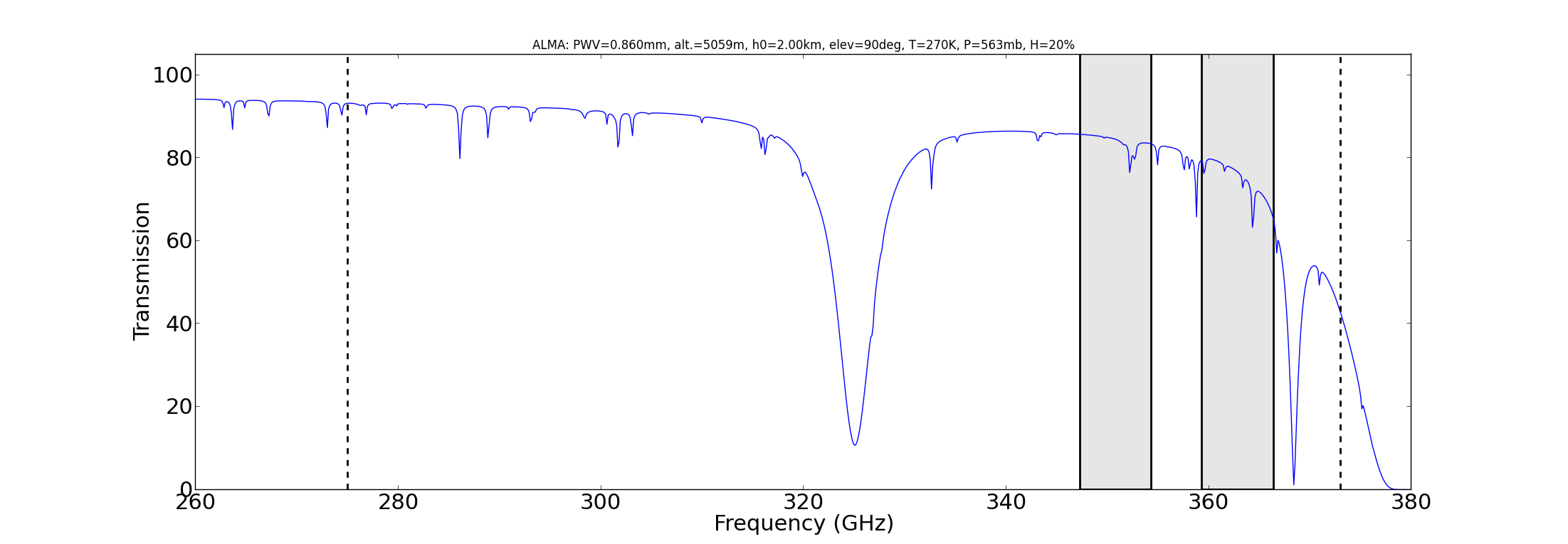}
\caption{\label{setupF} The spectral set-up for the observations. Band
  3 (left) for \COa and Band 7 (right) for dust continuum and
  \CIfull. The atmospheric curve is shown for the average pwv over the
  Band 7 observations (0.84mm). The solid shaded bands are the range
  of frequencies in which we observed lines and continuum, the
  vertical dashed lines represent the limits of the ALMA bands. There
  is only a {\em very} small range of redshift in which this
  experiment was possible due to the availability of receiver bands at
  the Cycle 1 call.}
\end{figure*}

\begin{table*}
\caption{\label{ALMAparamsT} Details of the ALMA observing.}  \centering
\begin{tabular}{lccccccccccc}
\hline
SB name & $\rm{\nu_{obs}}$ &  Date & EB & $N_{\rm ant}$ & $t_{\rm int}$ & p.w.v &  Phase Cal. & Flux Cal. & B.P. &  $\sigma_{\rm{cont}}$ & $\sigma_{\rm line}$\\
        & (GHz)           &        &        &              &  (min)    & (mm)   &               &          &      &   ($\mu$Jy) & (mJy)\\
\hline
B3  &  85.5 & 3.12.13   & Xe20 & 26 & 48.4 & 4.14 &  J$0854+20$ & Mars & J$0854+20$ & 40 & 0.9\\   
    &       & 3.12.13   & X1032 & 26 & 48.4 & 4.14 & J$0854+20$ & Callisto & J$0854+20$ & &  \\
\hline
LowV (163,  &  365.2  & 14.12.13  & X1644 & 26 & 50.2 & 0.84  & J$0854+20$ & Pallas  & J$1058+01$ & 90 &0.86 -- 1.2 \\ 
4104, 5347, 6418)       &         & 26.12.14  & X2543 & 40 & 33.3 & 0.84  & J$0909+01$ & Ganymede & J$0825+03$ & .. &  \\ 
SDP.5526  & 366.8 &  21.2.14  & X5fa & 27  & 12.0  & 0.46 & J$0909+01$ & Pallas & J$0825+03$ & 77 & 1.25\\
           &          & 26.12.14  & X23cb & 40 & 8.0 & 1.15 & J$0909+01$ & Ganymede & J$0825+03$ & .. & \\
SDP.3132   & 362.3   & 14.12.13 & Xf8d$^{\ast}$ & 30   & 5.5 & 1.10 &J$1007-02$ & Ganymede & J$0504-44$ & 70 & 0.87 \\
           &           & 2.1.15 & X1ba4 & 39  & 5.5 & 0.70 &J$0909+01$ & J$0750+12$ & J$0825+03$ & ..& \\
           &           & 2.1.15 & X1cf1 & 39  & 5.5 & 0.72 &J$0909+01$ & Callisto & J$0909+01$   & ..&  \\
HighV      &  363.5 & 21.2.14 & X3e5 & 28  & 35.3 & 0.64 &J$0914+02$ & Ganymede & J$0825+03$  & 65 & 0.84 -- 1.1\\
(1160, 2173, &       & 2.1.15 & X1e75 & 39  & 34.2 & 0.69 &J$0909+01$ & Callisto & J$0825+03$  & .. & \\
3366, 5323,  &       & 2.1.15 & X2365 & 39  & 15.4 & 0.82 &J$0909+01$ & Ganymede & J$0825+03$  & ..&  \\
6216, 6451)   &       & 2.1.15 & X2606 & 39  & 34.2 & 0.91 &J$0909+01$ & Ganymede & J$0909+01$  & ..&  \\
\hline
\end{tabular}
\flushleft{\small $\sigma_{\rm{cont}}$ is the average rms in
  $\mu$Jy/beam for the continuum. $^{\ast}$ M.S. not used as it did
  not improve the overall noise level or image quality.}
\end{table*}

Choosing an observing set-up to cover both the \COa and \CIfull lines
with only Band 3 and Band 7 receivers available led to a
severe restriction on the redshift of sources which could be observed
in Cycle 1. Unfortunately, due to this restricted receiver choice the
\CI(1--0) spectral window  had to be placed in the wings of an
atmospheric water line, thus the transmission in the \CI window is lower
than that in the other three used for the continuum (Figure~\ref{setupF}). The Band 3 base-bands
were placed as close to the edge of the band as possible in order to
maximise the transparency at the \CI(1--0) frequency. With a wider choice of
receivers, observations with much improved sensitivity to the \CI(1--0) line
are now possible.

\begin{figure}
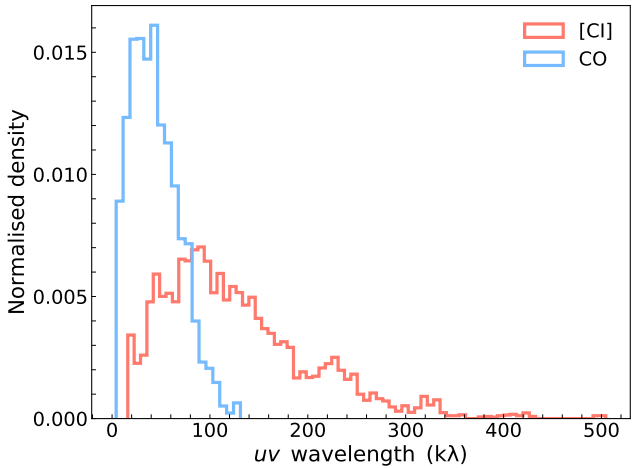

\begin{lpic}[l(0mm),r(0mm),t(0mm),b(0mm)]{UVdistance_hist(0.4,0.4)}
\end{lpic}
\caption{\label{uvF} The distribution of (u,v) separations in
  k$\lambda$ for the array configurations used to observe CO(1-0)
  (blue) and \CI(1-0) and dust (red). The (u,v) coverage is very
  different. The \CI u-v coverage contains high spatial frequency
  information (i.e. high resolution information) that CO lacks and,
  conversely, the CO(1--0) u-v coverage retains spatial sampling in
  the shortest spacings, meaning more sensitivity to larger spatial
  scales than \CI. The inner gap in the (u,v) coverage for \CI and
  dust at these small spacings results in a filtering of the larger
  spatial scales, which would result in a loss of \CI and dust continuum brightness
  distribution relative to that of CO(1--0) {\em if} there is extended
  gas and dust on scales larger than 25 kpc.}
\end{figure}

Observations were made with the ALMA-12~m array in the C32-2 and C32-3
configurations for Band 3 and the C43-1/2 configuration for Band
7. This produced beam sizes (with natural weighting) of
$\theta_3=2.4\asec \times1.8\asec$ and $\theta_7= 1.03\asec
\times0.64\asec$. The $u,v$ coverage for the two set-ups is shown in
Figure~\ref{uvF}, where there is clearly more sensitivity on longer
base-lines (smaller angular scales) in Band 7 compared to Band 3,
while the Band 3 observations have relatively more short baselines
(more sensitivity to extended structure). There are fewer missing
short-spacings in the $u,v$ coverage\footnote{which are always present
  for interferometric observations without total power measurements}
the CO(1--0) compared to the \CI(1--0) images, which results in better
sensitivity to larger angular scales in the CO(1--0) maps compared to those for
\CI and dust. The implications of this are discussed further in
Section~\ref{morphS}.

The Band-3 observations of the \COa line ($\nu_{\rm rest}=115.27$ GHz)
were made in Dec 2013 during Cycle 1, the spectral setup consisted of
four 2-GHz bandwidth spectral windows (spws): one spw with 976~MHz
channel width centered at 85.4~GHz covering the \COa line, and three
spw with 31.25~MHz channel width, centered at 87.4, 87.5 and 99.3 GHz
for the continuum. The total integration time was 96 minutes, giving
$\sigma_{\rm line}= 0.9$~mJy beam$^{-1}$ in a 50~\kms channel and
$\sigma_{\rm cont}= 40~\mu$Jy beam$^{-1}$. The Band-7 observations of
\CIfull and dust continuum were split across Cycles 1 and 2 over a
period from December 2013--January 2015. The spectral setup again
consisted of four 2-GHz bandwidth spectral windows, one with 976 MHz
channel width centered on the \CIfull line ($\nu_{\rm rest}=492.16$
GHz) and the other three with 31.25~MHz channel width at lower
frequencies offset by -2, -12, -14 GHz from the high resolution
spw. Four different central frequencies for the \CIfull spw were used,
depending on the redshifts of the sources: 366.8, 365.2, 363.5 and
362.3 GHz. The total integration time was 3.9 hours giving
$\sigma_{\rm line}=0.8-1.2$~mJy beam$^{-1}$ in a 50~\kms channel and
$\sigma_{\rm cont}=65-90~\mu$Jy beam$^{-1}$.

The 3-mm Band 3 data were reduced manually using the Common Astronomy
Software Applications (CASA) v4.5 package \citep{McMullin2007}. The
starting points were the observatory scripts delivered with the data,
but these needed to be adapted due to early issues in Cycle 1, such as
too much data being flagged in some cases and extra flagging needed in
other regions. Flux calibration was from Mars and the phase/bandpass
calibrator J0854+2006 was used. The two measurement sets (MS), which were
observed on the same day, were concatenated before imaging having been
set to a common flux scale. The \COa spectral line cubes were created
using {\sc clean} in CASA in a range of channel widths in order to
determine the best SNR for each source. Any emission detected was
cleaned on a channel-by-channel basis. Natural weighting was used in
order to maximise signal-to-noise. 

The Band 7 observations were set up in four Scheduling Blocks (SBs)
where the sources that could share a single tuning were grouped into a
given SB. This means that the \CIfull line is not always placed in the centre
of the spectral window. Occasionally both the line and neighbouring
continuum spw had to be imaged together to provide a good baseline for
line fitting. Due to some of these data being taking during Cycle 2,
the newer CASA v4.7 was used for the reduction. Some of the MS were
calibrated using the ALMA pipeline, while others were reduced manually,
depending on how the data were delivered. All MS were checked and
reprocessed allowing for tailoring of the calibration to the specific
issues in this data-set. For example, there are several atmospheric
lines which are evident in the Band 7 data and so the pipeline calibration
was modified to avoid flagging the Tsys response in some cases, and to
output the data-set from the pipeline after the generation of the water
vapour radiometer (WVR) and $\rm{T_{sys}}$ calibration tables. The
data were then manually processed from that stage so that the
atmospheric lines could be flagged in the bandpass calibrator and the
bandpass solutions interpolated in these regions. Additional manual
flagging was also applied where required.

Imaging of the \CIfull line and continuum was performed using the {\sc casa} task {\sc tclean}
with a Hogbom algorithm, using natural weighting. The dust continuum emission at $350-360$~GHz
(rest-frame $\sim$480 GHz) was imaged using the line-free portions of
the data. Images were made at full resolution ($\theta_7= 1.03\asec
\times0.64\asec$) and also tapered in the $uv$-plane to increase the
sensitivity to any low level extended emission. The optical diameter
of the galaxies is 4--9\asec, and so they are expected to be
resolved in our observations. Sources above $3\sigma$ in the dirty
image were masked and lightly cleaned (to 1.5$\sigma$).

Spectral line cubes were created using velocity channels chosen on the
basis of the lines found (ranging from 30-150\kms). Line emission
above 3$\sigma$ detected in the dirty image was masked in each channel
individually and cleaned down to 1.5$\sigma$. If the continuum
emission was strong, it was subtracted from the cube using the {\sc
  casa} task {\sc imcontsub}, this was only necessary in the case of
SDP.163 as in all other sources the peak continuum emission was much less
than the rms in the line channels. We searched near the optical
redshift for emission, also taking account of any neighbouring optical
sources which may have been at a similar redshift and interacting with
the main source (as turned out to be the case for four targets). The
cubes were collapsed to make a moment 0 map across the velocity range
where line emission was seen using the {\sc immoments} task in {\sc
  casa}.

Images are presented in Figure~\ref{163F}--\ref{6451F} and are not
corrected for the variable attenuation of the primary beam over the
field, being displayed instead as signal-to-noise contours.\footnote{It
  should be noted that a source nearer to the edge of the map will
  have a higher flux at the same contour level compared to a source in
  the centre, due to the fact that the primary beam attenuation
  increases the noise as a function of radius from the pointing
  centre.} 

\subsection{Flux measurements.}

Continuum fluxes were measured on either the full resolution or
tapered maps depending on the source size and morphology. For the most
accurate fluxes we fitted Gaussians, using the CASA tasks {\sc imfit}
and {\sc gaussfit} where possible, which meant using the tapered maps
for the more extended sources and the full resolution maps for any
point like sources. Where a source was reported to be unresolved or no
size could be measured, we opted to use the peak flux, which is
equivalent to that derived when fixing the size parameters to be equal to
those of the beam. When a source was resolved and had a peak SNR$>5$, we used
the integrated flux from the Gaussian fit, while for extended sources
where a Gaussian was not a good representation, or when the SNR was
less than 5, we used an aperture around the source emission \citep[e.g][]{Simpson2015counts,Oteo2016}

When fitting fluxes using {\sc imfit} or {\sc gaussfit}, we used only
a small region to actually solve the fit parameters as this makes the
fitting less unstable. However, the flux error returned by the
algorithm in {\sc casa} is usually underestimated due to the presence
of correlated noise in the map pixels \citep{Condon1997}. To deal with this, we first
measured $\rm{\sigma_{true}}$, the pixel rms noise on the {\em non}
primary-beam-corrected image, over a large enough region to be a true
representation. We then derived the correct fractional errors on the
flux output by the {\sc casa} fitting task using the following equation:\footnote{This is
  based on \citet{Condon1997}.}

\begin{equation}
\rm{\frac{\delta S}{S} = \sqrt{\left(\frac{\delta A}{A}\right)^2 \left(\frac{\sigma_{true}^2}{\sigma_{alg}^2}-1\right)+ \left(\frac{\delta I}{I}\right)^2}}
\end{equation}
where $\rm{\delta A/A}$ and $\rm{\delta I/I}$ are the fractional error on
the amplitude and integrated flux returned by the algorithm. 

The amplitude error is underestimated if too small a region is used by
the algorithm to determine the map rms, and so we correct this using
the ratio of our own measurement of the noise $\rm{\sigma_{true}}$ to
the rms used by algorithm in its determination of the errors
($\rm{\sigma_{alg}}$).\footnote{In {\sc imfit} one can simply supply
  $\rm{\sigma_{true}}$ to the task, thus overriding 
  the noise estimate from the smaller region over which the fit is to be performed.}

For the aperture measurements, we estimated the noise using the
formulation in \citet{Dunne2000} which accounts for correlated noise:
\[
\rm{\sigma_{ap}=\sigma_{pix} \times \sqrt{N_{ap}/N_{beam}}}
\]
where $\sigma_{\rm ap}$ is the noise in an aperture which has
$\rm{N_{ap}}$ pixels. $\sigma_{\rm pix}$ is the pixel to pixel rms
measured in large off-source regions of the map and ${\rm N_{beam}}$
is the number of pixels in the beam. This is equivalent to putting
down random apertures on the map and estimating the variance of their
sums (assuming the noise is Gaussian other than the beam scale
correlations).

We corrected the aperture measurements for the fraction of the beam
falling outside the aperture by comparing the fluxes within the same
aperture placed on the phase calibrator with that returned by the {\sc
  imfit} task in CASA, following the method outlined in \citet{Simpson2015counts}.

Integrated line fluxes were calculated in a similar
way to the continuum fluxes, either by fitting a Gaussian to the
moment 0 image, or using an aperture if that was more appropriate. No
thresholding was applied to the moment 0 maps to avoid biasing the
fluxes. All fluxes and errors were then corrected for the primary beam attenuation using the
primary beam model output by CASA during the {\sc clean} stage.
The integrated line intensities and other observational parameters for CO are
listed in Table~\ref{COdataT}, and those for \CI and dust are found in Table~\ref{CIdataT}.

\begin{table*}
\caption{\label{COdataT} Details of the Band 3 CO measurements}
\begin{tabular}{lccccccc}
\hline
Source  & R.A.       & Dec         & $\delta v_{\rm CO}$  & $\Delta v_{\rm CO}$ & $S_{10}$ & FWHM \\
      & (J2000)    & (J2000)     & (\kms )        & (\kms)     &     (Jy\kms)   & (\kms)    \\             
\hline
163  & 09:05:06.19 & $+$02:07:02.2 &  80 & $14\pm53$  &  $1.60\pm0.28^a$  & $728\pm129$ \\
1160 & 09:00:30.13 & $+$01:22:00.4  & 100 & $39\pm22$  &  $0.70\pm0.13^g$  & $296\pm53$  \\
2173 & 08:58:49.36 & $+$01:27:41.4 & 50   & $-34\pm8$  & $0.84\pm0.20^a$  & $147\pm19$     \\
3132 & 09:14:35.28 & $-$00:09:35.7 & 15  & $9\pm4$  & $0.70\pm0.12^g$ & $59\pm10$ \\
3366 & 09:04:50.10 & $-$00:12:03.6  & 50  & $54\pm11$  & $0.38\pm0.12^g$ & $205\pm 27$ \\ 
4104.a & 09:07:07.85 & $+$00:00:02.4 & 60   &  $-72\pm14$   & $0.85\pm0.25^g$  & $222\pm32$ \\
4014.b & 09:07:07.82 & $+$00:00:05.1 & 60  & $-366\pm18$  & $0.78\pm0.15^g$  & $176\pm43$ \\
5323 & 09:08:45.32 & $+$02:53:21.2 & 80  & $-114\pm31$  & $0.80\pm0.20^{gt}$ & $462\pm74$ \\
5347 &  09:06:58.22  & $+$02:02:45.73  & 80  & $218\pm17$  & $2.29\pm0.56^a$ & $417\pm40$  \\
5347(S) & 09:06:58.63 & $+$02:02:35.87 & 80  & $206\pm20$  & $1.90\pm0.45^a$ & $859\pm120$ \\
5526 & 09:04:44.88 & $+$00:20:48.2 & 50  &   $-7\pm7$   & $1.04\pm0.20^g$ & $148\pm16$ \\
6216 & 09:08:44.79 & $-$00:21:17.9 & 50      & $20\pm13$  & $0.70\pm0.11^{gt}$  & $259\pm30$  \\
6418 & 09:04:02.15 & $+$01:07:58.9 & 75  & $\mathbf{178\pm53}$ & $\mathbf{0.27\pm0.14^g}$  & $\mathbf{124\pm72}$\\
6451 & 09:08:49.2   & $+$02:25:57.8 & 50  & {\bf 0}   &  $\mathbf{0.27\pm0.11^{gp}}$  & $\mathbf{300}$  \\
     &              &               & 50  & {\bf -50} &  $\mathbf{0.40\pm0.19^a}$  &  $\mathbf{200}$          \\ 
\hline
\end{tabular}
{\flushleft \small{Each component found is listed on an individual
    line. $^a$ or $^g$ next to the flux indicate whether an aperture
    was used or a Gaussian fit, although where both methods could be
    used they gave comparable results within the errors. $t$ indicates
    the measurement was made on the tapered map. Bold font indicates
    measurements with integrated flux at $<3\sigma$. Details on
    individual source measurements can be found in
    Section~\ref{sourcesS}.\\SDP.5347 has two or three optical
    components and it is difficult to disentangle the CO emission. The
    sources appear to be merging and there is CO emission at a similar
    velocity outside the optical disks which may be tidal
    remnants. The first row includes emission from the three optical
    components which have similar velocities. The second row (S) is
    the clump of emission to the South which can be seen in
    Figure~\ref{5347F}. This southern emission is not used in any
    further analysis. 6451 has very weak CO emission over a velocity
    range $-150\kms\rightarrow -50\kms$}. The first entry in the table above is
  the peak flux from fitting a Gaussian to the brightest peak
  coincident with the \CI peak. The second entry is using an aperture
  which covers the same area as that used to measure the \CI flux in
  Table~\ref{CIdataT}.}\\
\end{table*}

\begin{table*}
\caption{\label{CIdataT} Details of the Band 7 CI and continuum measurements}
\begin{tabular}{ccccccccccc}
\hline
Source & R.A.       & Dec               & $\delta v_{\rm CI}$ &  $\Delta v_{\rm CI}$  & $S_{\rm CI}$       & FWHM &  $S_{\lambda}$ & $\lambda$ \\  
      & (J2000)    & (J2000)            & (\kms)           &   (\kms)           & (Jy\kms)        & (\kms)     & (mJy)         & (\mic) \\             
\hline
163  &  09:05:06.11 & $+$02:07:01.9 &  150     &  $-28\pm74$     & $8.3\pm1.3^a$ & $829\pm293$  & $3.05\pm0.34^a$ &  837.18 \\
 & & & & & & & &\\
1160 &  09:00:30.15 & $+$01:22:00.2 &  50      &  $2\pm18$         & $3.00\pm0.27^g$ & $383\pm43$ & $0.54\pm0.12^{gt}$ & 840.27 \\
 & & & & & & & &\\
2173 &  08:58:49.31 & $+$01:27:41.3  &  30       & $-30\pm5$         & $3.31\pm0.48^a$ &  $125\pm12$ & $0.74\pm0.20^{gt}$ & 840.69\\ 
 & & & & & & & &\\
3132 &  09:14:35.28 & $-$00:09:35.7 &  20      & $7\pm4$ & $2.32\pm0.38^a$ & $73\pm10$ &     $0.95\pm0.29^a$ &  843.98 \\
3132$^{\rm +ACA}$ &          &               &          &         & $2.57\pm0.38^a$ &           &     $1.51\pm0.39^a$ &         \\
 & & & & & & & &\\
3366.a.v1 & 09:04:50.13  &$-$00:12:03.6    &  100      &   $2\pm14$        & $1.77\pm0.49^{gt}$ & $196\pm33$    & $\mathbf{0.42\pm0.19^{gt}}$ & 840.29\\
(3366.a.v2) & 09:04:50.09  & $-$00:12:03.0  & 100 &   $-363\pm11$     & $1.06\pm0.28^{gt}$ & $305\pm27$    & ..                & ..\\
(3366.b) & 09:04:50.21 & $-$00:12:01.0   &  100  &  $-145\pm27$        &  $0.78\pm0.20^g$ & $727\pm63$   &   $0.20\pm0.054^g$ & 840.29 \\
 & & & & & & & &\\
4104.a & 09:07:07.85 & $+$00:00:02.4 &  50     & $-138\pm44$       &  $1.60\pm0.53^{a}$ & $354\pm102$     & $\mathbf{0.46\pm0.21^a}$ & 835.31\\
           &             &                &         &                   &  $2.41^c$         &    &                 &     \\
4104.b & 09:07:07.82 & $+$00:00:05.1 & 50   & $-61\pm33$        &  $1.02\pm0.30^g$ & $421\pm82$       &  $0.46\pm0.12^g$ & 835.31     \\
           &             &                &         &                   &   $1.13^c$        &    &                  &    \\ 
 & & & & & & & & \\
5323.v1 & 09:08:45.30 & $+$02:53:21.7 & 80 & $-273\pm34$    & $1.35\pm0.38^{gt}$ & $330\pm84$ & $0.89\pm0.24^a$ & 841.43 \\ 
5323.v2 & 09:08:45.30  & $+$02:53:21.7 &  80  &  $107\pm18$   &  $0.54\pm0.15^{g}$  & $390\pm95$  &  ..  & .. \\ 
(5323.v3)    & 09:08:45.29 & $+$02:53:20.9 &  80  & $1239\pm13$  & $2.37\pm0.46^{gt}$  & $325\pm30$  & ..  & .. \\
 & & & & & & & &\\
5347.a &  09:06:58.45  &  $+$02:02:45.2  &  80    &    $-219\pm19$ & $3.49\pm0.91^{at}$   &  $480\pm112$ & $0.39\pm0.13^{gt}$ & 849.38\\
5347.bc & 09:06:58.22    & $+$02:02:47.6  &  80    & $-790\pm16$ & $0.82\pm0.19^{gt}$     & $179\pm36$   & $0.38\pm0.11^g$ (b) & 849.38   \\
5347.bc &   `` ''             &  `` ''      &  80    & $110\pm28$  & $1.11\pm0.32^{gt}$     & $305\pm67$   & $1.16\pm0.29^{gt}$ (c)  & 849.38     \\
(5347.S) & 09:06:58.56 & $+$02:02:39.3  & 80      & $-47\pm20$   & $1.31\pm0.33^{gt}$               & $349\pm47$   & $\mathbf{0.54\pm0.24}^g$ & 849.38 \\
 & & & & & & & &\\
5526 & 09:04:44.88 & $+$00:20:48.2 &  50        & $-76\pm13$   & $2.07\pm0.28^{gt}$ & $217\pm30$ & $1.11\pm0.27^{gt}$ & 833.32\\
 & & & & & & & &\\
6216 & 09:04:02.15 & $-$00:21:17.9  & 50        & $-22\pm11$  & $2.02\pm0.36^{gt}$   &  $191\pm25$  & $1.49\pm0.25^a$ & 841.44 \\
(min)&             &                & 50        &             &                     &              & $1.02\pm0.26^a$ & .. \\
(max)&             &                & 50        &             &                     &              & $2.18\pm0.41^a$ & .. \\  
 & & & & & & & &\\
6418.a & 09:04:02.14 & $+$01:08:00.4 & 75      &  $249\pm19$    &  $4.62\pm0.64^{a}$    &  $290\pm44$ & $1.06\pm0.32^a$ & 835.33 \\
(6418.v1)    &             &               & 75  &   $-452\pm19$    & $1.23\pm0.29^{gt}$     &  $148\pm43$ &  ..     &  \\
6418.v2    &            &                & 75  &   $+249\pm17$    & $2.66\pm0.51^a$   &  $289\pm41$           &   ..    &  \\
(6418.v3)    &             &                & 75  & $+503\pm37$       & $0.73\pm0.26^{gt}$ &  $282\pm88$  &   ..     &  \\
(6418.b) & 09:04:02.37 & $+$01:07:58.5 & 75     &  $-369\pm14$      & $1.24\pm0.28^{gt}$  &  $314\pm34$       &  $\mathbf{0.24\pm0.11^{gt}}$ & 835.33 \\
 & & & & & & & &\\
6451  & 09:08:49.49 & $+$02:25:57.6 &  50                  &  $-24\pm13$     &  $4.46\pm0.97^a$ &  $200\pm30$  & $0.88\pm0.19^{gt}$ & 841.44 \\    
\hline
\end{tabular}
{\flushleft \small{The positions and fluxes of the various components
    are listed with the following nomenclature: Spatially distinct
    components which are identified with separate optical sources are
    given letters to denote them. Kinematically distinct components in
    the same source are denoted as $v_i$. The components which are
    used to derive the physical properties such as luminosity and gas
    mass in Tables~\ref{magphysT} and ~\ref{lumT} are shown without
    brackets, while those inside brackets are not used.\\ $^a$ or $^g$
    next to the flux indicate whether an aperture was used or a
    Gaussian fit, although where both methods could be used they gave
    comparable results within the errors. $t$ indicates the
    measurement was made on the tapered map. Measurements at
    $<3\sigma$ are in bold.\\For details on the individual entries
    please see notes in Appendix~\ref{sourcesS}.\\3132 had ACA data
    taken in Cycle 7 and the inclusion of this leads to the second row
    of fluxes denoted $^{\rm +ACA}$.\\3366 has three components: the
    main optical galaxy 3366.a has two associated velocity components,
    $v1$ and $v2$ while the small blue companion, 3366.b, has a broad
    line which overlaps the velocity profile of 3366.a.v2.\\4104
    (interacting pair) has spatially and kinematically resolved
    components in both CO and \CI.\\5323 has three velocity
    components: north of the nucleus $v_1 = -273$\kms, south of the
    nucleus ($v_2 = +100$\kms), and high velocity emission $v_3 =
    1239$\kms (Fig.~\ref{5323F}). There is no corresponding CO
    emission for the high velocity component, and we do not consider it further.\\5347 `a,b,c' refer to
    the main optical galaxy, the satellite seen to the west within the
    optical isophotes of the main galaxy with the strongest CO peak,
    and the satellite visible to the north-west respectively. 5347.S
    is a clump of emission to the South which overlaps in velocity
    with the clump seen in CO.\\6216 is an extended source with
    complex continuum morphology, three continuum fluxes are
    given. The first includes all the emission except for the Eastern
    arm-like extension. The second is a robust minimum flux which
    subtracts the compact clump discussed in Appendix~\ref{sourcesS},
    and does not include the Eastern extension. The third is a robust
    maximum which includes all three components.\\6418.a is the main
    optical target, and the next three rows are the three velocity
    components $v1,v2,v3$, the sum of which constitutes the flux for
    6418.a. The smaller optical companion, 6418.b, has very low SNR
    850\mic continuum flux but has \CI at the same $-400$\kms velocity
    as the 6418.a($v_1$).}}
\end{table*}

Spectral profiles were extracted either from the peak pixel of a point
source, or from the aperture which covered the region of interest. The
profiles were baseline subtracted, boxcar or Hanning smoothed in some
cases and fitted with a Gaussian line profile. The profiles are shown
in Figures~\ref{163F}-~\ref{6451F} and the fitted parameters are
listed in Tables~\ref{COdataT} and \ref{CIdataT}.

\section{Results}
\label{resS}

\begin{figure*}
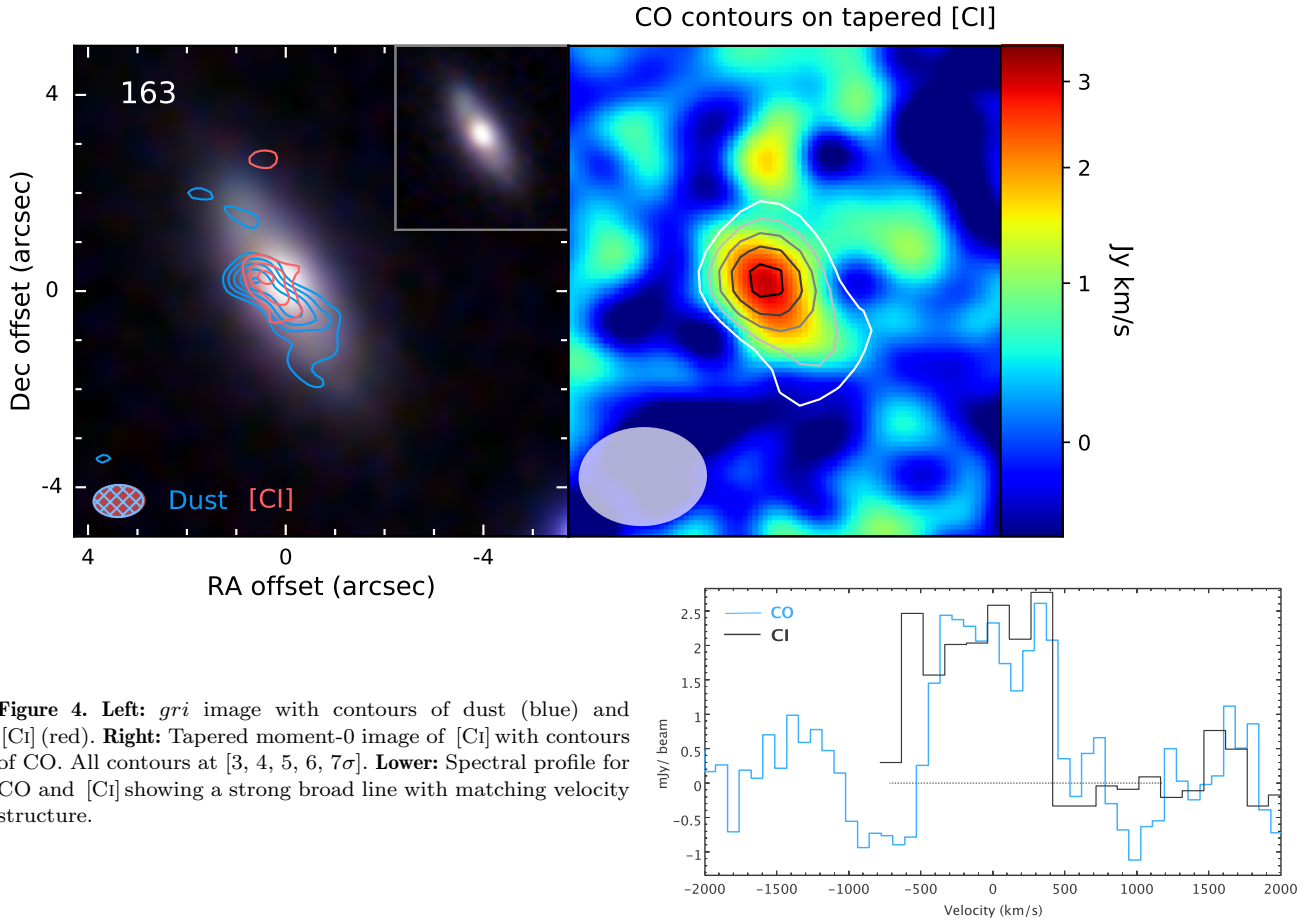

\begin{minipage}{0.98\linewidth}
\begin{lpic}[]{SDP163_gri_revised(0.8,0.8)}
\end{lpic}
\end{minipage}
\begin{minipage}{0.47\linewidth}
\caption{\label{163F} {\bf Left:} $gri$ image with contours of dust
  (blue) and \CI (red). {\bf Right:} Tapered moment-0 image of \CI
  with contours of CO. All contours at [3, 4, 5, 6, 7$\sigma$]. {\bf
    Lower:} Spectral profile for CO and \CI showing a strong broad
  line with matching velocity structure.}
\end{minipage}
\begin{minipage}{0.5\linewidth}
\begin{lpic}[l(0mm),r(0mm),t(-15mm),b(-10mm)]{163_CICO_profile_new(0.30,0.30)}
\end{lpic}
\end{minipage}
\end{figure*}

\begin{figure*}
\begin{minipage}{0.98\linewidth}
\begin{lpic}[]{SDP1160_gri_revised(0.8,0.8)}
\end{lpic}
\end{minipage}
\begin{minipage}{0.47\linewidth}
\caption{\label{1160F} {\bf Left:} $gri$ image with contours of dust (blue)
  and \CI (red). {\bf Right:} Moment-0 \CI image with contours of CO. All contours at [3,4,5,6,8,10,12$\sigma$]. {\bf Lower:} Spectral profile for CO and \CI
  showing a strong broad line with matching velocity structure.}
\end{minipage}
\begin{minipage}{0.5\linewidth}
\begin{lpic}[l(0mm),r(0mm),t(-15mm),b(-10mm)]{1160_CICO_bc2_profile_fit(0.30,0.30)}
\lbl{250,160;\large{\bf -- \CI}}
\lbl{250,140;\large{{\textcolor{cyan}{ -- CO}}}}
\end{lpic}
\end{minipage}
\end{figure*}

\begin{figure*}
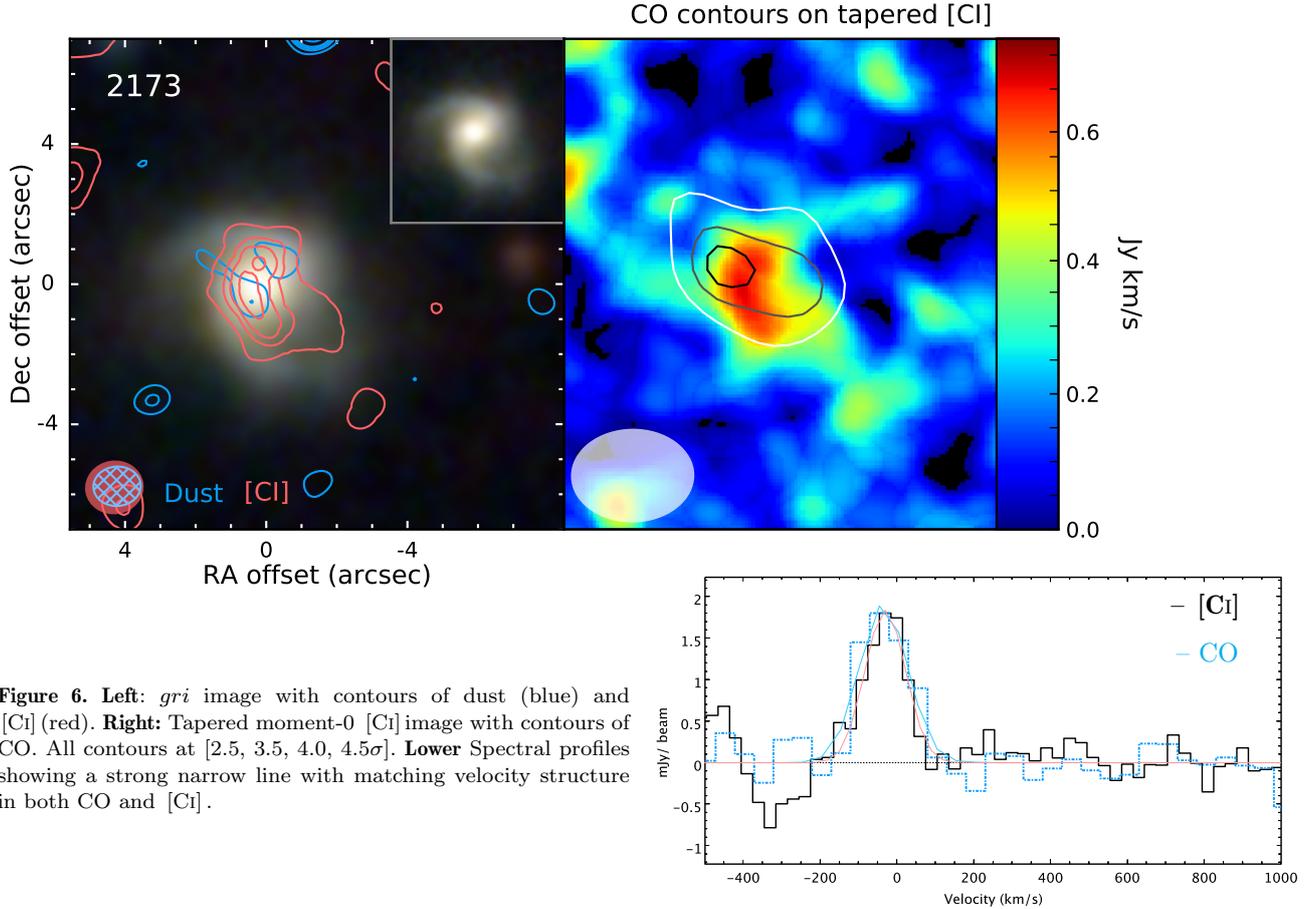

\begin{minipage}{0.98\linewidth}
\begin{lpic}[]{SDP2173_gri_revised(0.8,0.8)}
\end{lpic}
\end{minipage}
\begin{minipage}{0.47\linewidth}
\caption{\label{2173F} {\bf Left}: $gri$ image with contours of dust
  (blue) and \CI (red). {\bf Right:} Tapered moment-0 \CI image with
  contours of CO. All contours at [2.5, 3.5, 4.0, 4.5$\sigma$]. {\bf Lower}
  Spectral profiles showing a strong narrow line with matching
  velocity structure in both CO and \CI.}
\end{minipage}
\begin{minipage}{0.5\linewidth}
\begin{lpic}[l(0mm),r(0mm),t(-15mm),b(-10mm)]{2173_CICO_profile_fit_sm(0.30,0.30)}
\lbl{250,160;\large{\bf -- \CI}}
\lbl{250,140;\large{{\textcolor{cyan}{ -- CO}}}}
\end{lpic}
\end{minipage}
\end{figure*}

\begin{figure*}
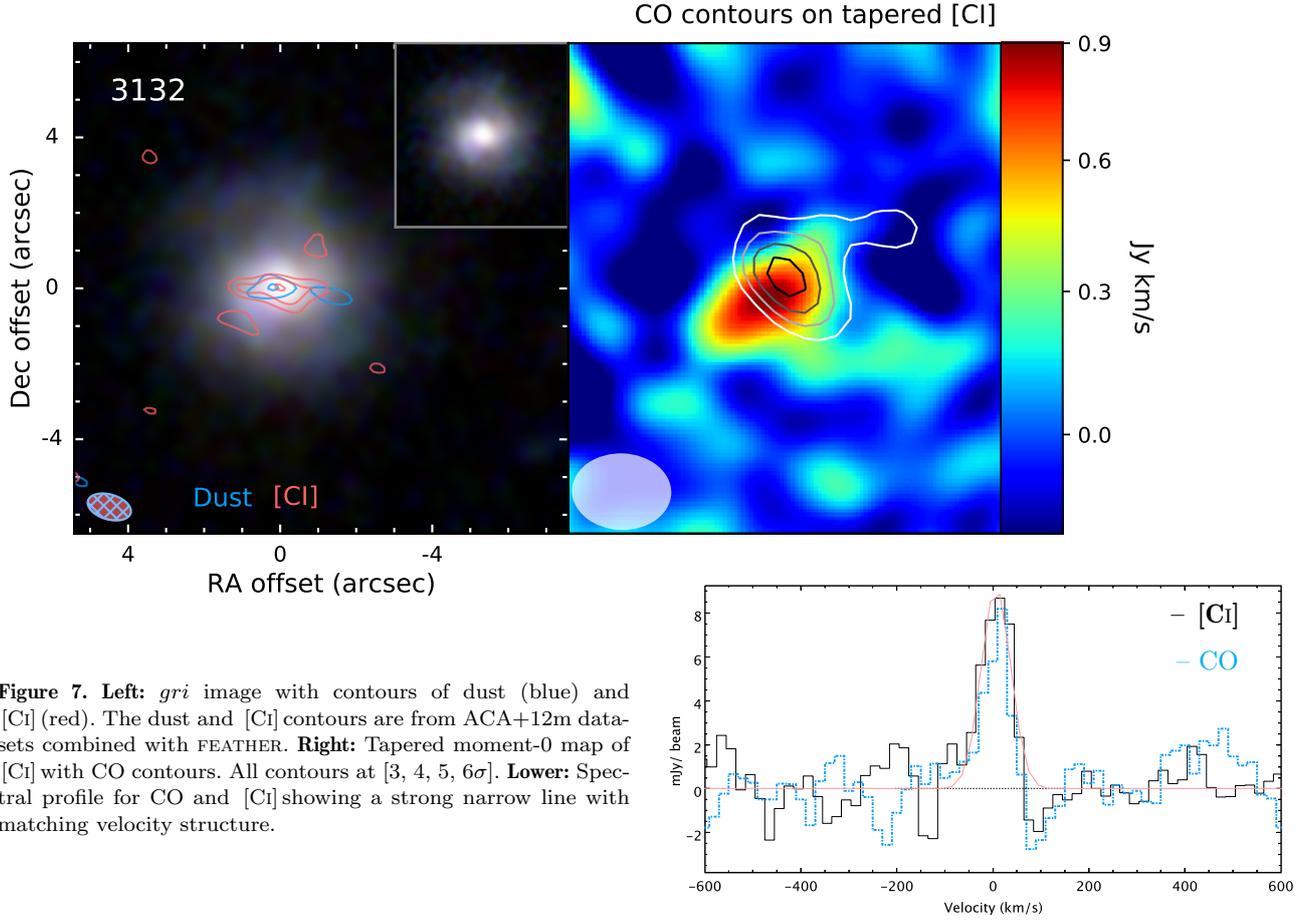

\begin{minipage}{0.98\linewidth}
\begin{lpic}[]{SDP3132_gri_feathered(0.8,0.8)}
\end{lpic}
\end{minipage}
\begin{minipage}{0.47\linewidth}
\caption{\label{3132F} {\bf Left:} $gri$ image with contours of dust
  (blue) and \CI (red). The dust and \CI contours are from ACA+12m
  data-sets combined with {\sc feather}.  {\bf Right:} Tapered
  moment-0 map of \CI with CO contours. All contours at [3, 4, 5,
    6$\sigma$]. {\bf Lower:} Spectral profile for CO and \CI showing a
  strong narrow line with matching velocity structure.}
\end{minipage}
\begin{minipage}{0.5\linewidth}
\begin{lpic}[l(0mm),r(0mm),t(-15mm),b(-10mm)]{3132_CICO_profile_fit(0.30,0.30)}
\lbl{250,160;\large{\bf -- \CI}}
\lbl{250,140;\large{{\textcolor{cyan}{ -- CO}}}}
\end{lpic}
\end{minipage}
\end{figure*}

\begin{figure*}
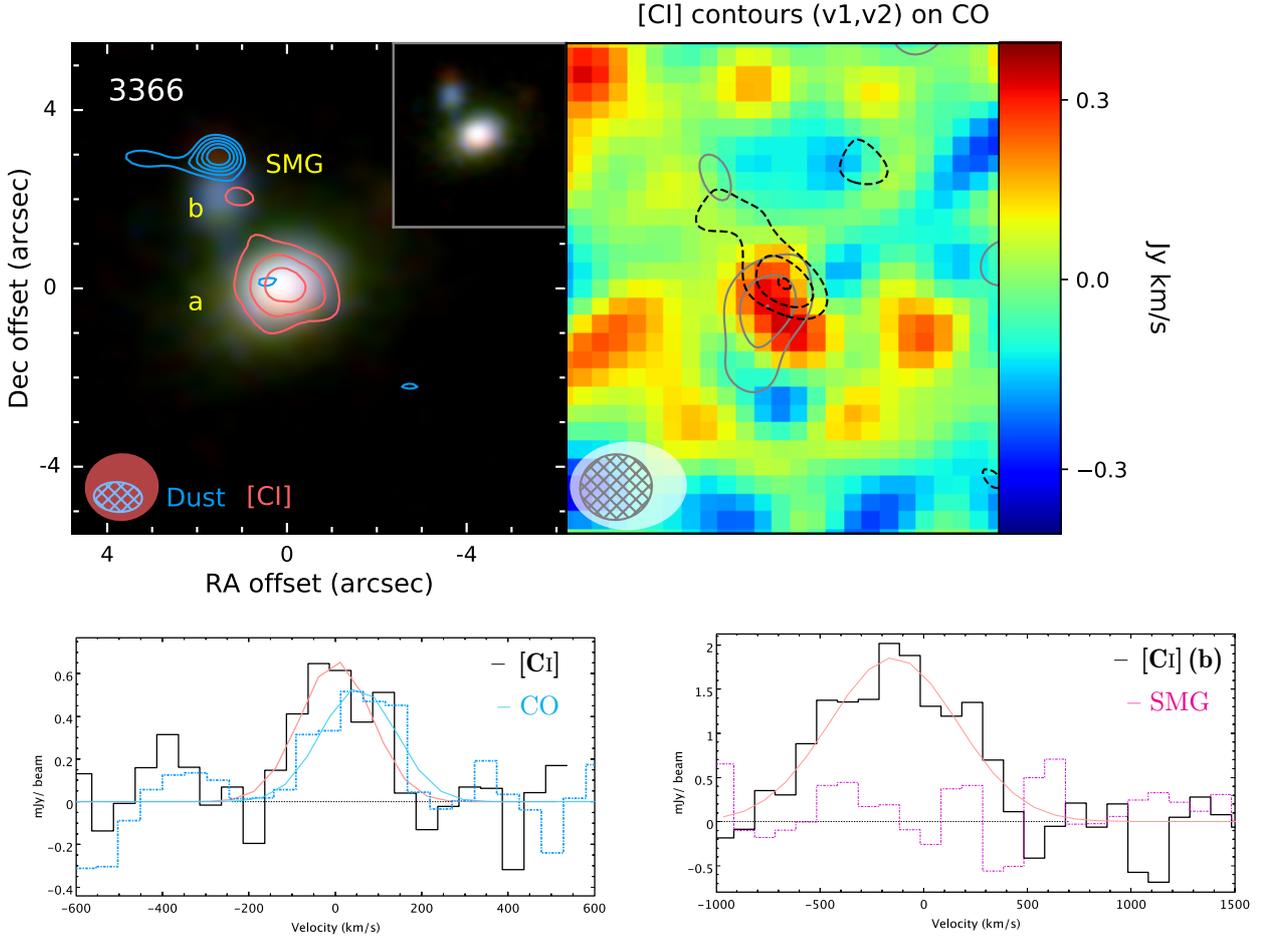

\begin{minipage}{0.98\linewidth}
\begin{lpic}[]{SDP3366_gri_revised(0.8,0.8)}
\end{lpic}
\end{minipage}
\begin{minipage}{0.47\linewidth}
\vspace{-0.6cm}
\begin{lpic}[l(0mm),r(0mm),t(-1mm),b(-3mm)]{3366_CICO_bc2_profile_fit(0.27,0.27)}
\lbl{250,160;\large{\bf -- \CI}}
\lbl{250,140;\large{{\textcolor{cyan}{ -- CO}}}}
\end{lpic}
\end{minipage}
\begin{minipage}{0.47\linewidth}
\begin{lpic}[l(0mm),r(0mm),t(-8mm),b(-3mm)]{3366_CIoffset_contoffset_profile_fit(0.27,0.27)}
\lbl{250,160;\large{\bf -- \CI (b)}}
\lbl{250,140;\large{{\textcolor{magenta}{-- SMG}}}}
\end{lpic}
\end{minipage}
\caption{\label{3366F} {\bf Left:} $Hrg$ image with contours of dust
  (blue) and \CI (red) at [3,4,5,6,7$\sigma$]. The bright continuum
  source to the NE of the target is an SMG coincident with an
  optically red source, clearly visible in the upper inset image. The
  target 3366.a and its satellite, 3366.b, have weak continuum
  emission (2-3$\sigma$ -- not all shown with these contours). The \CI
  contours on 3366.a are the full velocity range over which \CI was
  detected ($v1$ and $v2$) tapered with a 1.5\arcsec Gaussian, as
  shown by the beam in the lower left. The \CI contour for 3366.b is
  at full resolution (i.e. the same as the beam for the
  continuum). {\bf Right:} Moment-0 map of CO, which has a very weak
  detection (3$\sigma$) for 3366.a with a velocity profile matching
  that of the component $v1$ in \CI. There is no sign of any CO
  emission at the blue-shifted velocities associated with the
  companion. The contours are those of the two velocity components of
  \CI, tapered with a 1.5\arcsec Gaussian. Grey solid: the systemic
  component $v1$, black dashed: the blue-shifted component $v2$, which
  overlaps with the broad profile seen at the position of the
  companion galaxy 3366.b. Contours are at [2, 3, 4$\sigma$]. {\bf
    Lower left:} Spectral profile for CO and \CI in SDP.3366.a showing
  emission close to the systemic velocity (3366.a($v1$)). {\bf Lower
    right:} \CI profile for the companion source (3366.b) showing the
  much wider line which extends to the blue-shifted velocities from
  3366.a($v2$). Also shown is the spectrum extracted at the SMG
  position, the lack of signal further indicating its nature as a
  background source at a different redshift. Only the $v1$ component
  is considered in the analysis of gas calibration, though adding in
  the $v2$ component would not change any of our conclusions.}
\end{figure*}

\begin{figure*}
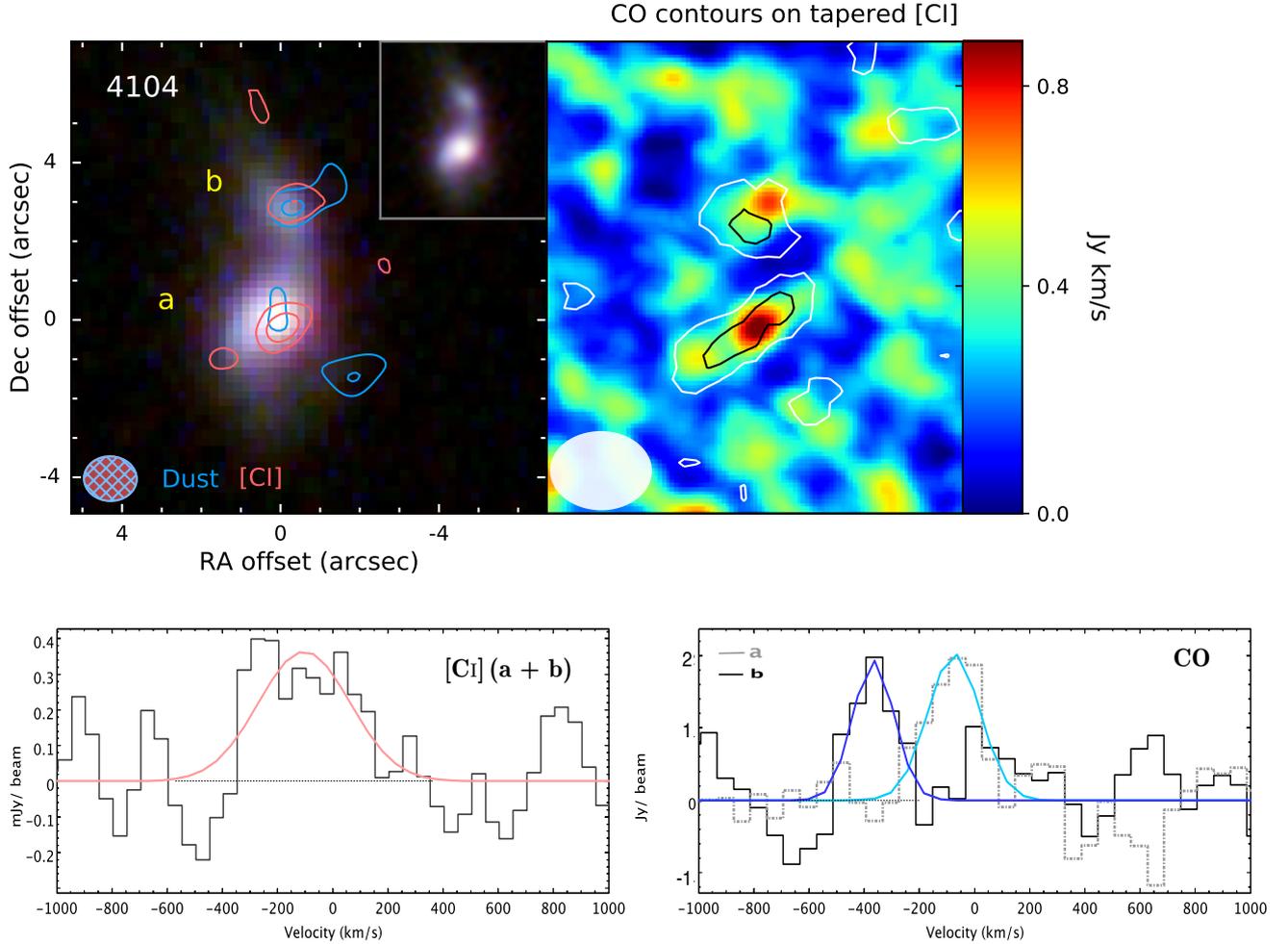

\begin{minipage}{0.98\linewidth}
\begin{lpic}[]{SDP4104_Zrg_reivsed(0.8,0.8)}
\end{lpic}
\end{minipage}
\begin{minipage}{0.49\linewidth}
\begin{lpic}[l(0mm),r(0mm),t(-12mm),b(-10mm)]{4104_CI50_both_bc3_profile_fit(0.3,0.35)}
\lbl{235,145;\large{\bf \CI (a + b)}}
\end{lpic}
\end{minipage}
\begin{minipage}{0.49\linewidth}
\begin{lpic}[l(0mm),r(0mm),t(-12mm),b(-10mm)]{4104_CO60_opt1_opt2_profile_fit(0.3,0.35)}
\lbl{255,150;\large{\bf CO}}
\end{lpic}
\end{minipage}
\caption{\label{4104F} {\bf Left:} $Zrg$ image with contours of dust
  (blue) and \CI (red), tapered with a 1\arcsec Gaussian. {\bf Right:}
  Moment-0 image of tapered \CI with CO contours. All contours at [2,
    3$\sigma$]. {\bf Lower left:} \CI line profile for both galaxies
  summed together. {\bf Lower right:} CO line profile for each source
  individually. See notes in Appendix~\ref{sourcesS} for details on the
  difficulties with measurement of \CI in this source.}
\end{figure*}

\begin{figure*}
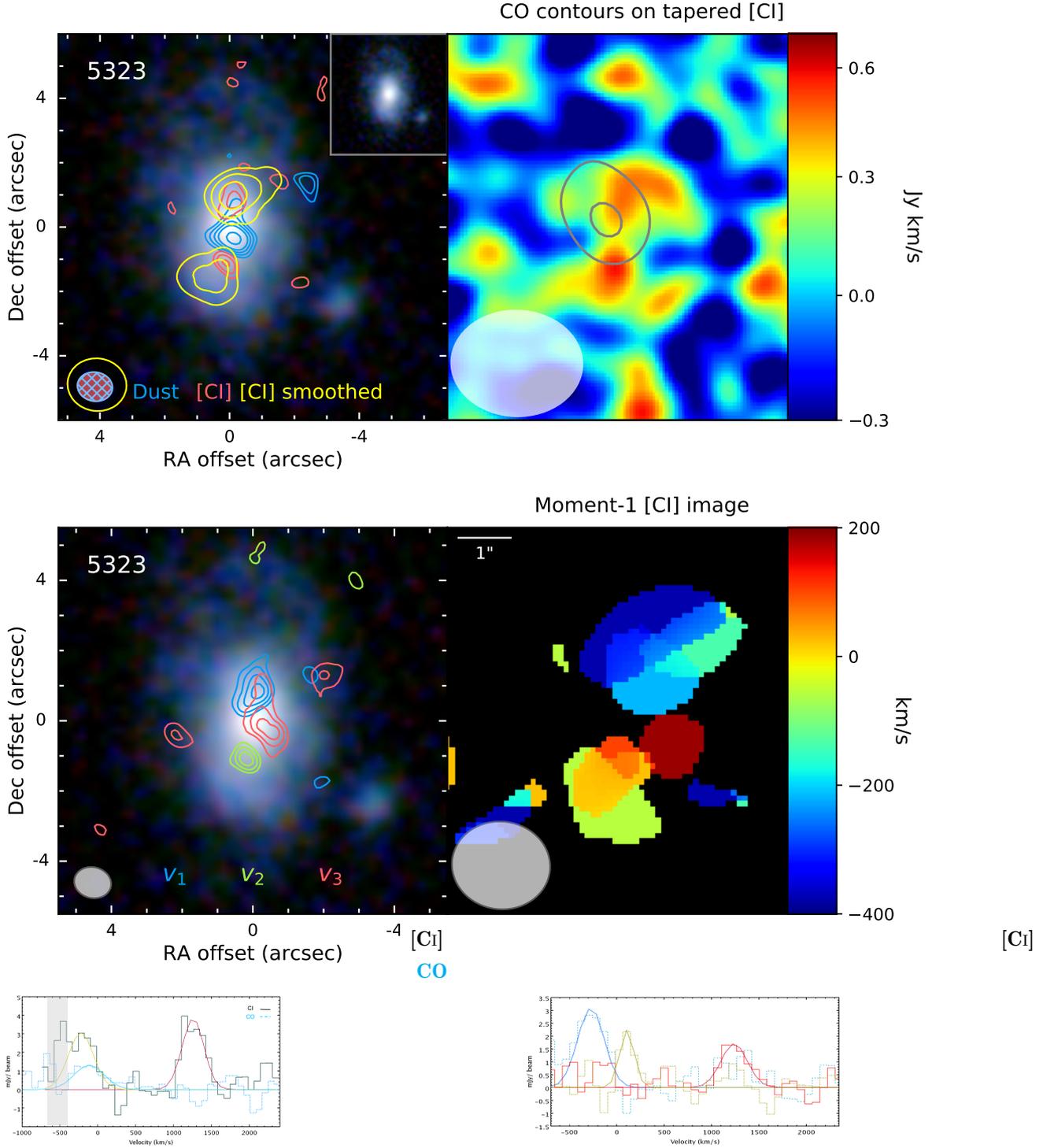

\begin{minipage}{0.98\linewidth}
\begin{lpic}[]{SDP5323_gri_revised(0.8,0.8)}
\end{lpic}
\end{minipage}
\begin{minipage}{0.98\linewidth}
\begin{lpic}[]{SDP5323_gri_allvel_revised(0.8,0.8)}
\end{lpic}
\end{minipage}
\begin{minipage}{0.49\linewidth}
\begin{lpic}[l(0mm),r(0mm),t(-3mm),b(-1mm)]{5323_4_11_tap3_plusCO(0.17,0.17)}
\lbl[W]{430,230;\large{\bf \CI}}
\lbl[W]{435,200;\large{{\textcolor{cyan}{{\bf CO}}}}}
\end{lpic}
\end{minipage}
\begin{minipage}{0.49\linewidth}
\begin{lpic}[l(0mm),r(0mm),t(-3mm),b(-1mm)]{5323_sep_CIcomps(0.19,0.17)}
\lbl{440,230;\large{\bf \CI}}
\end{lpic}
\end{minipage}
\caption{\label{5323F} {\bf Top left:} $irg$ image with contours of
  dust (blue) and \CI (red: same resolution, yellow: smoothed by
  1.5\arcsec).  {\bf Top right:} CO contours on tapered \CI moment-0
  map. Only the $v1$ and $v2$ \CI components are shown in this top
  row. Contours are [2.5, 3.0, 3.5, 4.0]$\sigma$ for \CI and dust and
  [2, 3]$\sigma$ for CO.  {\bf Middle left:} Velocity components of
  \CI, $v1:-415\rightarrow+15\kms$, $v2:-15\rightarrow+225\kms$ and
  $v3:+945\rightarrow+1645\kms$ ($v3$ is the high velocity component
  which has no CO counterpart and we cannot be sure of its
  origin). Contours are the same as the top row. {\bf Middle right:}
  velocity weighted moment map masked at 2$\sigma$ for the components
  $v1$ and $v2$ which we are certain are associated with SDP.5323.
  {\bf Lower left:} \CI and CO line profiles measured from a heavily
  tapered cube showing all three velocity features. {\bf Lower right:}
  \CI line profiles at the peak of each velocity feature. In the
  analysis of gas calibration only the $v1$ and $v2$ \CI components
  are used.}
\end{figure*}

\begin{figure*}
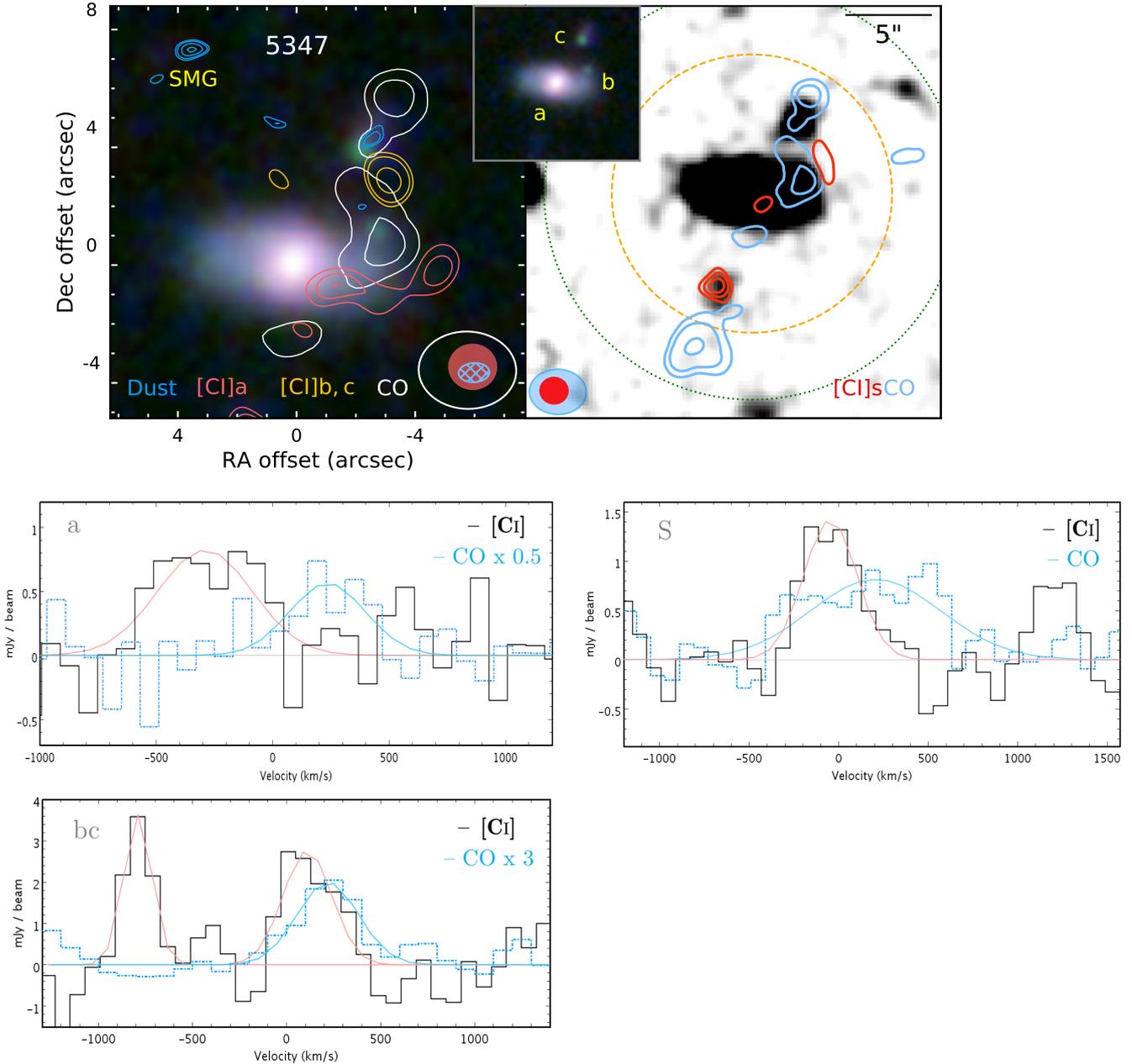

\begin{minipage}{0.98\linewidth}
\begin{lpic}[]{SDP5347_Krg_revised(0.8,0.8)}
\end{lpic}
\end{minipage}
\begin{minipage}{0.48\linewidth}
\begin{lpic}[l(-10mm),r(0mm),t(0mm),b(0mm)]{5347_COCI_a(0.18,0.23)}
\lbl{450,175;\large{\bf -- \CI}}
\lbl{440,155;\large{{\textcolor{cyan}{ -- CO x 0.5}}}}
\lbl{80,175;\Large{{\textcolor{gray}{a}}}}
\end{lpic}
\end{minipage}
\begin{minipage}{0.49\linewidth}
\begin{lpic}[l(-5mm),r(0mm),t(0mm),b(0mm)]{5347_COCI_Sblob(0.2,0.22)}
\lbl{400,180;\large{\bf -- \CI}}
\lbl{400,160;\large{{\textcolor{cyan}{ -- CO}}}}
\lbl{80,180;\Large{\textcolor{gray}{S}}}
\end{lpic}
\end{minipage}

\begin{minipage}{0.98\linewidth}
\begin{lpic}[l(-10mm),r(0mm),t(0mm),b(0mm)]{5347_COCI_bc(0.21,0.23)}
\lbl{380,160;\large{\bf -- \CI}}
\lbl{380,140;\large{{\textcolor{cyan}{ -- CO x 3}}}}
\lbl{80,160;\Large{\textcolor{gray}{bc}}}
\end{lpic}
\end{minipage}
\caption{\label{5347F} {\bf Top left:} $Krg$ image with contours of
  dust (blue), \CI (red is the $a$ component and orange is {\em bc})
  and CO (white). This is a very complex source (see
  Appendix.~\ref{sourcesS}). \CI is tapered by 1.5\asec, CO by
  2\asec. CO, dust and \CI{\em bc} (orange) contours at [3.0, 3.5,
    4.5$\sigma$], \CI$a$ (red) contours at [2.0, 2.5,
    3.0$\sigma$]. {\bf Right:} A highly smoothed KIDs $g$-band image
  over a wider area showing CO and \CI emission to the South. The \CI
  peak coincides with a low surface brightness patch of blue optical
  emission. The orange dashed circle denotes the radius at which the
  primary beam attenuation is a factor of 2, while the green dashed
  line shows the radius imaged in \CI. {\bf Center left:} \CI and CO
  profiles for the $a$ component. Apertures are used in each case to
  extract the spectra, as the emission is extended and diffuse. There
  is a significant detection in \CI but the CO is much weaker and also
  significantly shifted in velocity. {\bf Lower left:} Spectral
  profile of the \CI{\em bc} component, extracted at the peak of the
  \CI emission. The CO emission does not peak at the same location,
  and we show the CO profile from the full region spanning sources $b$
  and $c$ (scaled by a factor 3 for presentation). The velocity
  component at 100--200\kms is consistent in both CO and \CI but there
  is an additional velocity component at negative velocities which is
  only seen in \CI. We list the fluxes for each velocity component
  separately in Table~\ref{CIdataT}. {\bf Centre right:} Spectral
  profiles for the Southern emission, shown in the top right
  panel. The CO and \CI peaks are not in exactly the same location,
  however, the \CI field of view is not large enough to have detected
  comparable emission at the location of the CO peak, which lies out
  beyond the HPBW of the primary beam. We show the spectra together to
  illustrate that there is overlap in their spectral profiles, but the
  CO profile is much wider than that of the \CI. We use the \CI and CO
  emission from the a,b,c components in the analysis on gas calibration.}

\end{figure*}

\begin{figure*}
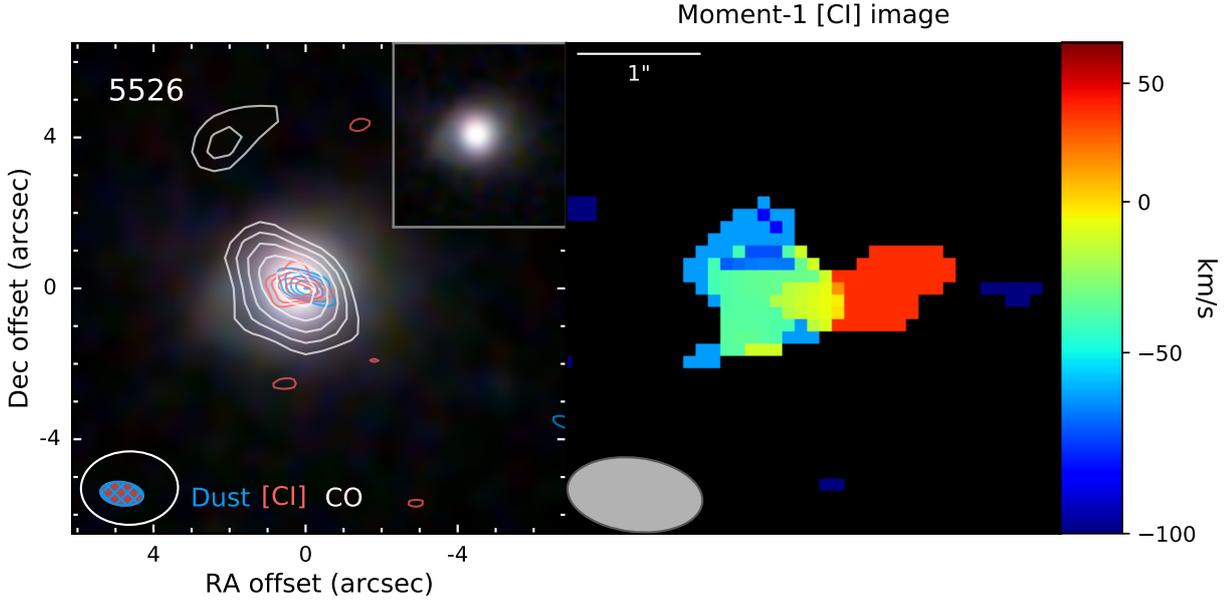

\begin{minipage}{0.98\linewidth}
\begin{lpic}[]{SDP5526_irg_revised(0.8,0.8)}
\end{lpic}
\end{minipage}
\begin{minipage}{0.47\linewidth}
\caption{\label{5526F} {\bf Left:} $gri$ image with contours of dust (blue), \CI (red)
  and CO (white) at [3, 4, 5, 6, 7$\sigma$]. {\bf Right:} Velocity-weighted \CI
  moment map masked at 3$\sigma$. {\bf Lower:} \CI (black) and CO (blue) line
  profiles with similar velocity structure. There is a hint that the
  \CI line may be asymmetric.}
\end{minipage}
\begin{minipage}{0.5\linewidth}
\begin{lpic}[l(0mm),r(0mm),t(-10mm),b(-5mm)]{5526_CICO_han_profile_fit(0.3,0.3)}
\lbl{250,160;\large{\bf -- \CI}}
\lbl{250,140;\large{{\textcolor{cyan}{ -- CO}}}}
\end{lpic}
\end{minipage}
\end{figure*}

\begin{figure*}
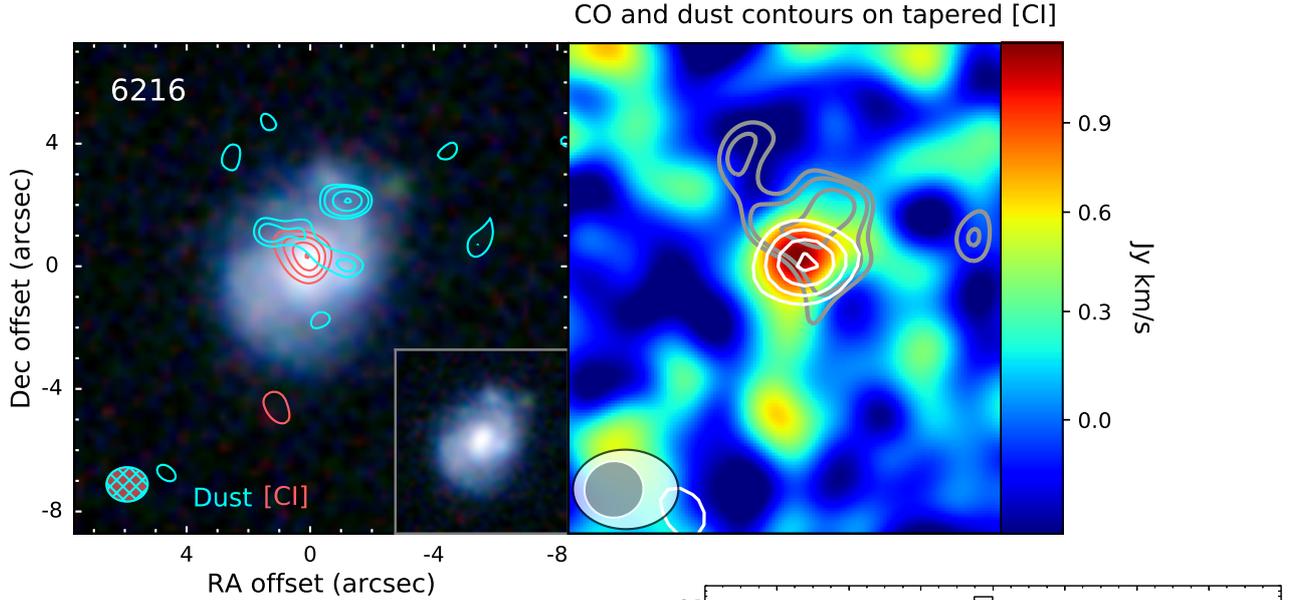

\begin{minipage}{0.98\linewidth}
\begin{lpic}[]{SDP6216_irg_revised(0.8,0.8)}
\end{lpic}
\end{minipage}
\begin{minipage}{0.47\linewidth}
\caption{\label{6216F} {\bf Left:} $gri$ image with contours of dust
  (blue) and \CI (red) tapered with a 1\asec Gaussian. The optical
  image shows a sharp stellar caustic to the SE and disturbed clumpy
  star forming regions to the North. {\bf Right:} CO (white) and dust
  (grey) on moment-0 image of \CI. The dust and \CI in this image are
  tapered with a 2\asec Gaussian. Contours of CO and \CI are at [3, 4,
    5, 6$\sigma$] with the continuum starting at 2.5$\sigma$. The CO and \CI
  emission are both peaked strongly in the centre of the galaxy. In
  contrast the 850\mic dust emission is spread around the disk in
  clumps. {\bf Lower:}  \CI (black) and CO (blue) spectral profiles.}
\end{minipage}
\begin{minipage}{0.5\linewidth}
\begin{lpic}[l(0mm),r(0mm),t(-15mm),b(-5mm)]{6216_CICO_profile_fit(0.3,0.3)}
\lbl{250,160;\large{\bf -- \CI}}
\lbl{250,140;\large{{\textcolor{cyan}{ -- CO}}}}
\end{lpic}
\end{minipage}
\end{figure*}

\begin{figure*}
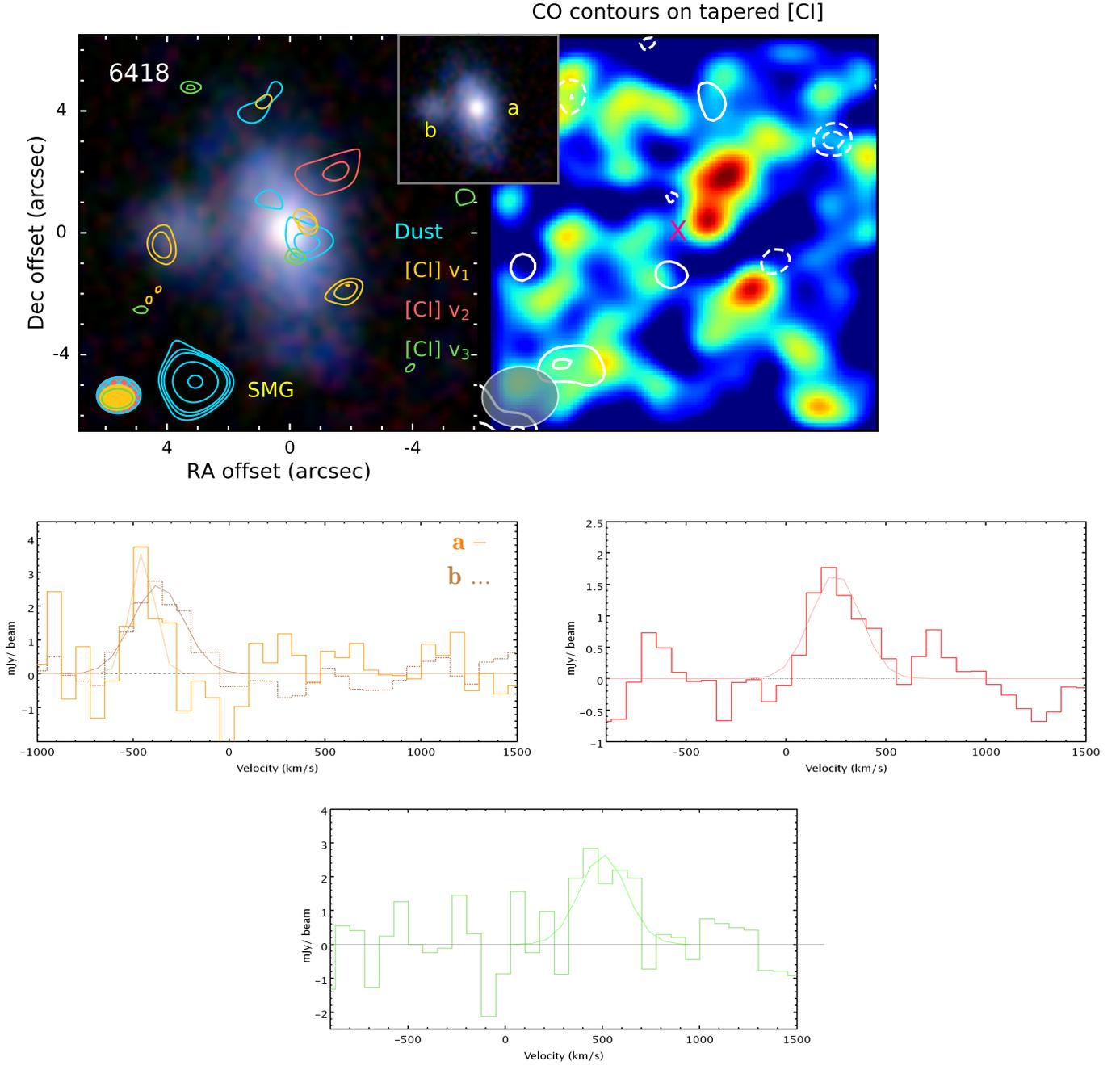

\begin{minipage}{0.98\linewidth}
\begin{lpic}[]{SDP6418_irg_excomps_revised(0.8,0.8)}
\end{lpic}
\end{minipage}
\begin{minipage}{0.49\linewidth}
\begin{lpic}[l(-5mm),r(0mm),t(-0mm),b(0mm)]{6418_minus500_both2(0.185,0.17)}
\lbl{430,230;\Large{{\textcolor{orange}{{\bf a} --}}}}
\lbl{430,200;\Large{{\textcolor{brown}{{\bf b} ...}}}}
\end{lpic}
\end{minipage}
\begin{minipage}{0.49\linewidth}
\begin{lpic}[l(0mm),r(0mm),t(-0mm),b(0mm)]{6418_plus200(0.185,0.17)}
\end{lpic}
\end{minipage}
\begin{minipage}{0.5\linewidth}
\begin{lpic}[l(0mm),r(0mm),t(0mm),b(0mm)]{6418_plus500(0.18,0.17)}
\end{lpic}
\end{minipage}
\caption{\label{6418F} {\bf Left:} $irg$ image with contours of dust
  smoothed by 2\asec (blue) and \CI tapered to 1\asec. The colours of
  the \CI contours denote the velocity components. Orange: $v1 =
  -450\kms$, red: $v2=+250\kms$, green: $v3 = +500\kms$. Dust and \CI
  contours are at $[3, 4, 5 \sigma]$. There is no line emission at the
  systemic velocity of the optical galaxy. {\bf Right:} CO contours at
  $[-2.5, -2.0, 2.0, 2.5\sigma$] on tapered moment-0 \CI image. The
  pink `X' marks the centre of the main optical galaxy `a'. Due to the
  difficulty of showing the three components together in a moment-0
  map, we have created this image by combining the velocity ranges of
  $v1$ and $v2$ but including only the positive pixels in the moment-0
  image. Thus this image is illustrative of where the \CI peaks are
  in these velocity components but flux cannot be measured from this
  image. {\bf Lower panels:} Spectral profiles for the three \CI
  velocity components, colour coded to match the contours. The
  $v1=-450\kms$ component is also present in the companion optical
  galaxy `b', shown as the dotted brown line in the first spectral
  profile. We use two values for the \CI luminosity in the gas
  calibration analysis, a conservative value just using the $v2$
  component in the main 6418.a galaxy, and a total value which uses
  all three components.}
\end{figure*}

\begin{figure*}
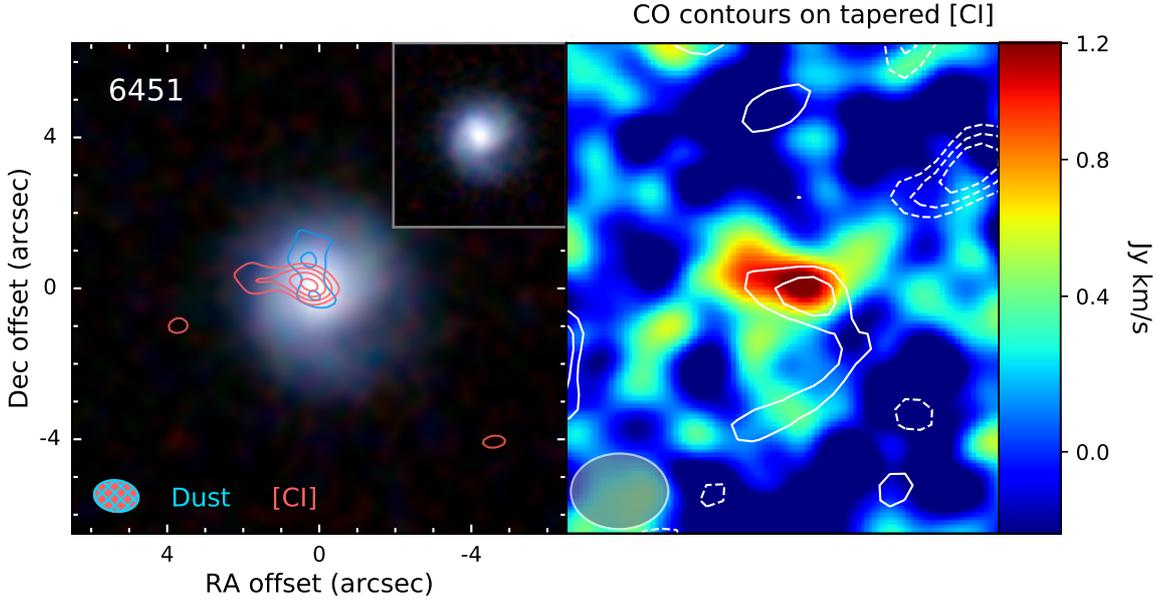

\begin{minipage}{0.98\linewidth}
\begin{lpic}[]{SDP6451_irg_revised(0.8,0.8)}
\end{lpic}
\end{minipage}
\begin{minipage}{0.47\linewidth}
\caption{\label{6451F} {\bf Left:} $irg$ image with contours of dust
  (blue) and \CI (red) at [3, 4, 5, 6$\sigma$]. {\bf Right:} Moment-0
  map of \CI smoothed with a 1\arcsec Gaussian and scaled so that the
  lower level emission is evident ($2\sigma=0.4$ Jy/\kms). CO contours
  are shown at intervals of
  [$-2.5$, $-2$, $-1.5$, 1.5, 2$\sigma$]. While the CO is very low
  significance, it does match the \CI reasonably well in position and
  morphology, even to the extent that both gas tracers appear to
  delineate a spiral arm to the South. {\bf Lower:} \CI (black) and CO (blue) line profiles.}
\end{minipage}
\begin{minipage}{0.5\linewidth}
\begin{lpic}[l(0mm),r(0mm),t(-10mm),b(-5mm)]{6541_CICO_profile_fit_nosm(0.3,0.3)}
\lbl{250,160;\large{\bf -- \CI}}
\lbl{250,140;\large{{\textcolor{cyan}{ -- CO}}}}
\end{lpic}
\end{minipage}
\end{figure*}

We detect a total \CI flux above 3$\sigma$ for all 12 sources, for
dust continuum the number is 11 and for CO we detect at this level in
10 sources. There are a wide variety of morphologies
and relative luminosities. The optical images with ALMA
signal-to-noise contours superimposed are shown in
Figures~\ref{163F}--\ref{6451F} along with the CO and \CI spectral
profiles. We comment in detail on each source in
Appendix~\ref{sourcesS}. 


Half of the fields contained one or more serendipitously detected high
redshift dusty star-forming galaxies, identifiable as point sources
with SNR$>5$ and having either no optical/NIR counterpart, or a very
red K-band detection in the VIKING imaging. In total 7 high redshift
dusty galaxies are confidently detected at $S_{850}>5\sigma$ in 6
fields. A full investigation of these sources, which constitute an
overdensity relative to the blank field counts of a factor 4--6 is
given by \citet{Dunne2020smg}.

\subsection{Physical parameters from SED fitting}
We have estimated physical parameters for the galaxies using the
energy balance SED fitting code {\sc magphys}, which uses libraries of
optical and infrared SEDs with parameters drawn stochastically from
physically motivated priors. Details of the fitting method are
presented in \citet{DaCunha2008}. We use extended infrared libraries
which have an ISM cold temperature range of 10--30~K
\citep{Rowlands2014,daCunha2015}, as several of the sources favoured
warmer ISM temperatures than allowed in the standard {\sc magphys}
infrared libraries. We fit the FUV--850\mic photometry (typically 22
or 23 bands), including an additional 7\% error in quadrature to
incorporate calibration, zero point and model scatter. The FUV-22\mic
photometry is taken from the LAMBDAR catalogue \citep{Wright2016},
while the {\em Herschel} photometry is listed in
Table~\ref{photomT}. We take the {\em IRAS} {\sc scanpi}
\citep{scanpi} upper limits at 60\mic and use them as a constraint on
the fitting. In several cases there is contamination of the SPIRE
photometry from background sources at higher redshift in the {\em
  Herschel} beam. Sometimes the contaminant source is visible in the
ALMA continuum images, other times it is evident from the
250\mic/350\mic and 350\mic/500\mic colours that another higher
redshift source must be contributing to the SPIRE fluxes. We describe
in Appendix~\ref{boostS} and in \citeauthor{Dunne2020smg} how we
estimate the range of possible contamination of the {\em Herschel}
fluxes, in order to correct them. Where there is evidence for
contamination, we have fitted the photometry with and without making
this correction to check the impact on the SED. {\em We find that,
  while in all cases making these corrections improves the fit to the
  SED, the changes to the estimated parameters are very small and
  within the original 1$\sigma$ uncertainties in all cases. Therefore
  the choice of whether to correct, or how to correct has no bearing
  on our findings in the next section.} Our companion paper provides
further detail and also investigates the wider implications of this
boosting for cases where ALMA data are not present.

We show the best fit SED and parameter PDFs in
Figure~\ref{A163F}-\ref{A6418F}. Alongside the best-fit SED in red we
also show 200 of the SEDs which are strongly contributing to the
PDFs. The median parameters and their uncertainties are listed in
Table~\ref{magphysT}.

\begin{table*}
\caption{\label{photomT} FIR photometry for the galaxies from {\rm
    Herschel-ATLAS} DR1 and our ALMA measurements.}
\begin{tabular}{lcccccccccccc}
\hline
Source & $S_{100}$ & $\sigma_{100}$ & $S_{160}$ & $\sigma_{160}$ & $S_{250}$ & $\sigma_{250}$ & $S_{350}$ & $\sigma_{350}$ & $S_{500}$ & $\sigma_{500}$ & $S_{850}$ & $\sigma_{850}$\\
\hline
163 & 73.9 & 20.7 & 102.1  & 24.1 & 107.6 & 7.3 & 50.7 & 8.1 & 23.9 & 8.5 & 3.05 & 0.34\\
1160$^\dag$ & 49.8 & 17.5 & 57.1 & 19.7 & 48.4 & 7.2 & 32.4 & 8.1 & 21.6 & 8.7 & 0.54 & 0.12 \\
2173$^\dag$ & 59.0$^M$ & 34.0 & 68.0$^M$ & 38.0 & 46.2 & 6.5 & 21.4 & 7.5 & 11.1 & 7.8 & 0.74 & 0.20\\
3132 & 64.3$^M$ & 26.5 & 65.1$^M$ & 20.0 & 40.6 & 6.4 & 23.6 & 7.4 & 13.1 & 7.8 & 0.95 & 0.29\\
3366$^\dag$ & 19.7 & 24.5 & 114 & 37.6 & 40.3 & 7.3 & 26.3 & 8.0 & 16.5 & 8.8 & 0.42 & 0.019\\
4104$^\dag$ & 77.9 & 17.6 & 53.3 & 26.4 & 46.2 & 7.2 & 28.3 & 8.1 & 12.1 & 8.8 & 0.92 & 0.24\\
5323 &  ..  & ..   & ..   & ..   & 28.6 & 7.1 & 30.0 & 8.0 & 9.6 & 8.4 & 0.89 & 0.24\\
5347$^\dag$ & 33.0 & 41.2 & 68.0 & 17.7 & 32.7 & 7.5 & 29.8 & 8.2 & 17.2 & 8.7 & 1.84 & 0.35\\
5526 & 62.0$^M$ & 32.6 & 56.9$^M$ & 40.7 & 31.2 & 7.3 & 20.0 & 8.2 & $-$10.7 & 8.5 & 1.11 & 0.27\\
6216 & 35.0$^M$ & 36.6 & 34.0$^M$ & 40.7 & 36.2 & 7.3 & 19.9 & 8.1 & 3.8 & 8.8 & 1.49 & 0.25\\
6418$^\dag$ & 40.4$^M$ & 19.8 & 27.0$^M$ & 33.3 & 31.6 & 7.3 & 18.5 & 8.0 & 16.6 & 8.4 & 1.06 & 0.32\\
6451$^\dag$ & 69.4 & 41.7 & 59.5 & 47.8 & 33.7 & 7.3 & 29.7 & 8.2 & 19.5 & 8.6 & 0.88 & 0.19\\
\hline
\end{tabular}
\flushleft{\small{\bf Notes:} Fluxes are all in mJy. $^{\dag}$
  indicates that there is evidence for contamination of these {\em
    Herschel} fluxes by high-z SMG in the beam. Before SED fitting we
  subtract from these fluxes the estimated contamination from high-z SMG listed in
  Table~\ref{SMGcontamT}. See Appendix~\ref{boostS} for details. $^M$ indicates a
  PACS flux re-measured from the DR1 maps. }
\end{table*}

\begin{table*}
\caption{\label{magphysT} Luminosities and derived physical properties from {\sc magphys} SED fitting.}
\centering
\renewcommand{\arraystretch}{1.4}
\begin{tabular}{lcccccccc}
\hline
Source &
\multicolumn{1}{c}{Log \lci} &
  \multicolumn{1}{c}{Log \lcoa} &
  \multicolumn{1}{c}{Log \lsub} &
Log \Ms & Log \Lir & Log \Md & log sSFR & $T_C$\\ 
   &
 \multicolumn{1}{c}{($\rm{K\,kms^{-1}\,pc^2}$)} &
 \multicolumn{1}{c}{($\rm{K\,kms^{-1}\,pc^2}$)} &
\multicolumn{1}{c}{($\rm{W\,Hz^{-1}}$)} &
  \multicolumn{1}{c}{(\msun)} &
  \multicolumn{1}{c}{(\lsun)} &
  \multicolumn{1}{c}{(\msun)} &
  \multicolumn{1}{c}{($\rm{yr^{-1}}$)} &
  \multicolumn{1}{c}{(K)} \\       
\hline
163 & $9.46\pm0.07$ & $10.01\pm0.08$ & $23.51\pm0.05$ &  $11.46^{+0.10}_{-0.10}$ & $11.66^{+0.04}_{-0.04}$ & $8.60^{+0.05}_{-0.06}$ & $-10.12^{+0.1}_{-0.2}$ & $22.5^{+0.7}_{-0.8}$\\
1160$^\dag$ &  $9.02\pm0.04$ & $9.65\pm0.08$ & $22.75\pm0.10$ & $11.04^{+0.11}_{-0.10}$ & $11.35^{+0.06}_{-0.10}$ & $7.73^{+0.08}_{-0.08}$ & $-9.98^{+0.2}_{-0.2}$ & $28.3^{+1.1}_{-1.8}$\\
2173$^{\dag}$ &  $9.06\pm0.06$ & $9.73\pm0.10$ & $22.91\pm0.12$ & $10.89^{+0.09}_{-0.07}$ & $11.30^{+0.10}_{-0.08}$ & $7.94^{+0.12}_{-0.12}$ & $-9.58^{+0.1}_{-0.1}$ & $24.5^{+2.5}_{-2.3}$ \\ 
3132 & $8.91\pm0.07$ & $9.65\pm0.07$ & $23.03\pm0.13$ & $10.74^{+0.09}_{-0.09}$ & $11.49^{+0.08}_{-0.07}$ & $8.01^{+0.08}_{-0.07}$ & $-9.33^{+0.15}_{-0.15}$ & $25.9^{+2.0}_{-1.5}$\\
3366$^\dag$ & $8.79\pm0.12$ & $9.38\pm0.14$ & $22.68\pm0.20$ & $11.21^{+0.12}_{-0.13}$ & $11.13^{+0.10}_{-0.15}$ & $7.71^{+0.15}_{-0.14}$ & $-10.67^{+0.15}_{-0.3}$ & $26.5^{+2.3}_{-2.7}$ \\
4104$^{\dag}$ & $8.96\pm0.10$ & $10.02\pm0.08$ & $22.98\pm0.11$ & $10.93^{+0.08}_{-0.09}$ & $ 11.45^{+0.11}_{-0.09}$ & $7.87^{+0.11}_{-0.11}$ & $-9.68^{+0.2}_{-0.1}$ & $27.7^{+1.3}_{-2.3}$\\
5323 &  $8.91\pm0.08$ & $9.71\pm0.11$ & $23.00\pm0.12$ & $11.05^{+0.05}_{-0.05}$ & $11.17^{+0.12}_{-0.08}$ & $8.05^{+0.18}_{-0.17}$ & $-9.88^{+0.15}_{-0.1}$ & $21.8^{+3.6}_{-2.7}$\\
5347$^{\dag}$ & $9.08\pm0.10$ & $10.05\pm0.10$ & $23.34\pm0.08$ & $11.17^{+0.09}_{-0.10}$ & $11.11^{+0.11}_{-0.12}$ & $8.38^{+0.14}_{-0.16}$ & $-10.32^{+0.2}_{-0.15}$ & $19.9^{+2.2}_{-1.9}$ \\
5526 & $8.86\pm0.06$ & $9.80\pm0.08$ & $23.15\pm0.09$ & $10.82^{+0.03}_{-0.09}$ & $11.33^{+0.07}_{-0.11}$ & $7.98^{+0.13}_{-0.11}$ & $-9.63^{+0.15}_{-0.1}$ & $24.4^{+2.5}_{-2.8}$ \\
6216 & $8.85\pm0.08$ & $9.65\pm0.07$ & $23.22\pm0.07$ & $10.58^{+0.07}_{-0.06}$ & $11.32^{+0.08}_{-0.10}$ & $8.25^{+0.18}_{-0.18}$ & $-9.18^{+0.1}_{-0.2}$ & $20.0^{+3.5}_{-2.4}$ \\
6418$^{\dag}$ & $8.97\pm0.08$ & $9.23\pm0.23$ & $23.07\pm0.13$ & $10.79^{+0.12}_{-0.08}$ & $11.28^{+0.10}_{-0.15}$ & $8.00^{+0.19}_{-0.18}$ & $-9.58^{+0.15}_{-0.1}$ & $23.4^{+3.5}_{-2.9}$  \\
6451$^{\dag}$ & $9.19\pm0.09$ & $9.18\pm0.26$ & $22.98\pm0.09$ & $10.75^{+0.08}_{-0.07}$ & $11.36^{+0.11}_{-0.12}$ & $7.96^{+0.15}_{-0.13}$ & $-9.48^{+0.15}_{-0.15}$ & $25.0^{+3.0}_{-2.9}$ \\
\hline
\end{tabular}
\flushleft{\small{\bf Notes on Luminosities} 3366: \CI is from
  3366.a.v1 only. 4104: the sum of both members of the pair as
  measured with the missing channels. 5323: \CI is the sum of the two
  velocity components $v1$ and $v2$ from Table~\ref{CIdataT}. 5347:
  sum of the massive galaxy and the two satellites which have dust
  and/or \CI and CO emission (5347a--c in Table~\ref{CIdataT}). 6216:
  flux at 850\mic is that from the main disk including the compact
  clump discussed in Appendix.~\ref{sourcesS}. 6418: \CI flux is only
  that of the $v2$ component in the main galaxy. {\bf Notes on {\sc
      magphys} parameters:} $^\dag$ Fits use sub-mm fluxes from
  Table~\ref{photomT} which are then corrected for high-z
  contamination where required (see Appendix~\ref{boostS}). The
  differences in the fitted parameters using fluxes which are
  corrected for estimated contamination by dusty high redshift
  galaxies detected by ALMA within the SPIRE beam is well within the
  $1\sigma$ error and so the consideration of contamination has a
  negligible effect on the properties derived here. The dust mass
  estimates using the ALMA data in addition to the {\em Herschel}
  fluxes are considerably lower than when using the {\em Herschel}
  data alone.}
\end{table*}

\subsection{Kinematics}
The kinematics of the sources were more complex than anticipated in
many cases. In seven cases, there is a single optical galaxy where CO
and \CI have similar line-widths and velocities and similar
morphology.\footnote{As far as we can tell with the resolution
  differences between 85 GHz and 365 GHz.} We call these the Simple Kinematic
(SK) sources. The other five sources have disturbed kinematics (DK),
four of which (SDP.3366, SDP.4104, SDP.5347, SDP.6418) have a smaller
neighbouring optical source which shows either CO, \CI or both at a
similar velocity, indicating tidal interaction. SDP.5323 has a very
complex velocity field in \CI with three components, and may be
hosting an outflow or be involved in an interaction with a satellite.

\subsection{Gas and dust morphology}
\label{morphS}
The ALMA Band 7 data provide 850\mic (rest-frame 620\mic) dust and \CI line
imaging at a common angular resolution ($0.6\times 1.1\asec$) and with the same ($u,v$)
coverage, so that morphological comparisons between the \CI and dust
are meaningful. The CO imaging is at a very different frequency, which
translates into different sensitivity to spatial scales within the
interferometer. The smallest spatial scales probed are set by
the longest baselines, and the synthesised beam with natural weighting gives a physical
resolution of 10 kpc for CO(1--0), and 4 kpc for \CI(1--0) and
dust. The CO is unresolved or very marginally resolved in all sources,
apart from SDP.5347, where there appears to be some large scale tidal debris.

The Maximum Recoverable Scale (MRS) for the CO imaging is of order
20\asec (100 kpc), while for \CI and dust it is only $\sim 5$\asec (25
kpc), where we are quoting the quantity $0.5\times \lambda/\sqrt{u_m^2
  + v_m^2}$).\footnote{At the time of the Cycle 1 call, the proposal
  guide gave an MRS of 25\asec and 7.1\asec respectively for the two
  bands. This, plus the severe restriction on observing time for Cycle
  1 meant that we did not request ACA data in the first instance.} The
optical diameter of the galaxies ranges from 4-7\asec (20--35 kpc at
this redshift), and we did not expect the molecular gas to be
distributed smoothly on scales this large.  However this may not be so
for the CI-bright $\rm H_2$ distribution as CO-poor gas will be
preferentially be at larger galactocentric distances because of
metallicity gradients and/or lower gas densities allowing FUV
radiation and/or cosmic rays to shift the $\rm{[C^0/CO]}$ relative abundance
towards higher values in outer regions. Moreover cold dust is
typically found well-beyond optical disks and even well-concomitant
with HI \citep{Alton1998,Thomas2002,smith16disc,casasola17}. A
subsequent ACA proposal to image those sources with the most extended
emission was approved in the Cycle 7 supplementary call, however, only
one source had data taken before ALMA ceased operations in March
2020. This source (SDP.3132) shows a small (10 percent) increase in
\CI flux once the ACA data are included, but a much larger (60
percent) increase in the dust flux, from $\rm{S_{12m}=0.95\pm0.29}$ to
$\rm{S_{12m+ACA}}=1.51\pm0.39$.\footnote{While the fractional increase
    is large, the uncertainties mean that the actual difference is not
    statistically very significant.} This is not unexpected as the
  fainter dust continuum is affected more by the filtering effect of
  the missing short spacings, because there is no frequency dependence
  in the emission morphology. Other sources which appear to have
  diffuse dust emission (predominantly the disks which are face-on,
  SDP.2173, SDP.6216, SDP.6418) may also have substantial fractions of
  the dust flux resolved out, although there is no indication of
  anything larger than the effect on SDP.3132 given that none of these
  sources are outliers in the \CI--dust or CO--dust relations.

Ultimately, only higher sensitivity  (u,v)-matched imaging of CO(1-0),
dust  continuum and  \CI line  emission can  resolve these  issues, an
observing setup  that is not  trivial, especially between  CO(1-0) and
\CI(1-0) lines, given their  very different frequencies. Still, should
significant tracer brightness  be resolved out by  our current imaging
setup,  this will  certainly impact  any  extended \CI  line and  dust
continuum distribution much  more than that of CO(1-0)  because of the
differential  (u,v)  coverage  between  the  low  and  high  frequency
observations filtering-out  extended emission from  the high-frequency
tracers (see Figure~\ref{uvF}).


Within the limitations  of the data (both in terms  of resolution, $(u,v)$
coverage, and signal-to-noise),  we can say that three  of the sources
show  evidence   for  some  difference   between  the  dust   and  \CI
distributions. SDP.5323 has a very compact dust morphology centered on
the nucleus while  the \CI is found  in two lobes outside  of the dust
core. The CO  appears to be more  similar to the dust than  to the \CI
but the  poorer angular  resolution and  SNR of the  CO imaging  means we
cannot be confident. SDP.6216  has centrally concentrated \CI emission
while the  continuum morphology is  diffuse and complex.  SDP.6418 has
dust emission at low level near to the optical centre while the \CI is
located  in  several  clumps  at  different  velocities.  The  \CI  is
distributed over a much wider area  than the dust, although the SNR is
very low  for the dust  and so differences here  could just be  due to
there being a diffuse distribution and low SNR.

In summary, there are some very interesting morphological differences
between the \CI and dust, however, the SNR is not high enough for us
to be confident that these are real in all cases. The CO(1-0)
observations are at a very different angular resolution and do not
provide much information on the morphology. To undertake a meaningful
morphological comparison of the three tracers requires further
observations to (a) improve the sensitivity in all three tracers and
(b) crucially match the resolution and the ($u,v$) coverage between
the Band 3 and Band 7 observations.

\subsection{Relationships between the tracers}
The fluxes which are not in parentheses in Tables~\ref{COdataT} and
\ref{CIdataT} are used to estimate the line luminosities
\citep{Solomon2005} and rest-frame 850\mic monochromatic luminosity as
follows:

\begin{equation}
L^{\prime} = \rm{\frac{3.25\times10^7}{\rm{\nu_{rest}^{2}}}\left(\frac{D_{L}^2}{1+z}\right)\,S\Delta v} \,\,\,\,\rm{[K\,kms^{-1}\,pc^2]}
\end{equation}
Where $\rm{S\Delta v}$ is the velocity-integrated flux density in Jy\kms, $D_L$ is the luminosity
distance in Mpc and $\rm{\nu_{rest}}$ is the rest frequency of the
transition in GHz.

\begin{equation}
\lsub= 4\pi S_{\rm{\nu(obs)}}\times K \left(\frac{D_L^2}{1+z}\right) \,\,\,\,\, [\rm{W\,Hz^{-1}}]
\end{equation}
where $D_L$ is the luminosity distance, $S_{\rm{\nu(obs)}}$ is the
observed flux at 353 GHz, and $K$ is the K-correction to rest-frame 850\mic,
defined as
\begin{equation}
\label{KcorE}
K=\left(\rm{\frac{353\,GHz}{\nu_{rest}}}\right)^{3+\beta}\,\frac{e^{\rm{h\nu_{rest}/kT_d}}-1}{e^{16.956/T_d}-1}
\end{equation}
Here $\rm{\nu_{rest}= \nu_{obs}(1+z)}$, \td is the luminosity weighted
dust temperature (from an isothermal fit to the SED) with
$\beta=1.8$. K-corrections for this sample ranged from 2.6--3.0,
therefore making this extrapolation to rest frame 850\mic in order to
be consistent with the literature \citep{Scoville2016} does not add
significant uncertainty to what we are doing. These luminosities are
reported in Table~\ref{magphysT}.

\begin{figure*}
\includegraphics[width=0.45\textwidth,trim=0.2cm 0cm 0cm 0cm, clip=true]{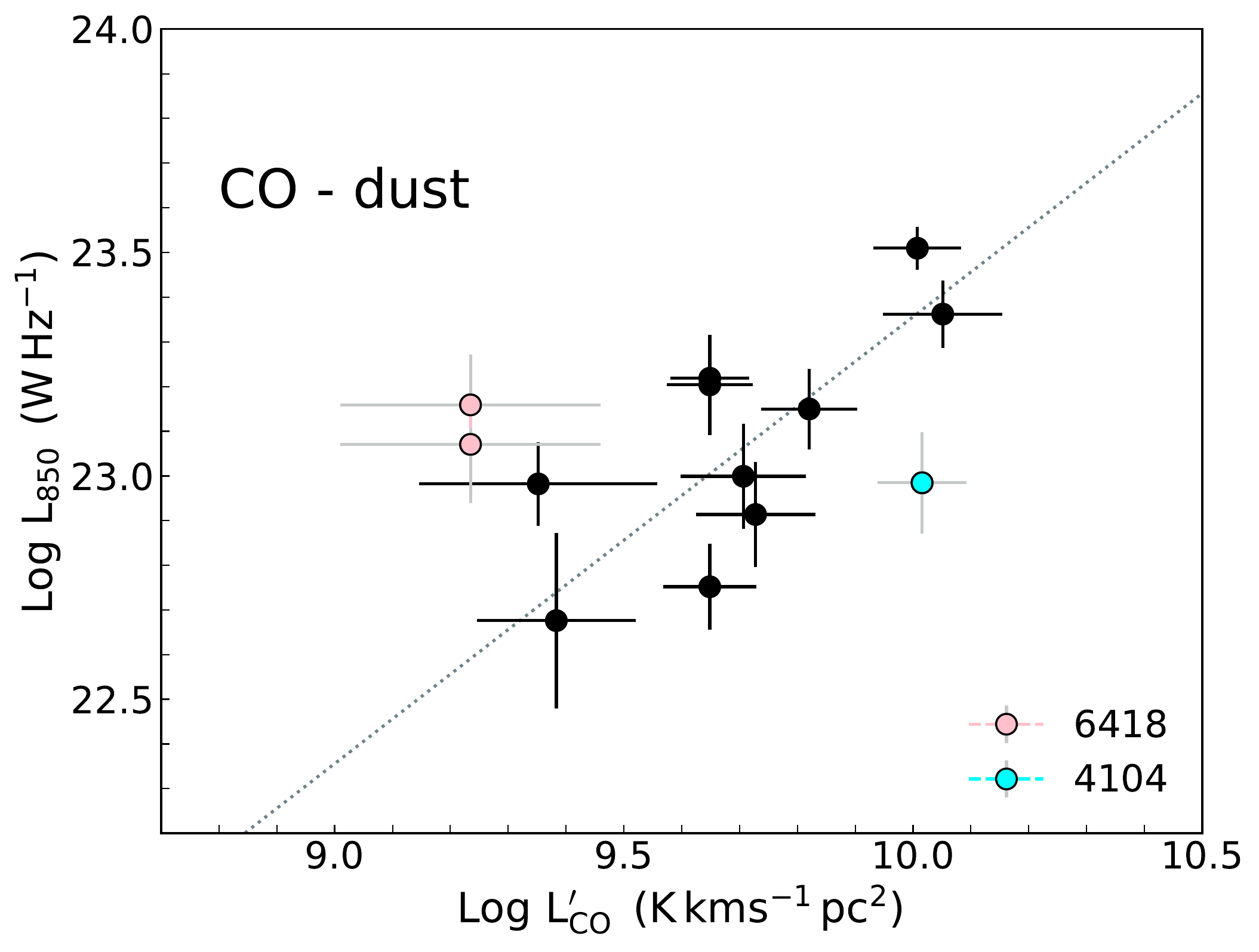}
\includegraphics[width=0.47\textwidth,trim=0cm 0cm 0cm 0cm, clip=true]{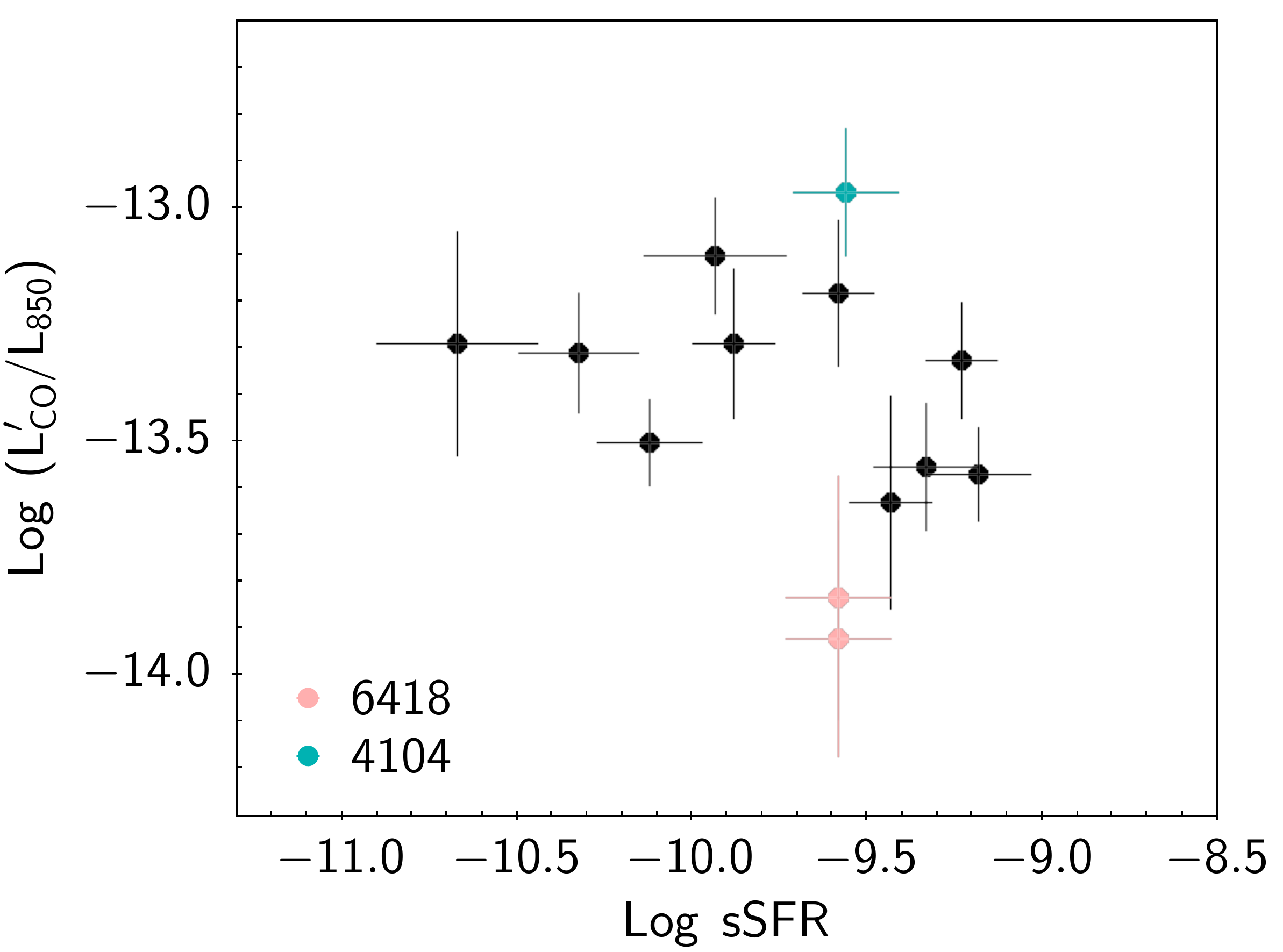}\\
\includegraphics[width=0.45\textwidth,trim=0.2cm 0cm 0cm 0cm, clip=true]{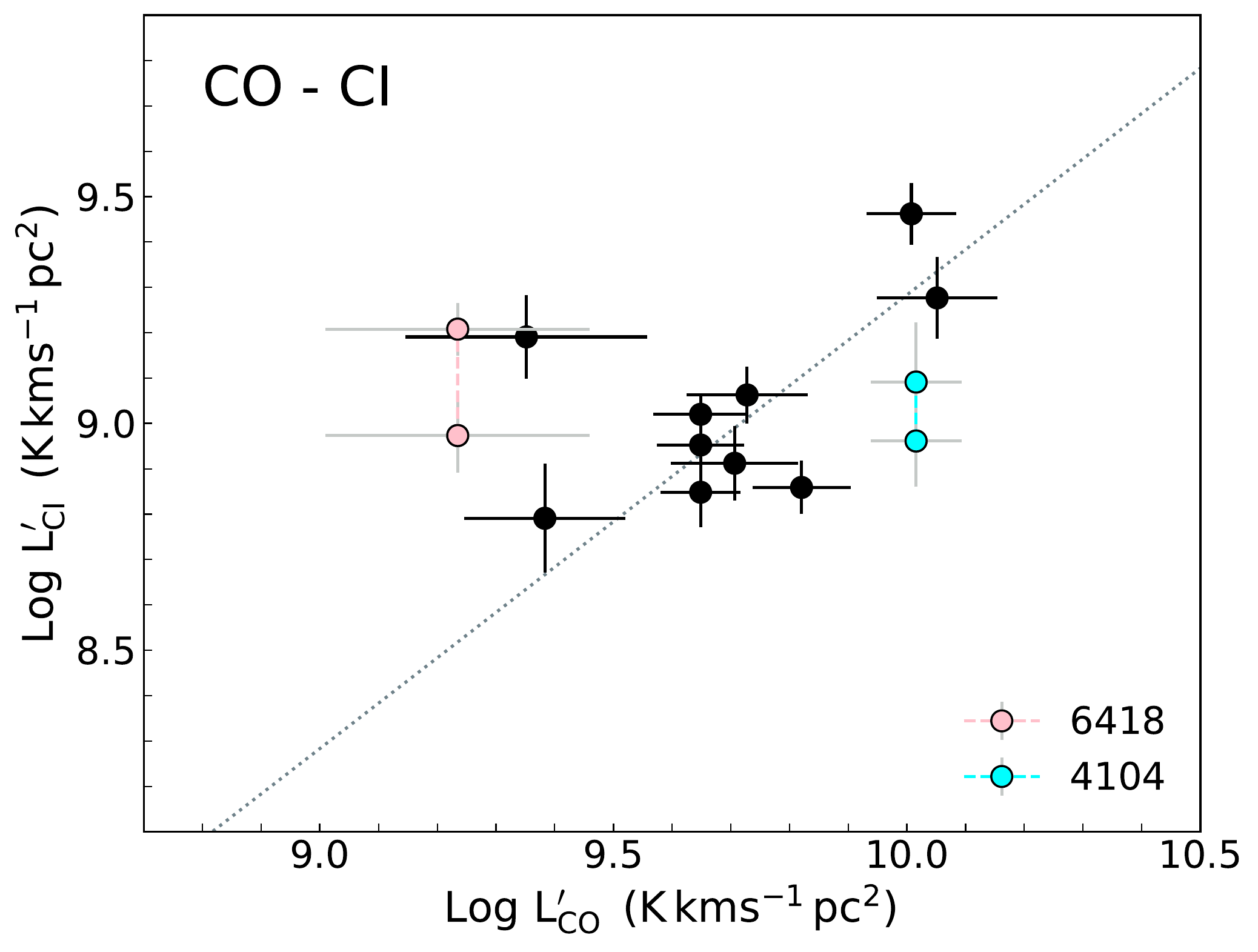}
\includegraphics[width=0.47\textwidth,trim=0cm 0cm 0cm 0cm, clip=true]{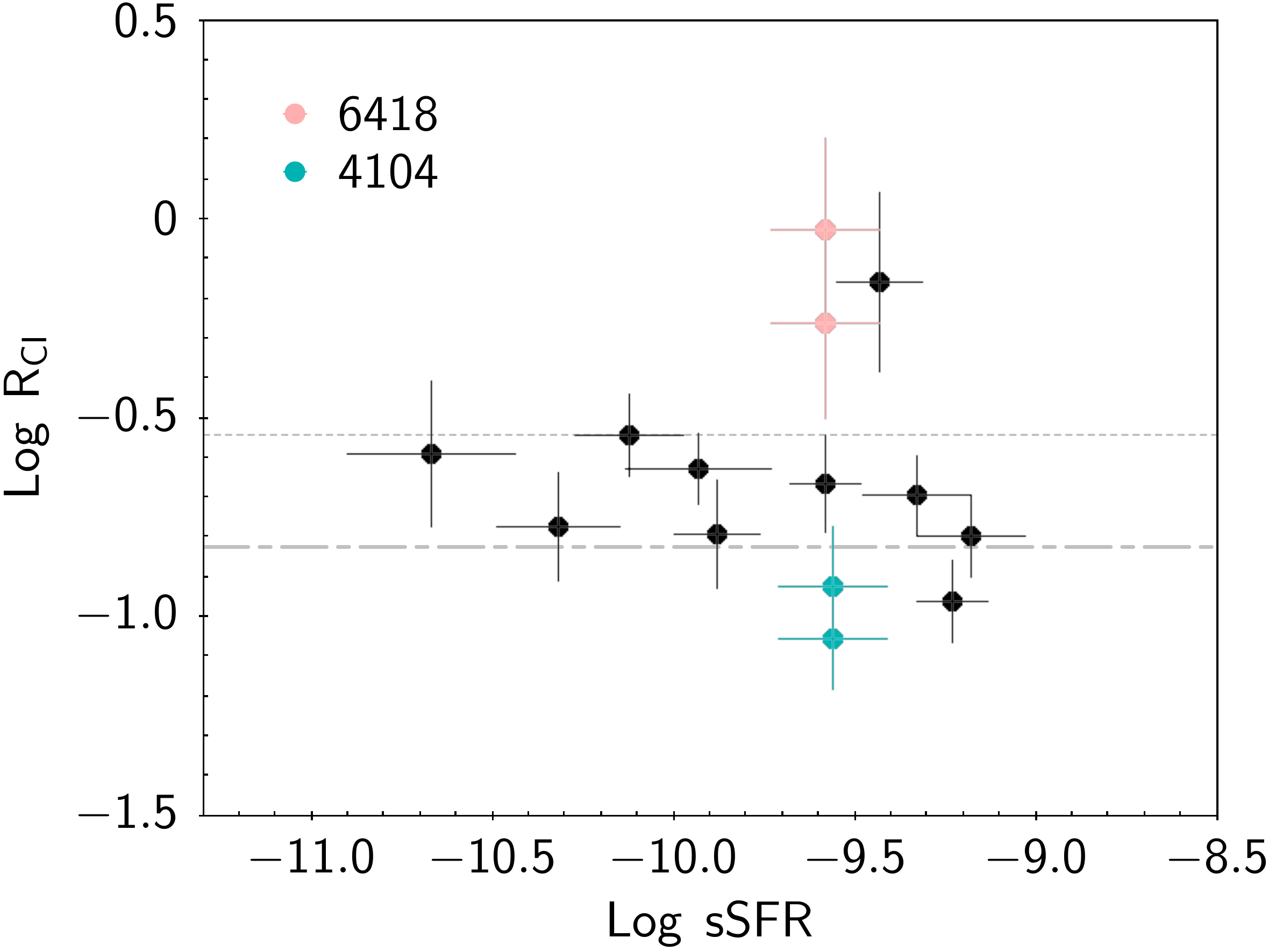}\\
\includegraphics[width=0.45\textwidth,trim=0.2cm 0cm 0cm 0cm, clip=true]{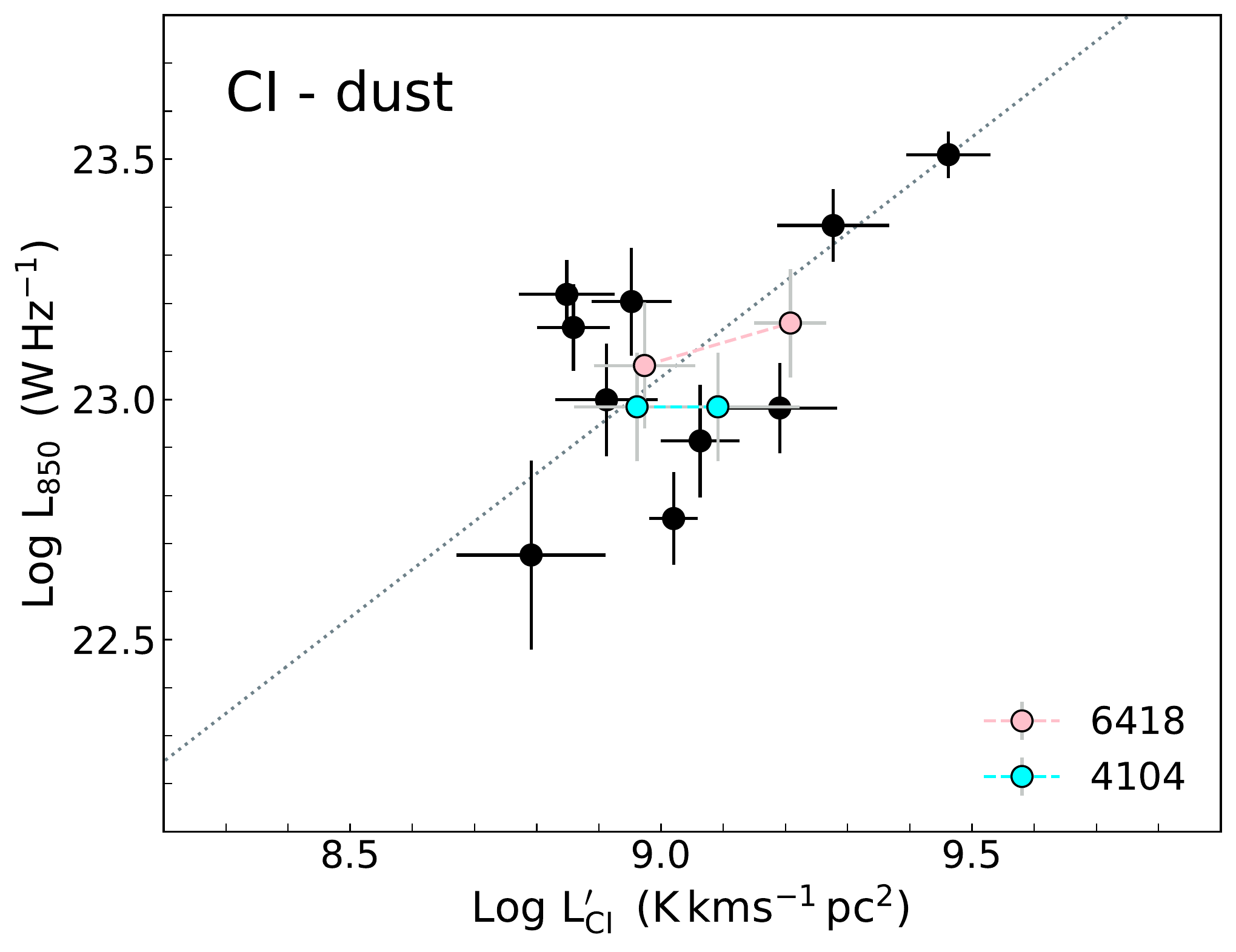}
\includegraphics[width=0.47\textwidth,trim=0cm 0cm 0cm 0cm, clip=true]{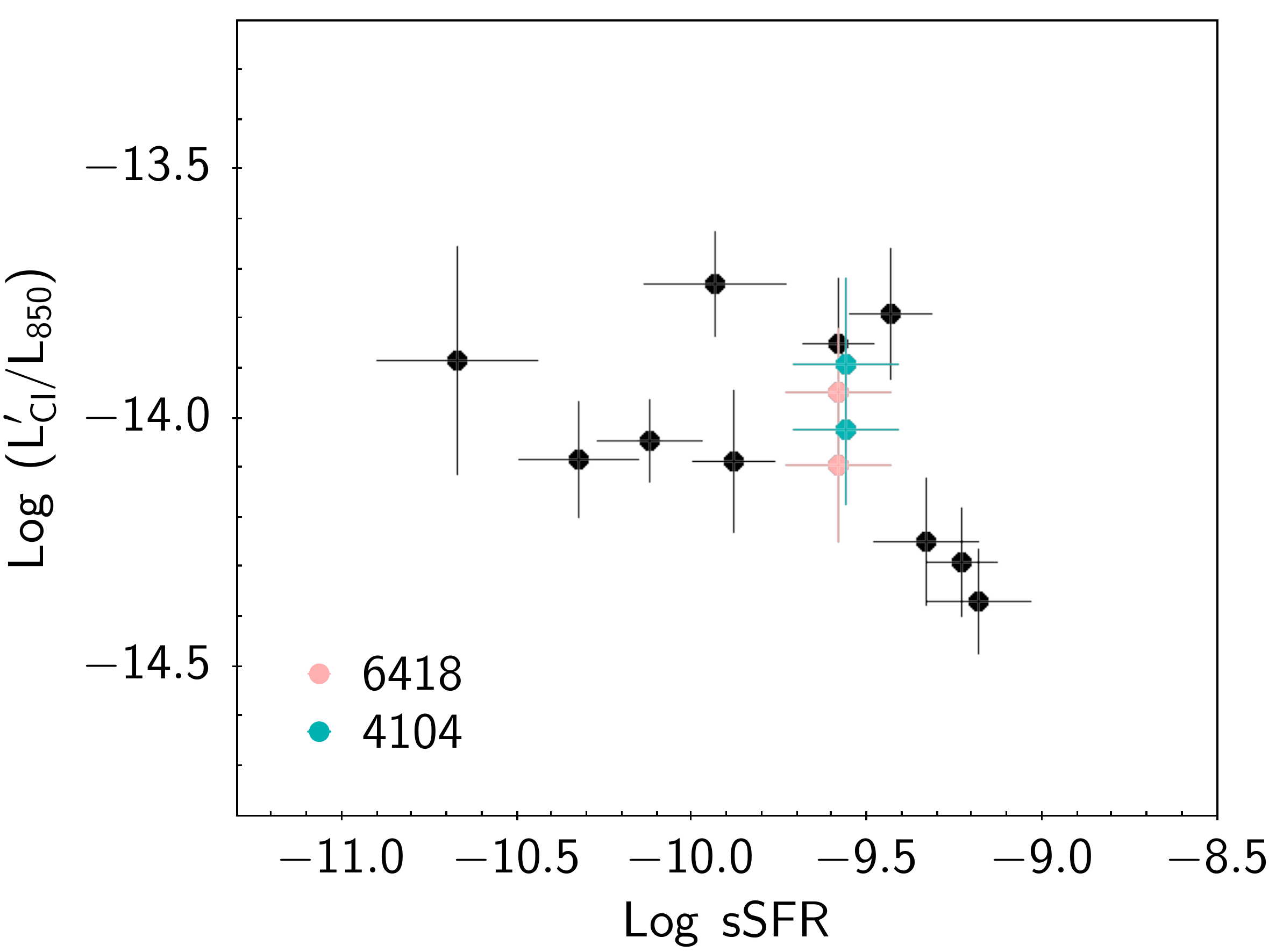}
\caption{\label{corF} {\bf Left column:} Relationships between the
  three gas tracers for our sample. The dotted line shows the best
  linear fit (slope fixed to 1) and. The two sources with alternative
  \CI measurements are shown as pink and cyan circles connected by
  dashed lines. {\bf Right column:} The three tracer ratios as a
  function of sSFR. The pale dotted and dot-dashed lines in the middle
  row show the average $\rm{R_{CI}}$ ratio measured for high redshift
  QSO \citep{Walter2011} and the average for Milky Way clouds
  \citep{Frerking1989} respectively. Quantities for each source are
  derived using their measured flux and error, regardless of whether
  they are `detected' at $>3\sigma$. }
\end{figure*}

The correlations between the three observables, \lci, \lcoa and \lsub
are shown in Figure~\ref{corF}, along with the best linear 1:1 fit in
order to visually compare the consistency of the various tracers. The
two sources which have alternative \CI estimates (SDP.4104 and
SDP.6418) are shown as cyan and pink circles with the two values
joined by a dashed line. The choice of which \CI flux to use has no
material impact on the conclusions or results. This sample is too
small and has too limited a dynamic range in luminosity to make
meaningful fits to these relationships, and we defer this process to a
much more comprehensive analysis which combines this sample with
others from the literature (Dunne et al. {\em in prep}). Suffice to
say that the trend noted here, for \CI--dust to have the lowest
scatter, is borne out by the analysis on a much larger sample of $70+$
sources.

We also show the ratios of the tracers as a function of sSFR in the
right column of Figure~\ref{corF}. The top row shows the \lcoa/\lsub
ratio which shows no apparent trend. The middle row
shows $\rm{R_{CI}=\lci/\lcoa}$, with the average values for the Milky
Way \citep{Frerking1989} and high-z QSO \citep{Walter2011} shown as the
grey dot-dashed (lower) and dotted (upper) horizontal lines
respectively. With two exceptions (SDP.6418 and SDP.6451), the
galaxies fall in a reasonably tight band with a spread of $\sim 0.5$
dex, and a trend to lower values of $\rm{R_{CI}}$ with increasing
sSFR. The two exceptions have much higher $\rm{R_{CI}}$ for their
sSFR, although the significance of the offset would need to be
confirmed with more sensitive CO observations.  The bottom row shows
\lci/\lsub, which has the smallest dispersion. The very highest sSFR
galaxies have the lowest \lci/\lsub ratios, but given the sample size
and error bars there is no significant trend.
 
The means and medians (in brackets) and 1$\sigma$ ranges for the ratios are: log $(\lsub/\lci) =
14.04\, (14.07) \pm 0.19$, $\log (\lsub/\lcoa) = 13.39\, (13.32) \pm
0.25$ and $\log(\lci/\lcoa) = -0.66\, (-0.68) \pm 0.25$. The
relationship with the least variance is \lsub--\lci which implies that
these two tracers are better correlated with each other than either of
them are with CO. We will return to the implications of these
relationships for calibrating the $\rm H_2$ tracers in
Section~\ref{gasS}.

\subsubsection{Potential CO-`dark' galaxies}
\label{codarkS}

The two sources (SDP.6418 and SDP.6451) which are outliers in the \CI--CO
relationship (Figure~\ref{corF}), have very weak CO emission {\it
  despite being detected both in dust and \CI.} The sources are
unremarkable in the \CI--dust relationship indicating that this seems
to be a deficiency in CO relative to the other tracers rather than an
enhancement in \CI.  There is nothing particularly special about these
sources, their stellar masses are log $\Ms\sim 10.8\,\msun$, and
neither show evidence for an AGN in the optical spectra, using the
criteria of \citet{Lamareille2010}; they have O[{\sc ii}] flux greater
than O[{\sc iii}], and O[{\sc iii}] is generally weak. As pointed out
in Section~\ref{morphS}, any filtering of flux on extended scales is
going to affect \CI more than CO and so this cannot be an explanation
for the high \CI/CO ratio. More data would be needed to confirm this
unusual \lci/\lcoa ratio, both to increase the sensitivity in the line
observations and to match the resolution between the tracers. The
ratio of line luminosity \lci/\lcoa ($=\rm{R_{CI}}$) in these galaxies
ranges from $0.54<\rm{R_{CI}}<0.94$, using the measured CO fluxes
which are only 2$\sigma$ significance. Using a 3$\sigma$ upper limit
on the CO flux gives a lower limit of $\rm{R_{CI}}> 0.35\,\, (0.42)$ for
SDP.6418 and SDP.6451 respectively. The average value in Milky Way
clouds is $\rm{R_{CI}} \sim 0.15$ \citep{Frerking1989}, while the
average value seen in QSO is $\rm{R_{CI}}\sim 0.29$
\citep{Walter2011}.

Understanding how reasonably normal star forming galaxies like these
can be so deficient in CO is very important for understanding the C/CO
dependencies in the ISM. One mechanism for reducing CO emission
relative to \CI is cosmic ray destruction of CO molecules
volumetrically in the clouds, something discussed at length in
\citet{Bisbas2015,Glover2016,Bisbas2017,Papadopoulos2018}. The more
recent works show that conditions do not have to be extreme in terms
of $\zeta_{\rm CR}$ ($\propto\rho_{\rm SFR}$) in order to produce a
high $\rm{[C^0/CO]}$ abundance ratio, providing that the average gas density is
moderate to low. We note that it is not unheard of to find global $\rm
H_2$ reservoirs dominated by low-density gas, even in vigorous SF
environments, and can be the result of coherent SF feedback on the
$\rm H_2$ gas reservoir. This may indeed be so for Arp\,193, the most
extreme ULIRG in the local Universe after Arp\,220, where only small
($\sim 5\%$) gas mass fractions at $\rm n\ga 10^4\,cm^{-3} $ are
inferred from multi-J CO, $^{13}$CO, HCN and CS observations
\citep{PPP6240}. We will discuss the line luminosity $\rm{R_{CI}}$
  ratio and how observations compare to theory in more detail with a
  much larger sample in (Dunne et al. {\em in prep}).\footnote{An
    example of a very {\em low} \lci/\lcoa ratio in a local LIRG has been recently
    reported by \citet{Michiyama2020}, which further indicates the
    need to better study a diverse range of galaxies with ALMA in
    order to understand the effects of the $\rm{[C^0/CO]}$ ratio on global
    $\rm{H_2}$ tracers.}
%

Finally we note that since: a) C-rich gas may extend further out in
galactocentric distance than CO-rich gas, and b) differential
($u,v$)-sampling will filter out extended \CI line emission more than
the much-lower frequency CO(1-0) line, we expect that ($u,v$)-matched
CO, \CI line imaging of this sample will not only preserve our results
about CO-`dark' galaxies, but it may even `push' some other galaxies
towards a global CO-deficient/\CI-rich state by revealing more \CI
line emission over larger areas.  In that regard, ACA imaging
observations, that include total power, of the CO(1-0), \CI(1-0) and
dust continuum, can provide a definitive test.


\section{Calibration of gas mass.}
\label{gasS}\label{optS}
We estimate the molecular gas mass of the galaxies in our sample using
each of the tracers in turn following the method described in detail
in Dunne et al. {\em in prep}. We differentiate between \Mh - the mass
of molecular hydrogen, and \Mmol - the mass of molecular gas including
a factor 1.36 contribution from He. Briefly, to determine \Mh we need
a value for the calibration parameter for each tracer: for \CI we need
to know the abundance $\rm{\Xci=[C^0/H_2]}$, for CO we need the \aco
factor and for dust we need to know the gas-to-dust ratio \gdr. The
molecular hydgrogen gas mass (in \msun units) then depends on these
calibration factors as follows:\footnote{Other
    literature works define quantities such as \aci and \asub which
    are simply related to our calibration factors, and will be
    presented at the end of this section.}

\begin{multline}
{\rm \Mh (\CI) = \frac{9.51\times 10^{-5}}{\Xci\, Q_{10}}\,\lci} \label{H2E} \\
{\rm \Mh (CO) = \aco \lcoa} \\
{\rm \Mh (dust) = \gdr \Md} \\
\end{multline}

The conversion of \lci to \Mh also requires an estimate of $Q_{10}$,
which is an excitation function. This is not a strong function of
kinetic temperature (see \citealp{PPP2004}, Dunne et al. {\em in
  prep}) and so we will simply take a single value as is common in the
literature. We choose to use the value $Q_{10}=0.43$ based on the more
detailed analysis of a larger sample in Dunne et al. {\em in
  prep}. The dust mass is calculated from \lsub using the mass
weighted dust temperature from the {\sc magphys} fits and assuming a
value for the dust mass opacity at 850\mic of
$\kappa_{850}=0.065\,\rm{m^2\,kg^{-1}}$ which is consistent with the
results from \citet{Planck2011xix} (see Dunne et al. {\em in prep} for more
detailed descriptions of the dust opacity factors).

For each source, we have four unknowns, \Mh,
\aco, \gdr, \Xci,
and three observables \Md, \lcoa, \lci. If we had an
independent measure of the true \Mh, then the observed values of
\lcoa, \lci and \Md would provide direct estimates of the
three calibration parameters, \aco, \Xci and
\gdr. But, of course, we do not know the value of \Mh,
so we need to use a probabilistic argument based on the fact the
observations do provide constraints on the relative values of
\aco, \Xci, and \gdr for each source.

For each source there is a set of  self-consistent calibration factors
which link the observed \Md, \lcoa, \lci to the true
\Mh, with a common, but unknown constant factor.  We can write
down the products and ratios of these calibration factors in terms of
the observed luminosity ratios:

\begin{multline}
\mathrm{X_{CI}}\, \delta_{\mathrm{GDR}} = 2.212 \times 10^{-4}
\frac{L'_{\mathrm{CI}}}{M_d} \\ 
\mathrm{ X_{CI}}\, \alpha_{\mathrm{CO}} = 2.212 \times 10^{-4}
\frac{L'_{\mathrm{CI}}}{L'_{\mathrm{CO}}} \\ 
\frac{\alpha_{\mathrm{CO}}}{\delta_{\mathrm{GDR}}} = \frac{M_d}{L'_{\mathrm
                                           CO}}\\
\label{ratioE}
\end{multline} 

From our parallel study of a larger sample of 72 sources all with CO,
\CI and dust measurements (Dunne et al. {\em in prep}), we have measured the
intrinsic variance for each pair of factors in
Eqn.~\ref{ratioE}. We assume that the covariance between the
calibration factors is zero, and use the three pair variances to
estimate the intrinsic variance of each individual calibration
factor, the population standard deviations of these are denoted $\sigma_{X}$, $\sigma_{\alpha}$ and $\sigma_{\delta}$ for \Xci, \aco and \gdr respectively.  


With the constraints given by Eqn~\ref{ratioE} we are in a position to
determine self-consistent values for each of the calibration factors,
\aco, \Xci and \gdr, except for one thing: we do not have any
independent measure of the gas mass on which to tie the normalisations
of the our calibration (4 unknowns but only 3 measurements). We
therefore have to assume a {\em sample average} for one of the unknown
calibration factors, but having done so the relative values of all
three will be optimised, and it is trivial to re-scale the solutions
to a different normalisation if desired. We choose to normalise to a
sample mean $\delta_\mathrm{GDR}=135$ which is the Milky Way value for
Hydrogen (no He) from the THEMIS dust framework
\citep{Jones2018}.\footnote{More precisely the combination of \gdr/\kd
  we use is 2077, which is the quantity that calibrates dust emission
  to gas mass. This is equvialent to $\sigma_{850}/\rm{m_H}$ in
  S.I. units, where $\sigma_{850}$ is the dust opacity at 850\mic per
  H atom.}

Dunne et al. {\em in prep} use their sample of 72 sources which have
observations for all three gas tracers\footnote{That sample includes
  the 12 galaxies from this analysis.} to estimate the mean values
of \Xci and \aco. This gives us the 3 calibration factors normalised
to the Milky Way value of $\delta_\mathrm{GDR}$. Using these mean
values together with the population standard deviations,
$\sigma_{X}$, $\sigma_{\alpha}$, $\sigma_{\delta}$, we can estimate the
probability of finding a particular set of calibration factors for any
given source.  Using $x_i, i=1,2,3$ to denote the logarithms of the
three calibration factors, we can write the means and effective
standard deviations as $\bar x_i$, and $\sigma_i$ respectively, where
the effective standard deviation is the intrinsic scatter on each
parameter added in quadrature to the measurement error for that gas
tracer. Assuming these follow Gaussian distributions, the probability
of finding calibration factors $x_i$ for any source is given by

\begin{multline}\label{eqn:chi2}
P \propto \displaystyle\prod_{i=1}^3 \exp\left(-\frac{(x_i-\bar x_i )^2}{2\sigma_i^2}\right)\\
= \exp \left(- \displaystyle\sum_{i=1}^3 \frac{(x_i-\bar x_i
  )^2}{2\sigma_i^2} \right) \\
\end{multline} 

\begin{figure*}
\includegraphics[width=\textwidth,trim=0.0cm 0cm 0cm 0cm, clip=true]{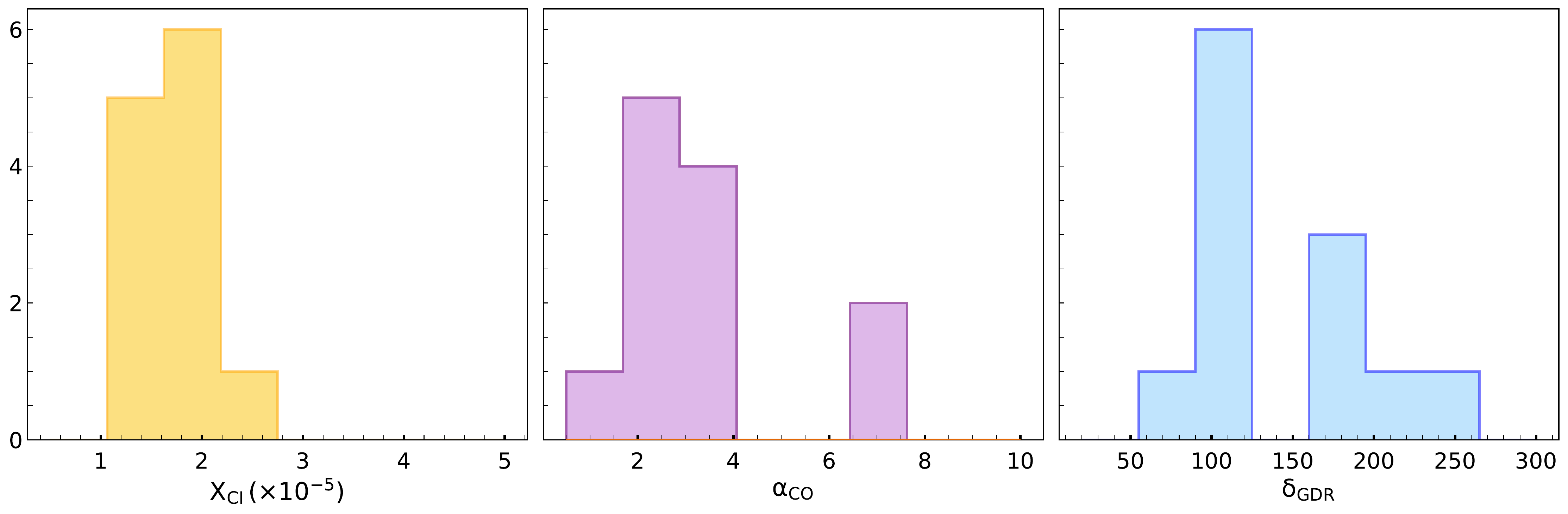}\\
\caption{\label{calhistF} Distribution of the calibration parameter
  values which are listed in Table~\ref{lumT}.}
\end{figure*}

For any given source, we use the observed luminosity ratios in
Eqn.~\ref{ratioE}, to fix the ratios of calibration factors, and
choose the common scaling factor that maximises the probability in
equation \ref{eqn:chi2}. The resulting optimised parameters are listed
in Table~\ref{lumT}, and shown in Figure~\ref{calhistF}, while the average values for the sample are listed in Table~\ref{calT}. By design,
each tracer for a given source will produce the same gas mass when
used in Eqn.~\ref{H2E}. The errors are quoted as the values which give
$\Delta \chi^2=1$. We list these gas masses in Table~\ref{lumT} along
with the gas fraction, \fg, which is defined as
$\Mmol/(\Ms+\Mmol)$. \footnote{If we use the median calibration factors
  from the sample, listed in Table~\ref{calT}, to derive gas masses
  from each tracer independently (as one would do if only one
  measurement were available) there can be some dispersion in the
  three gas mass estimates depending on how far away a particular
  source lies in the distribution from the median. For most sources,
  the mass difference between any pair of tracers is within $\pm0.2$
  dex, except for the two sources with high \aco (SDP.6418 and
  SDP.6451).}

\begin{table*}
\centering
\caption{\label{lumT} Calibration and gas mass parameters. \Mh, \Xci, \aco and \gdr do not include a contribution from He. \aci and \asub along with $f_g$ do include a factor 1.36 for He.}
\begin{tabular}{lcccccccc}
\hline
  \multicolumn{1}{c}{Source} &  
  \multicolumn{1}{c}{Log $\Mh$} &
  \multicolumn{1}{c}{\Xci} &
  \multicolumn{1}{c}{\aci} &
  \multicolumn{1}{c}{\aco} &
  \multicolumn{1}{c}{\gdr} &
  \multicolumn{1}{c}{\asub} &
  \multicolumn{1}{c}{$f_g$} &
  \multicolumn{1}{c}{$\tau_{\rm dep}$}\\
  \multicolumn{1}{c}{} &
  \multicolumn{1}{c}{[\msun]} &
  \multicolumn{1}{c}{[$\times 10^{-5}$]} &
  \multicolumn{2}{c}{[\aunit]} &
  \multicolumn{1}{c}{} &
  \multicolumn{1}{c}{[$\rm{W\,Hz^{-1}\,\msun^{-1}}$] } &
  \multicolumn{1}{c}{} &
  \multicolumn{1}{c}{[Gyr]}\\
\hline
  163     & $10.58\pm0.11$ & $1.7\pm0.4$ & 17.9 & $3.8\pm0.9$      &  $111\pm28$ & $6.2\times10^{12}$ & $0.15$  & 2.4 \\
  1160 & $10.06\pm0.10$ &  $2.0\pm0.5$ & 14.9 & $2.6\pm0.6$ & $261\pm61$ & $3.6\times10^{12}$ & $0.13$  & 1.1\\
  2173 & $10.14\pm0.11$ &  $1.8\pm0.5$ & 16.4 &  $2.6\pm0.6$ & $181\pm47$ & $4.3\times10^{12}$ & $0.19$ & 0.9\\
  3132     &  $10.11\pm0.11$ &  $1.5\pm0.4$ & 19.9 & $2.9\pm0.8$  & $94\pm24$ & $9.0\times10^{12}$ & 0.24   & 0.7 \\
  3366 &  $9.85\pm0.15$ & $1.9\pm0.7$ & 15.7  &  $3.0\pm1.0$  & $177\pm62$ &  $4.9\times10^{12}$ & 0.06  & 2.8\\
  4104 &  $10.16\pm0.13$ & $1.4\pm0.4$ & 21.4  & $1.4\pm0.4$  & $186\pm55$ & $4.9\times10^{12}$ & 0.19   & 0.9 \\
  4104$^{c}$ & $10.22\pm0.14$ & $1.6\pm0.5$ & 18.4 & $1.6\pm0.5$ & $217\pm69$ & $4.2\times10^{12}$ & 0.21 & 1.0 \\
  5323 &  $10.09\pm0.12$ & $1.5\pm0.4$ & 20.4 & $2.4\pm0.7$  & $110\pm31$ & $6.0\times10^{12}$ & 0.13    & 1.1  \\
  5347 &  $10.48\pm0.12$ & $1.3\pm0.4$ & 21.5 & $2.7\pm0.7$   & $103\pm29$ & $5.7\times10^{12}$ & 0.22  & 5.7\\       
  5526 &  $10.09\pm0.11$ & $1.3\pm0.3$ & 23.3 & $1.9\pm0.5$  & $92\pm23$ & $8.4\times10^{12}$ & 0.14 & 1.7\\
  6216 & $10.13\pm0.11$ &  $1.2\pm0.3$ &26.0   &  $3.0\pm0.8$ & $64\pm17$ & $9.0\times10^{12}$ & 0.32   & 0.7 \\
  6418 & $10.07\pm0.13$ & $1.8\pm0.5$  & 17.1 & $6.9\pm2.1$   & $100\pm30$ & $7.3\times10^{12}$ & 0.20  & 1.0\\
  6418$^t$ & $10.26\pm0.12$ & $2.0\pm0.5$ & 15.3 & $10.6\pm2.9$ & $125\pm34$ & $5.8\times10^{12}$ & 0.28 & 1.5 \\
  6451 &  $10.17\pm0.13$ & $2.3\pm0.7$  &  13.0 & $6.6\pm2.0$  & $168\pm50$ & $4.8\times10^{12}$ & 0.26  & 0.9  \\
\hline
\end{tabular}
\flushleft{\small{Gas mass in \msun using the optimised calibration
    method described in Section~\ref{gasS}. The optimised calibration
    parameter for each gas tracer, \Xci, \aco and \gdr. Gas fraction \fg
    defined as \Mmol/(\Ms+\Mmol), using the value of \Ms from the {\sc
      magphys} fitting in Table~\ref{magphysT}. Gas depletion
    timescale, \Mmol/SFR, in Gyr. SDP.4104$^c$ uses the \CI fluxes
    corrected for flagged channels within the line profile as
    described in Appendix A. SDP.6418$^t$ uses all three of the \CI velocity
    components, as opposed to just the $v2$ component. }}
\end{table*}
\npnoround

\begin{table}
\centering
\caption{\label{calT} Average calibration factors derived from this
  sample. Note that these calibration factors do not include the
  correction for He.}  \renewcommand{\arraystretch}{1.4}
\begin{tabular}{llcccc}
\hline
Parameter & mean & $\sigma_{m}$ & median  \\
\hline
\Xci ($\times10^{-5}$)  & $1.6^{+0.3}_{-0.2}$ & 0.1 &  1.6\\
         & $1.7^{+0.3}_{-0.3}$ & 0.1 & 1.7 \\
\aco     & $3.0^{+1.4}_{-0.7}$  &  0.5  & 2.8  \\
         & $3.1^{+1.3}_{-0.8}$  & 0.7   &  2.8 \\
\gdr     & $128^{+54}_{-35}$  & 16  & 111\\
         & $132^{+57}_{-39}$  & 16  & 118\\
\hline
\end{tabular}
\flushleft{\small{The sample averages for the calibration parameters,
    using the optimisation procedure described in
    Section~\ref{optS}. The first line represents the values using the
    conservative \CI estimates for 6418 and 4104. The second row
    includes all the velocity components for 6418 and corrects the \CI
    flux for 4104 for the missing channels as described in the
    text. The errors quoted on the mean are the 16th and 84th
    percentiles representing the $\pm 1\sigma$ scatter. $\sigma_m$ is
    the formal error on the mean (i.e. $1/\sqrt{12}$ times the
    scatter.}}
\end{table}

\section{Discussion}

\subsection{Calibration factors}
We now place the values of our gas calibration factors into
context with other recent work and other nomenclature for describing them.

We find that the mean and standard deviation for $\Xci=(1.6\pm0.3)\times
  10^{-5}$, which is similar to the values found in clouds in the
  Milky Way \citep{Frerking1989} and to the recent determination of
  \Xci using GRB/QSO absorber systems using is a totally independent
  method \citep{Heintz2020}. Several authors have adopted the use of
  an emprical quantity \aci, with similar units of
  $\rm{\msun\,(K\,kms^{-1})^{-1}}$ to that of \aco, which is used as
  standard in extragalactic studies. \aci is related to carbon
  abundance, \Xci as follows, where we now include a factor of 1.36 for He:

\[
\aci=1.293\times 10^{-4} (\Xci \rm Q_{10})^{-1}
\]
Thus for our average $\Xci = 1.6\times 10^{-5}$ and $Q_{10}=0.43$, we find that $\aci=18.8 \aunit$.

The only other source which has an estimate of \aci independent of a
blind assumption on the value of \aco is the AGN influenced CND region
of NGC 7469, for which \citet{Izumi2020} find a value of $\aci=4.4
\,\msun (\rm{K\,kms^{-1}\,pc^2)^{-1}}$, which is equivalent to
$\Xci=7\times 10^{-5}$. They attribute their much higher value of \Xci
to the effects of an X-ray dominated region (XDR) on the ISM, which
increases the [$\rm{C^0/H_2}$] ratio. High values for \Xci have also been
inferred for other extreme sources in the high redshift universe such
as QSO \citep{Walter2011}, however in these cases an assumption is
made that $\aco\sim 0.8$ in such sources, which then produces a higher
\Xci compared to what would be inferred using a standard $\aco\sim
4$. The \citeauthor{Izumi2020} work is novel because the calibration
factors are estimated using a dynamical mass estimate for $\rm{H_2}$
from very well sampled ALMA data. A comprehensive comparison of
calibration factors with those in the literature is undertaken in
Dunne et al. {\em in prep}.

We find an average $\aco =
  3.0^{+1.4}_{-0.7}\,\aunit$, not including He,\footnote{To compare to other work which does include He and heavier elements, our \aco values should be multiplied by 1.36.} which is typical of
  star forming galaxies with an ISM dominated by self-gravitating
  clouds \citep{Papadopoulos2012,Bolatto2013,Sandstrom2013}.

While we began with a prior that the gas-to-dust ratio \gdr would be
similar to that in the Milky Way (given these are massive, metal rich
star-forming galaxies), our method allows the \gdr to vary slightly in order to
maximise the likelihood when considered along with the other two gas
tracers. With this we find an average $\gdr=129\pm57$ with a range of
\gdr for individual sources ranging from 60--260, as expected for such
a sample. Once again this \gdr does not include the contribution from
He. It is now common to refer to the calibration between dust
luminosity and gas mass as \asub, following the work of
\citet{Scoville2014,Scoville2016}. This is defined as \asub =
\lsub/\Mmol with units of $\rm{W\,Hz^{-1}\, \msun^{-1}}$, where now He
is included in the definition of \Mmol. We describe this parameter and
its measurements in the literature in great detail in Dunne et
al. {\em in prep}, but here just note that it is related to \gdr (assuming our $\kd=0.065$ and including a factor 1.36 for He) as:
\[
\asub=7.78\times 10^{14} (24.5/\mwtd)^{-1.4} \gdr^{-1}
\]
where \mwtd is the mass-weighted dust temperature,
  commonly assumed by \citet{Scoville2016} and others to be $\sim
  25$~K in the sources they studied (typically high redshift star
  forming galaxies). Our empirically derived \asub, using the optimal
  gas mass calculated using the above method, is
  $\asub=(5.9\pm1.8)\times10^{12}\, \rm{W\,Hz^{-1}\, \msun^{-1}}$.
For our
average $\gdr=129$ and a $\mwtd=25$K (which agrees well with the
results of our {\sc magphys} SED fits), we find a comparable value of
$\asub=6.0\times 10^{12} \,\rm{W\,Hz^{-1}\,\msun^{-1}}$. This compares
with the value initially suggested by \citep{Scoville2016} of
$10.1\times10^{12}$ and more recent literature studies which indicate
$\asub = 3.8-8.8\times 10^{12}$ \citep{Orellana2017, Hughes2017,
  Kaasinen2019}.

Overall, this sample of 250\mic selected galaxies at $z=0.35$ have \Mh
calibration factors for CO(1--0), \CI(1--0) and dust which are
comparable to normal spiral star forming galaxies in the local
Universe, and conditions in the Milky Way.

\subsection{Gas fractions and evolutionary status.}
The mean gas fraction for the sample is $0.19\pm0.07$ where the error
is the 1$\sigma$ standard deviation. This narrow range is rather
surprising as 250\mic selection from H-ATLAS at $z=0$ produces samples
with a huge range of gas fraction (0.1-0.9), and a mean value of
$f_g\sim 0.5$ \citep{Clark2015}. This $z=0$ selection was volume
limited and dust mass selected, and was therefore dominated by lower
mass, blue and H{\sc i} rich galaxies. In contrast,
Figure~\ref{MSplotF} suggests that by $z=0.35$ H-ATLAS is only
sensitive to the most massive dust sources, which presumably restricts
the range of gas fractions, assuming that the anti-correlation between
gas fraction and stellar mass holds also at $z=0.35$
\citep{deVis2017a,saintonge17}. At a given stellar mass, the gas
fractions we find at $z=0.35$ are significantly higher than those
found in the local Universe by the xCOLD GASS survey
\citep{saintonge17}, who find $\fg = 0.02-0.04$ for galaxies in our
stellar mass range. However, in terms of specific star formation rate,
the scaling found by xCOLD GASS seems to hold at $z=0.35$, where for
$\rm{sSFR} = -10.5$ the expectation is $\fg=0.039$, while at log
$\rm{sSFR/yr^{-1}}=-9.5$ this increases to $\fg=0.15$. Thus the
increased sSFR 3.9 Gyr in the past is completely consistent with the
extra gas available without any changes to the efficiency of how it is
converted into stars. This is noteworthy since the observables
providing the measures of gas fraction (CO, \CI and dust) and sSFR
(UV--optical SED) are completely independent in this study.

It is common in the literature for authors to define a gas depletion
timescale (also the inverse of the star formation efficiency) as
$\rm{\tau_{dep}= \Mmol/SFR}$, which is a crude way of measuring the
relative speed at which the current molecular reservoir is being
converted into stars. We list this value for our sources in
Table~\ref{lumT}. There are limitations to this idea, principally that
neither \Mmol nor SFR are fixed quantities, \Mmol being only a
snapshot of the molecular mass at the observed epoch, and SFR being
only the current SFR (averaged over $\sim 100$ Myr). The SFR will
decline as the gas is consumed (thus $\rm \tau_{dep}$ is a lower limit
to any physical depletion timescale), while in addition more gas may
be accreted at later times, from either the cosmic web, mergers or
simply the infall of atomic gas from the outer regions of the
disk. {\textcolor{purple}The median of this quantity $\rm{\tau_{dep} =
  1.1^{+1.3}_{-0.2}}$~Gyr where the error is the 16-84th percentile
range. This is consistent with the range of $\tau_{dep}$ in the most
massive gas-rich galaxies today, which is 1--1.6 Gyr over the same
mass range as our sample \citep{saintonge17}.} The time elapsed from
$z=0.35$ to $z=0$ is 3.9~Gyr, which means that all of these galaxies
would be passive today if they continued to form stars at the same
rate without fresh gas supply. Certainly there must be some
transformation from dust and gas rich galaxies at z=0.35 to passive
galaxies today \citep{Dunne2011,Beeston2018}, but it is not plausible
that all of them will become passive further high-lighting the
limitations of the simple $\rm{\tau_{dep}}$ parameter.

\section{Conclusions}
We present ALMA observations of sub-mm dust continuum at 850$\mu m$, \COa and \CIfull
for a complete and volume limited sample of 12 sources selected at 250\mic from the
H-ATLAS survey. We detected dust and either CO or \CI in all of the targets, finding:
\begin{itemize}
\item{While the galaxies in the sample are reasonably `normal' in terms
  of their \Lir and sSFR at these redshifts, a very high proportion
  (40 percent) of the sample are found to be interacting or
  kinematically disturbed systems. The interactions are mostly minor
  mergers, and hence maybe it is not surprising that they do not
  induce major star formation events.}
  
\item{There is a large variety of morphology in the gas and dust,
  and in some cases the \CI and dust do not appear to have the same
  spatial structure. The CO resolution is too poor to allow a spatial comparison but
  kinematically there are also differences between \CI and CO in the
  most disturbed systems.}
  
\item{The range of \lci/\lcoa in these sources is very large, covering
  the full range observed in local clouds in the Milky Way up to those
  seen in high-z QSO. We find two potential CO-`dark' candidates,
  detected in \CI and dust but not in CO.}


\item{We estimate the gas calibration parameters, \Xci, \aco and \gdr
  for each source using a likelihood approach. The sample averages
  are: Carbon abundance (required to calibrate \CI as a gas tracer)
  $\Xci=1.6^{+1.3}_{-0.2} \times 10^{-5}$, while
  $\aco=3.0^{+1.4}_{-0.7}\,\aunit$ and $\gdr=128^{+54}_{-35}$, where
  we quote the sample log means and 16-84th percentile ranges. These
  factors do not include a correction for He. The starting assumption
  is that the average \gdr is similar to that for the Milky Way in
  these metal-rich massive galaxies. In other commonly used
  calibration units, which do account for He, these translate to
  $\aci=18.8\,\aunit$ and $\asub=5.9\times10^{12}\,\rm{W\,Hz^{-1}
    \msun^{-1}}$.}

\item{The gas fractions in these galaxies range from 0.06--0.32, with
  an average of $\fg=0.19$ and a scatter of 0.07. This is a very small
  range of gas fraction, which is surprising given the very diverse
  galaxies in a volume limited 250\mic selected sample at $z=0$.}


\end{itemize}

We conclude this work by noting the supreme importance of
(u,v)-matched imaging in all three prime $\rm H_2$ gas mass tracers
explored here, in order to fully and accurately reveal the
similarities and differences of their distributions in galaxies.
ALMA+ACA+(total power) imaging could provide the key for such
comparisons in the future, by allowing the most unbiased imaging for
the high-frequency \CI lines and sub-mm dust continuum.

\section*{Data Availability}
The data underlying this article are publicly available from the ALMA archive
http://almascience.eso.org/aq/ using the project code 2012.1.00973.S. 

\section*{Acknowledgements} The authors thank the anonymous referee for their time in helping to improve the manuscript. LD thanks P. Papadopoulos for superb cooking, enlightening conversations and a careful proof-read and improvement of the draft manuscript.
LD and SJM acknowledge support from the European Research Council
Advanced Investigator grant, COSMICISM and Consolidator grant, COSMIC
DUST. HLG acknowledges support from the European Research Council
Consolidator grant, COSMIC DUST. This paper makes use of the following
ALMA data: ADS/JAO.ALMA\#2012.1.00973.S. ALMA is a partnership of ESO
(representing its member states), NSF (USA) and NINS (Japan), together
with NRC (Canada), MOST and ASIAA (Taiwan), and KASI (Republic of
Korea), in cooperation with the Republic of Chile. The Joint ALMA
Observatory is operated by ESO, AUI/NRAO and NAOJ. The National Radio
Astronomy Observatory is a facility of the National Science Foundation
operated under cooperative agreement by Associated Universities,
Inc. The {\it {\em Herschel}}-ATLAS is a project with {\it {\em
    Herschel}}, which is an ESA space observatory with science
instruments provided by European-led Principal Investigator consortia
and with important participation from NASA. The H-ATLAS web-site is
http://www.h-atlas.org. GAMA is a joint European-Australasian project
based around a spectroscopic campaign using the Anglo- Australian
Telescope. The GAMA input catalogue is based on data taken from the
Sloan Digital Sky Survey and the UKIRT Infrared Deep Sky
Survey. Complementary imaging of the GAMA regions is being obtained by
a number of independent survey programs including GALEX MIS, VST KIDS,
VISTA VIKING, WISE, {\em Herschel}-ATLAS, GMRT and ASKAP providing UV
to radio coverage. GAMA is funded by the STFC (UK), the ARC
(Australia), the AAO, and the participating institutions. The GAMA
website is: http://www.gama-survey.org/. This publication has made use
of data from the VIKING survey from VISTA at the ESO Paranal
Observatory, programme ID 179.A-2004. Data processing has been
contributed by the VISTA Data Flow System at CASU, Cambridge and WFAU,
Edinburgh.

\bibliographystyle{mnras/mnras}
\bibliography{Masterbib}

\bsp

\appendix

\section{Notes on individual sources}
\label{sourcesS}
\subsubsection*{SDP.163 (SK)}
This is the most massive galaxy with $\Ms=2.3\times10^{11}\,\msun$,
and $f_g = 0.13$, residing in a GAMA group (ID104290) with estimated
dynamical mass of $4.3\times10^{13}\,\msun$ \citep{Robotham2011}. The
GAMA spectrum (Fig.~\ref{A163F}) has good SNR, and strong O[{\sc
    ii}] and O[{\sc iii}]. It is classed as a Sy 2 using the
\citet{Lamareille2010} blue classification. The source, which is
strongly detected in CO, \CI and dust continuum, is a highly inclined
massive disk galaxy with $\rm{FWHM=700-800}$\kms in both CO and
\CI. Continuum subtraction was performed in the $u,v$-plane for the B7
cube. There is an extra kinematic component in the \CI cube at the
northern end of the galaxy which is not present in the CO map, but
which is significant in the \CI cube with $\rm{S_{CI}=1.7\pm0.5}$
Jy~\kms. The flux reported in Table~\ref{CIdataT} includes this
component. This suggests that the CO/C{\sc i} ratio is not constant
across the galaxy. Matched resolution imaging is needed to further
study the spatial variation in the \CI/CO ratio.
\subsubsection*{SDP.1160 (SK)}
\begin{figure*}
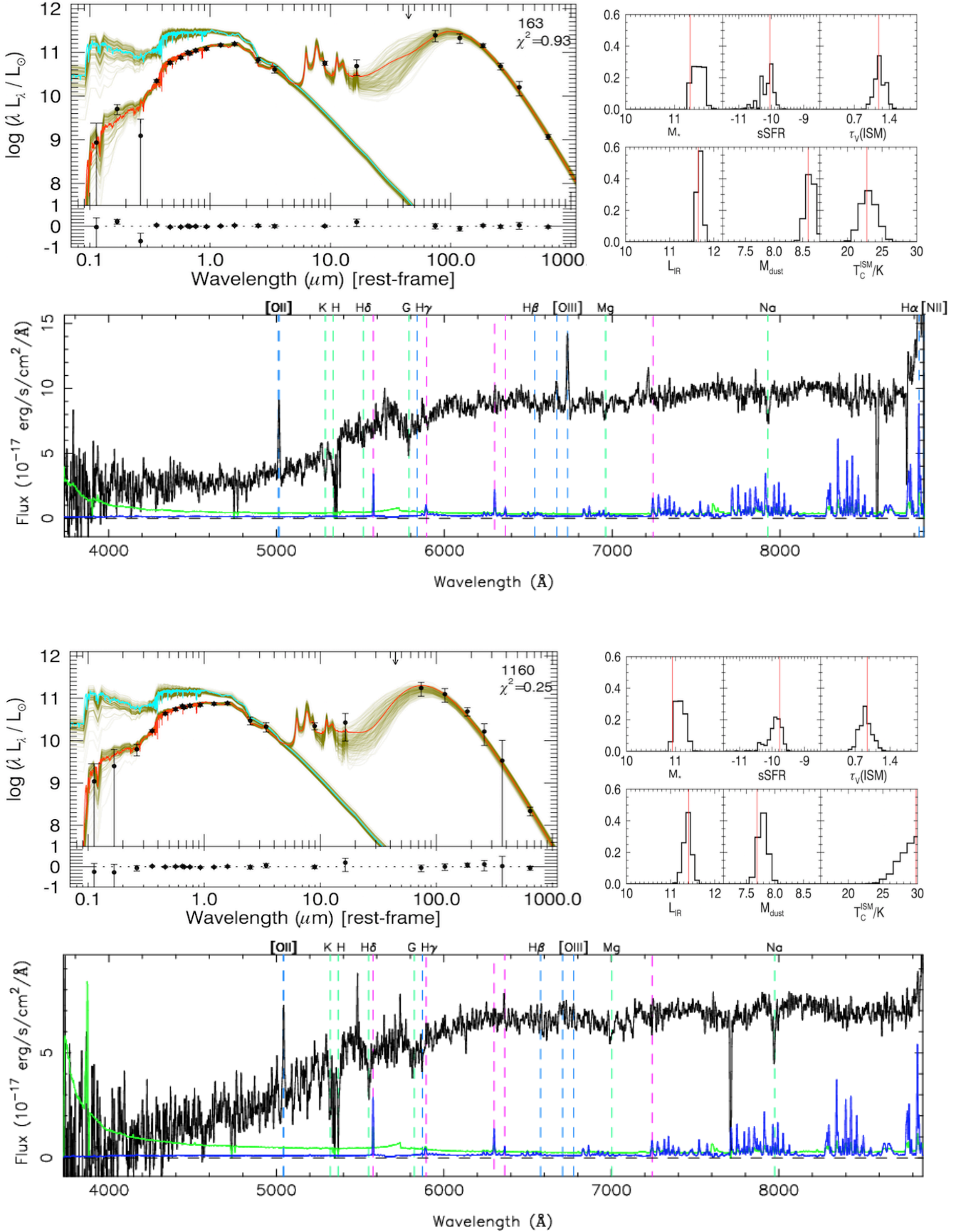

\vspace{-1cm}
\begin{minipage}{0.62\linewidth}
\begin{lpic}[]{163_sed(0.365,0.34)}
\end{lpic}
\end{minipage}
\begin{minipage}{0.37\linewidth}
\begin{lpic}{163_pdfs_sm(0.14,0.17)}
\end{lpic}
\end{minipage}
\begin{minipage}{0.98\textwidth}
\vspace{-0.8cm}
\begin{lpic}[]{SDP163_zspec(0.55,0.52)}
\end{lpic}
\end{minipage}
\begin{minipage}{0.62\linewidth}
\begin{lpic}[]{1160_sed(0.365,0.34)}
\end{lpic}
\end{minipage}
\begin{minipage}{0.37\linewidth}
\begin{lpic}{1160_pdfs_sm(0.14,0.17)}
\end{lpic}
\end{minipage}
\begin{minipage}{0.98\linewidth}
\vspace{-0.8cm}
\begin{lpic}[]{SDP1160_zspec(0.55,0.52)}
\end{lpic}
\end{minipage}
\caption{\label{A163F} {\sc magphys} best fit SED in red with 200 of
  the next best fits in olive. {\em top:} Optical spectrum for SDP.163 showing broad O[{\sc
      iii}] and O[{\sc ii}] in emission on a reddened
  continuum. {\em bottom}: Optical spectrum for SDP.1160 showing O[{\sc ii}]
  in emission on a reddened continuum.}
\end{figure*}

This galaxy has $\Ms = 1\times10^{11}\,\msun$ and $f_g = 0.07$. It is the brightest member of a GAMA group (ID103864) with a
second galaxy at the same velocity a distance of 0.076 Mpc/h away. The
GAMA spectrum (Fig.~\ref{A163F}) shows O[{\sc ii}] in emission on a
reddened continuum. This source has bright and compact \CI and dust
emission (Fig.~\ref{1160F}). 
\subsubsection*{SDP.2173 (SK)} 
This is a face-on spiral galaxy with $\Ms=7.9\times10^{10}\,\msun$ and $f_g=0.14
$. The optical spectrum (Fig.~\ref{A2173F}) shows broad lines of
O[{\sc iii}] lines and narrow O[{\sc ii}] and H$\beta$ emission, the
continuum is quite blue and there is weak dust attenuation. This is
the brightest galaxy in the r-band, with an absolute magnitude of $R =
-21.7$, and while not a member of a GAMA group, it has three dwarf
companions with compatible photo-z ($\rm{z_{ph}=0.37-0.38}$) at
projected distances of 40--90 kpc. The dust, \CI and CO emission are
spatially extended, with a narrow profile consistent with the face-on
orientation (Fig.~\ref{2173F}). 
\subsubsection*{SDP.3132 (SK)}
\begin{figure*}
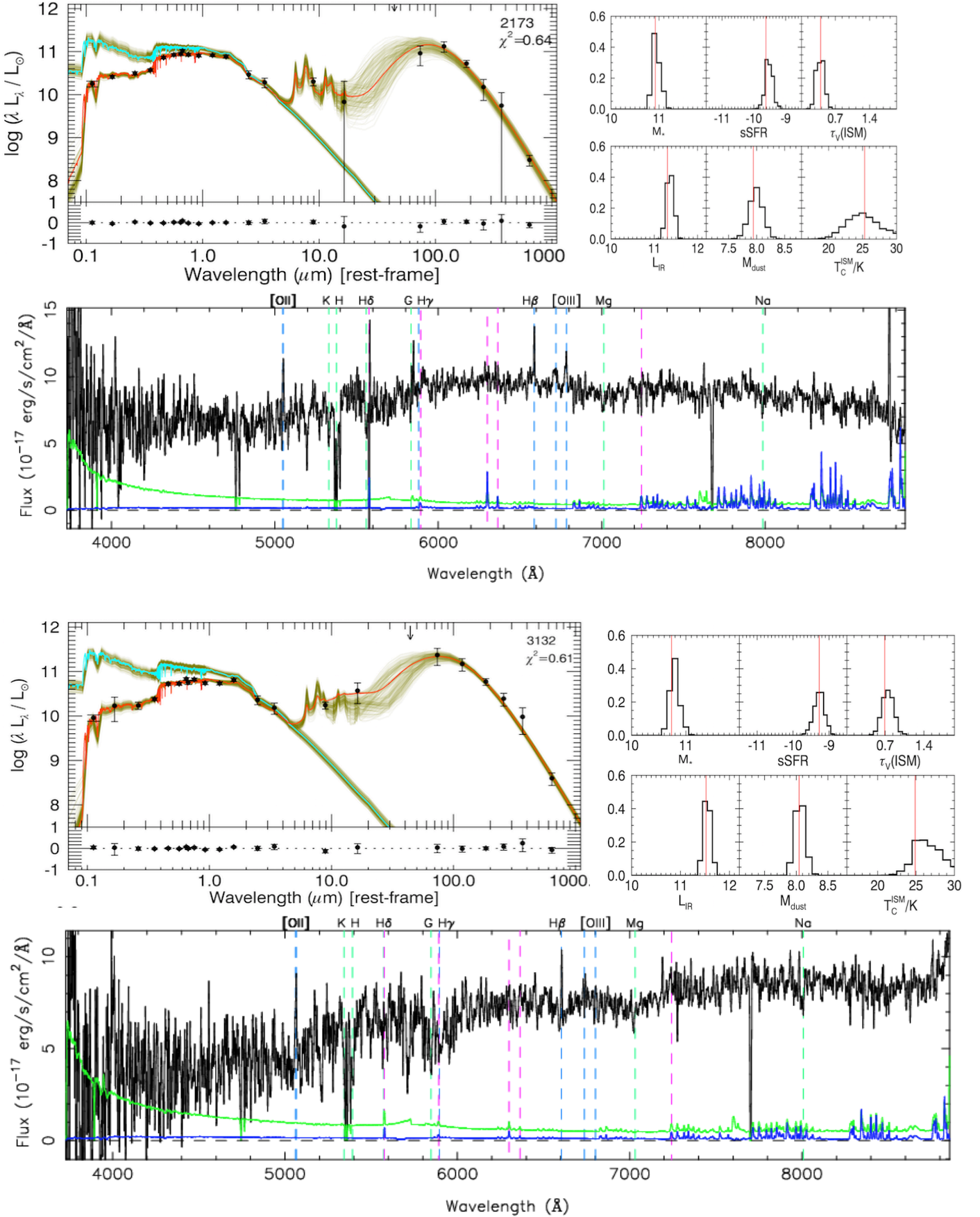

\vspace{-1cm}
\begin{minipage}{0.62\linewidth}
\begin{lpic}[]{2173_sed(0.36,0.34)}
\end{lpic}
\end{minipage}
\begin{minipage}{0.37\linewidth}
\begin{lpic}{2173_pdfs_sm(0.14,0.17)}
\end{lpic}
\end{minipage}
\begin{minipage}{0.98\textwidth}
\vspace{-0.8cm}
\begin{lpic}[]{2173_zspec(0.55,0.53)}
\end{lpic}
\end{minipage}
\begin{minipage}{0.62\linewidth}
\vspace{-0.3cm}
\begin{lpic}[]{3132_sed(0.375,0.35)}
\end{lpic}
\end{minipage}
\begin{minipage}{0.37\linewidth}
\begin{lpic}{3132_pdfs(0.25,0.29)}
\end{lpic}
\end{minipage}
\begin{minipage}{0.98\textwidth}
\vspace{-0.8cm}
\begin{lpic}[]{SDP3132_zspec(0.58,0.55)}
\end{lpic}
\end{minipage}
\caption{\label{A2173F} {\sc magphys} best fit SED in red with 200 of
  the next best fits in olive. SDP.2173 optical
  spectrum showing weak O[{\sc ii}], H$\beta$ and broad O[{\sc iii}]
  in emission. The attenuation for this source is relatively low as it
  is a face on spiral.SDP.3132 optical spectrum showing O[{\sc ii}]
  and H$\beta$ in emission, O[{\sc iii}] is very weak. }
\end{figure*}

This is a face-on spiral galaxy with $\Ms=5.4\times10^{10}\,\msun$ and
$f_g= 0.21$. The GAMA spectrum (Fig.~\ref{A2173F}) is noisy with
emission lines of O[{\sc ii}] and H$_{\beta}$. The dust, \CI and CO
emission are extended. The inclusion of ACA data taken in Cycle 7
shows that as much as 60 percent of the continuum flux of this source
was resolved out in the original 12-m configuration. The \CI emission
was not so affected, with a 10 percent increase once the ACA data were
included. The CO and \CI lines are very narrow with consistent
profiles.
\subsubsection*{SDP.3366 (DK)}
This galaxy is massive with $\Ms=1.3\times10^{11}\,\msun$ and
$f_g=0.04$. The GAMA spectrum (Fig.~\ref{A3366F}) shows strong and
broad lines of O[{\sc iii}] and strong O[{\sc ii}]. H$_{\delta}$ and
H$_{\beta}$ are strong in absorption, Ca H is stronger than Ca K and
there is no strong 4000\AA break, indicating a recent significant star
formation event. The line ratios put it in the Sy 2 region of the
\citet{Lamareille2010} blue diagram, however, with such strong
absorption underlying the H$\beta$ line this will be a very uncertain
classification. The Sersic fits from GAMA \citep{Kelvin2012} are steeper in the
K-band ($n_K=4.6\pm0.2$) compared to the Z-band ($n_Z=2.6\pm0.1$),
suggesting that the bluer emission is coming from a rejuvenated disk
while the older stellar population is consistent with a
spheroid. There is a small, blue neighbouring source in the
optical imaging which has photometric redshift $z_{ph}=0.26-0.38$ from
the KIDS survey \citep{KIDSMLPz,Wright2018}. This is denoted as
3366.b in Table~\ref{CIdataT} and Fig.~\ref{3366F}.

The dust and gas morphology and kinematics are complex. Firstly,
Fig.~\ref{3366F} shows strong 850\mic continuum emission from a SMG
coincident with a very red source ($\rm{K_{AB}=19.0}$) located
3.4\asec to the north-east of the $z=0.35$ galaxy. There is no line
emission associated with this continuum (see spectral profile in
Fig~\ref{3366F}).

Both the bluer satellite galaxy and the massive central target galaxy
show weak dust emission. The 850\mic flux for the satellite was fitted
at the same time as fitting a point source to the position of the SMG
using the 2-D Gaussian fit task {\sc imfit}. There is also broad \CI
emission spatially coincident with the blue satellite galaxy (3366.b),
shown as the black spectral profile in the lower right panel of
Fig~\ref{3366F}. There is no CO counterpart to the satellite visible
in the depth of our 3-mm data.

The massive central galaxy has \CI emission in two velocity
components: 3366.a($v1$) at the optical redshift and a blue-shifted
component, 3366.a($v2$), which overlaps in velocity with the broad \CI
emission from the small blue satellite galaxy (3366.b). These two
components are shown in the upper right panel of Fig~\ref{3366F}. This
suggests an interaction between the two sources, probably responsible
for the recent star formation activity in an otherwise older stellar
population, as indicated by the optical spectrum and Sersic fits. The
spectral profiles for the velocity component centered on the optical
redshift, 3366.a($v1$), are shown in the lower left panel of Fig~\ref{3366F}
with CO in blue and \CI in black, showing reasonable agreement between
the CO and \CI kinematics for this component. CO is only detected at
the $v1$ velocity, the blue-shifted \CI emission associated with the
main target, 3366.a($v2$), and that of the satellite galaxy (3366.b)
have no corresponding CO emission.

To compute the \CI, CO and dust luminosities used in the comparison of
the tracers and gas masses (Section~\ref{gasS}) we use the values for
the 3366.a($v1$) component only.

\subsubsection*{SDP.4104 (DK)}
\begin{figure*}
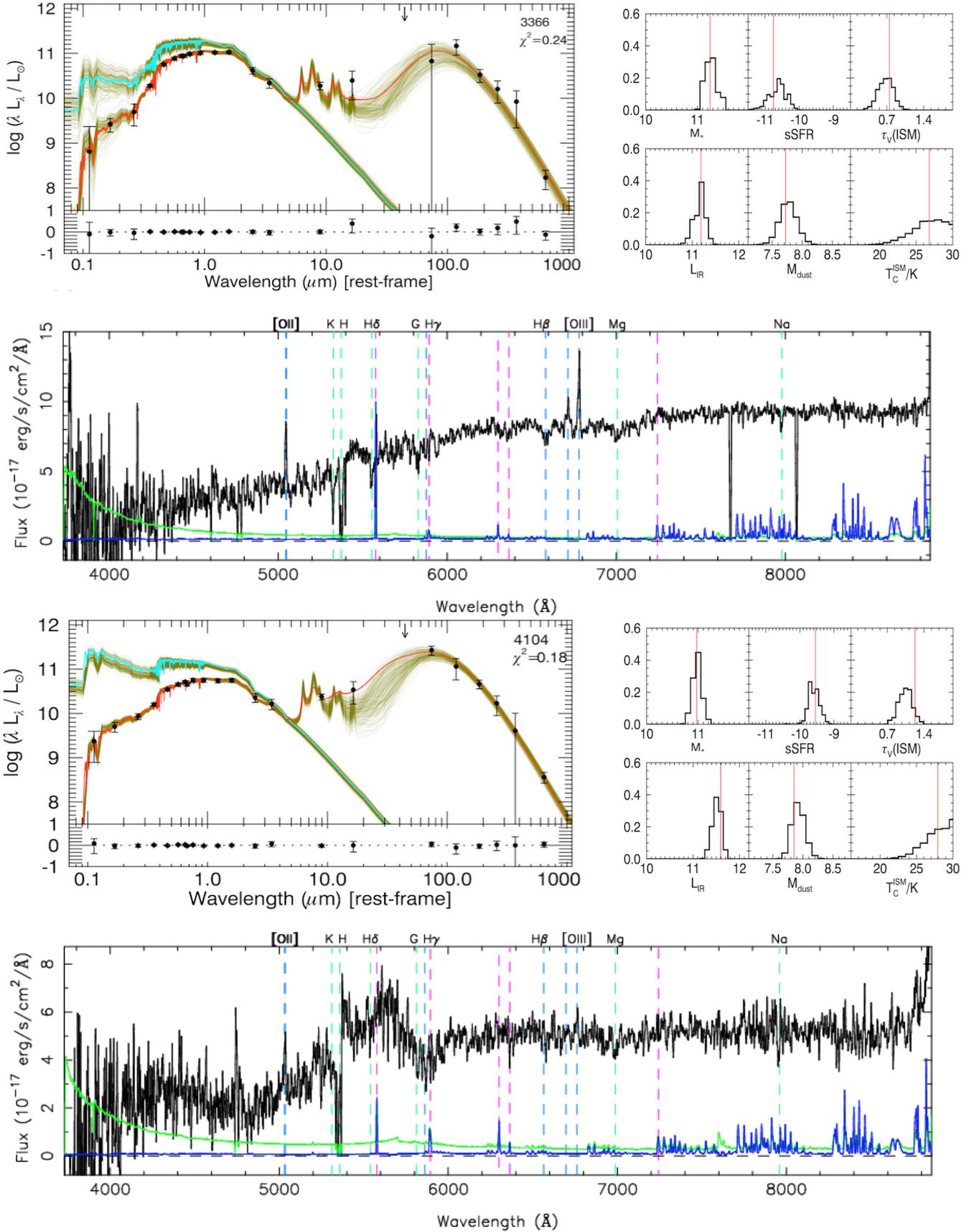

\begin{minipage}{0.62\linewidth}
\vspace{-1.2cm}
\begin{lpic}[]{3366_sed(0.36,0.34)}
\end{lpic}
\end{minipage}
\begin{minipage}{0.37\linewidth}
\vspace{-1.2cm}
\begin{lpic}{3366_pdfs(0.23,0.27)}
\end{lpic}
\end{minipage}
\begin{minipage}{0.98\textwidth}
\vspace{-0.5cm}
\begin{lpic}[]{SDP3366_zspec(0.55,0.53)}
\end{lpic}
\end{minipage}
\begin{minipage}{0.62\linewidth}
\vspace{-1.0cm}
\begin{lpic}[]{4104_sed(0.36,0.34)}
\end{lpic}
\end{minipage}
\begin{minipage}{0.37\linewidth}
\vspace{-1.2cm}
\begin{lpic}{4104_pdfs(0.23,0.27)}
\end{lpic}
\end{minipage}
\begin{minipage}{0.98\textwidth}
\vspace{-0.5cm}
\begin{lpic}[]{SDP4104_zspec(0.55,0.53)}
\end{lpic}
\end{minipage}
\caption{\label{A3366F} {\sc magphys} best fit SED in red with 200 of the
  next best fits in olive. SDP.3366 optical spectrum showing strong
  Balmer absorption with broad O[{\sc iii}] and O[{\sc ii}]. SDP.4140 optical spectrum which is noisy with
  weak O[{\sc ii}] and O[{\sc iii}] emission and a large 4000\AA
  break.}
\end{figure*}

The H-ATLAS target is the southern source (4104.a) of a galaxy pair,
with $\Ms=7.8\times10^{10}\,\msun$ and $f_g=0.13$. The GAMA optical
spectrum (Fig.~\ref{A3366F}) has a noisy continuum with detection of
O[{\sc ii}], and very weak O[{\sc iii}]. The neighbouring optical
source to the north (4104.b) has $\rm{z_{ph}=0.37}$ from KIDs
\citep{KIDSMLPz} but was below the $r$-band limit for inclusion in the
GAMA spectroscopic sample. There is a relatively narrow CO line
feature close to the optical velocity of the target ($\rm{\Delta
  V_{CO}=-72}$\kms) which appears to be elongated along the major axis
of the galaxy (see lower right inset panel in Fig.~\ref{4104F}). A
second, blue-shifted CO component is detected at the position of the
northern companion, with a velocity offset of $-360$\kms, confirming
that the two galaxies are interacting. The dust continuum image also
shows emission at the location of both sources, albeit at very low SNR
in the main target (which is extended). The \CI data for this target
are affected by an atmospheric absorption line in the middle of the
expected velocity profile, which was flagged before making the
spectral cubes. This resulted in two 50\kms channels being unusable
due to the large noise from mostly flagged data. The moment 0 maps
were made excluding these affected channels ($-50\rightarrow+50$\kms)
resulting in a lower limit to the integrated line flux (listed in the
first row of Table~\ref{CIdataT}). Notwithstanding this, both galaxies
are detected in \CI with positions and velocities consistent with the
CO data. The elongation seen in the CO emission along the major axis of the
southern galaxy is also seen in \CI.
 
In an attempt to correct for the missing flux, we have added a second
measurement in Table~\ref{CIdataT} denoted $^c$ which includes a
contribution from the two missing channels assuming that their flux is
the average of the neighbouring good channels on either side. For the
northern source (4104.b) with a velocity offset of $\Delta
V_{CO}=-360$\kms, this is a negligible addition but for the main
optical target (4104.a), the extra flux estimated this way is 50\% of
the measurement with the missing channels. We use both sets of fluxes
(corrected and uncorrected) in the comparison of gas mass tracers in
Section~\ref{gasS} summing together 4104.a and 4104.b (since both will
be contributing equally to the H-ATLAS photometry). There is only a
modest change in the parameters when the correction is made, and so
the decision to correct or not is not critical to the conclusions in
the paper.

\subsubsection*{SDP.5323 (DK)}
\begin{figure*}
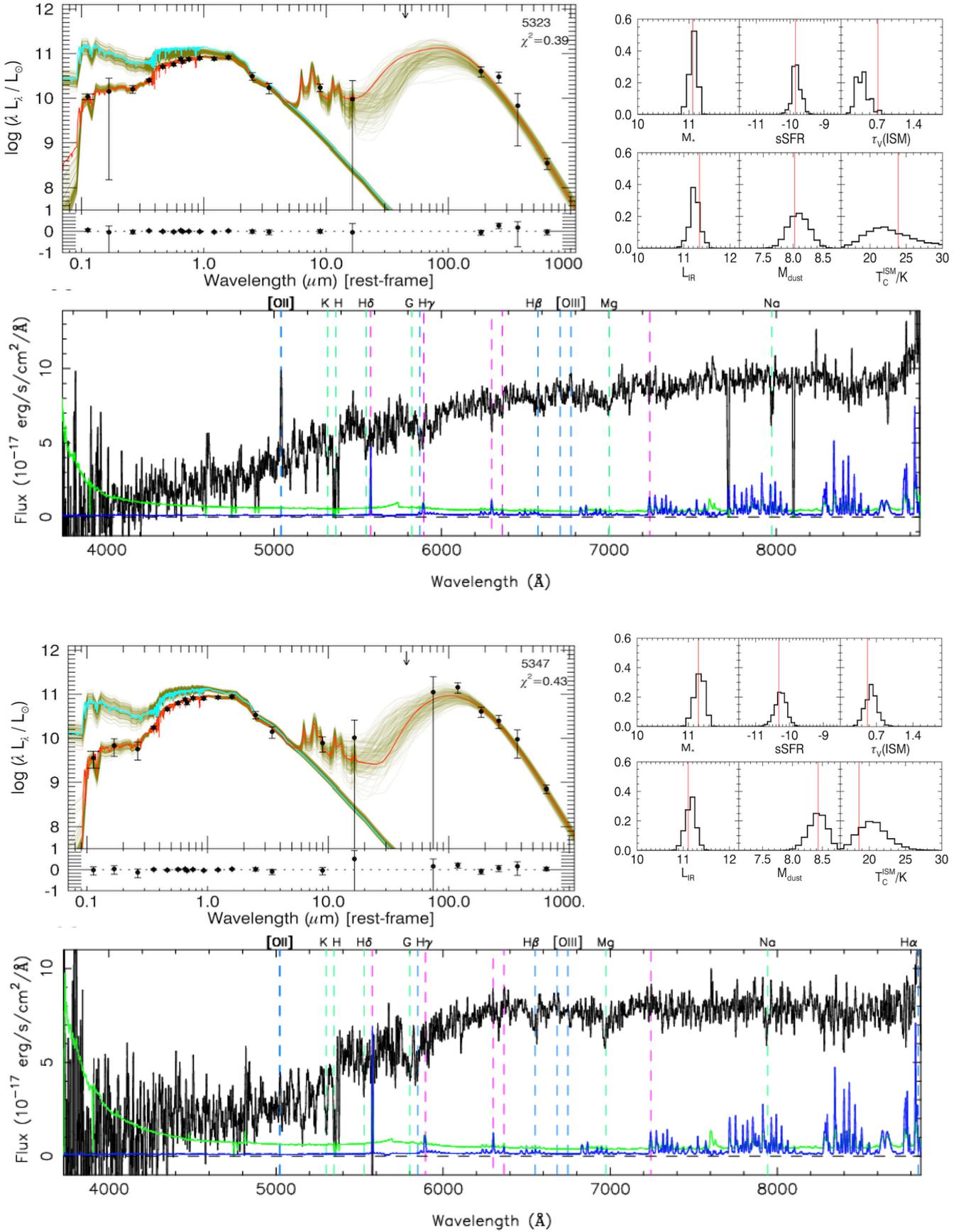

\begin{minipage}{0.62\linewidth}
\vspace{-1cm}
\begin{lpic}[]{5323_sed(0.365,0.34)}
\end{lpic}
\end{minipage}
\begin{minipage}{0.37\linewidth}
\vspace{-0.9cm}
\begin{lpic}{5323_pdfs(0.23,0.27)}
\end{lpic}
\end{minipage}
\begin{minipage}{0.98\textwidth}
\vspace{-0.8cm}
\begin{lpic}[]{SDP5323_zspec(0.55,0.53)}
\end{lpic}
\end{minipage}
\begin{minipage}{0.62\linewidth}
\begin{lpic}[]{5347_sed(0.365,0.34)}
\end{lpic}
\end{minipage}
\begin{minipage}{0.37\linewidth}
\vspace{-1.0cm}
\begin{lpic}{5347_pdfs(0.23,0.25)}
\end{lpic}
\end{minipage}
\begin{minipage}{0.98\textwidth}
\vspace{-0.8cm}
\begin{lpic}[]{SDP5347_zspec(0.55,0.53)}
\end{lpic}
\end{minipage}
\caption{\label{A5323F} {\sc magphys} best fit SED in red with 200 of the next
  best fits in olive. SDP.5323 optical spectrum with O[{\sc ii}]
  and weak but broad O[{\sc iii}] emission. SDP.5347 optical spectrum which has no emission
  lines, a large 4000\AA break and strong Balmer absorption features.}
\end{figure*}

This galaxy has $\Ms=1.1\times10^{11}\,\msun$ and $f_g=0.11$. The GAMA
spectrum (Fig.~\ref{A5323F}) has strong O[{\sc ii}] emission and weak
but broad O[{\sc iii}]. The Balmer series is visible in absorption
with CaH being stronger than CaK and there is some residual H$\beta$
emission in the absorption trough, indicating a recent episode of star
formation (A-star signatures). The low level optical emission is
suggestive of a disturbed disk in the young stellar population and
there is a small satellite of similar optical colour to the south west
which has a 9-band photo-z of $z_{ph}=0.27$
\citep{Wright2018}. Figure~\ref{5323F} ({\em top left}) shows a very
complex morphology. Continuum dust emission is detected along the
major axis of the galaxy with a peak located just south of the
nucleus, while the \CI appears in two peaks either side of the
dust. The line emission is clearly resolved and so tapering was
employed to improve the visibility of low level extended flux (yellow
contours in Fig.~\ref{5323F} ({\em top left}) are \CI smoothed by
1.5\asec). The top right panel shows CO contours on a tapered \CI
moment-0 map, and it can be seen that the CO appears to be more
similar to the dust in its position and morphology but matched
resolution observations are required to investigate further the
correspondence between the CO and \CI and dust.

This source also has a complex velocity field, with a velocity offset
between the northern and southern \CI peaks as shown in
Fig.~\ref{5323F} ({\em centre right}). The green and blue contours
denote the two velocity components: v1=-273\kms (blue) and v2=+107\kms
(green). The {\em centre right} panel shows the velocity weighted
moment map masked at 2$\sigma$ for these main two components which we
are certain are associated with SDP.5323. The two velocity peaks are
quite clearly distinguished although the SNR is not high enough to say
if this is consistent with rotation. A third \CI component at much
higher velocity ($v3=+1239\,\kms$) is shown as red contours in
Fig.~\ref{5323F} ({\em centre left}) with an extended and complex
morphology. This gas may be associated with tidal disruption
experienced by the galaxy. The total flux for the main galaxy
components ($v1$ and $v2$) was measured by making two moment 0 maps in
the velocity ranges shown by the blue and green contours in
Fig.~\ref{5323F}, measuring the fluxes on each and then summing
them. These are the measurements which are used in the analysis of the
gas mass and calibration. The CO shows no complex morphology or
kinematics and has a velocity profile similar to the combined $v1$ and
$v2$ components in \CI, as shown in Fig.~\ref{5323F} ({\em lower
  left}). The {\sc magphys} SED fit (Fig.~\ref{A5323F}) is not well
constrained for dust properties because there was no PACS coverage
for this source.
\subsubsection*{SDP.5347 (DK)}
This galaxy has $\Ms=1.3\times10^{11}\,\msun$ and $f_g=0.17$. The GAMA
spectrum (Fig.~\ref{A5323F}) shows a strong 4000\AA break and Balmer,
Ca-H and K and Mg absorption features, but no emission lines are
seen.\footnote{H$\alpha$ does appear right at the edge of the spectrum
  but there is often a problem with the throughput calibration in this
  region so we do not have confidence in this as the only line.} The
$\rm{Krg}$ image (Fig.~\ref{5347F} {\em top left}) shows the massive central galaxy
`5347.a' and two small sources to the west and north (`b' and `c')
which have bluer optical colours. The small source to the west is not
de-blended by KIDs and the one to the north has $\rm{z_{ph}}=0.62$
\citep{KIDSMLPz,Wright2018}, however there is a colour gradient across
this source, which may make the fluxes used for the photo-z
inaccurate. 

The CO moment-0 map is integrated over the velocity range
($-200\rightarrow +500$\kms) and shows an extended structure
stretching N--S between the two small optical galaxies and the disk of
the target. 

There is dust emission at the position of the northern optical
companion (5347.c), and also a background SMG (see Dunne et al. {\em
  submitted}) to the North-East of the system. The dust emission
elsewhere is very weak.

Tapering the continuum subtracted \CI cube reveals a structure between
the two small satellites to the west at similar velocity to the CO
emission in the same region (Fig~\ref{5347F} {\em top left} orange
contours: component `b,c', and {\em lower left}). There is an
additional component at large blue-shifted velocity ($-750\kms$) which
has no counterpart in CO. This blue-shifted emission peaks at exactly
the same location as the main velocity component between the optical
satellites $b$ and $c$, and also has a second clump just north of the
main optical galaxy. There is a second \CI component associated with
the main galaxy reaching out in the direction of $b$, this is shown as
red contours in Fig~\ref{5347F}({\em top left}) with the spectral
profile shown in the {\em centre left} panel. There is no CO component
at the same velocity.

Based on the CO and \CI morphology and kinematics, we assume this is
an interacting system, though there is no optical verification for
this. There is very little gas or dust associated with the centre of
SDP.5347.a, which is not so surprising given the rather passive nature
of the optical spectrum. 

There is also a clump of CO emission further South which is shown in
Fig~\ref{5347F}({\em top right}), which may also be related to tidal
disruption \citep[e.g.][]{Thomas7465,Zhu2007}. It has a very broad
velocity range (Fig~\ref{5347F}:{\em centre right}). The band 7 field
of view is much smaller and so this CO clump falls beyond the
half-power radius for the primary beam (shown as an orange dashed
circle in the {\em top right} panel) which means we would not have the
sensitivity to detect any counterpart in \CI. A \CI clump is however
detected significantly slightly north of the CO clump but at
overlapping velocity ranges. The \CI clump is coincident with a faint
blue patch of emission in the $g$-band image (smoothed in
Fig~\ref{5347F} {\em top right}). The southern clumps are listed as SDP.5347(S) in
Tables~\ref{COdataT} and ~\ref{CIdataT} but are not used further in any analysis.

h\subsubsection*{SDP.5526}
This isolated galaxy has $\Ms=6.6\times10^{10}\,\msun$ and
$f_g=0.2$. The optical GAMA spectrum (Fig.~\ref{A5526F}) shows strong
O[{\sc ii}] and also H$\beta$ in emission. At the very red end of the
spectrum the N[{\sc ii}] and H$\alpha$ lines can be seen, though the
spectrum is very noisy in this region. This source is well detected in
CO, C{\sc i} and dust continuum, with a narrow line-width and the
signature of rotation in the velocity weighted moment map
(Fig.~\ref{5526F}). The dust and gas are centrally concentrated,
however there is lower level continuum emission which contributes
significantly to the integrated flux in the tapered image. The WISE
22\mic flux for this source seems very high relative to the
permissible SEDs from {\sc magphys} in Figure~\ref{A5526F}.
\subsubsection*{SDP.6216}
\begin{figure*}
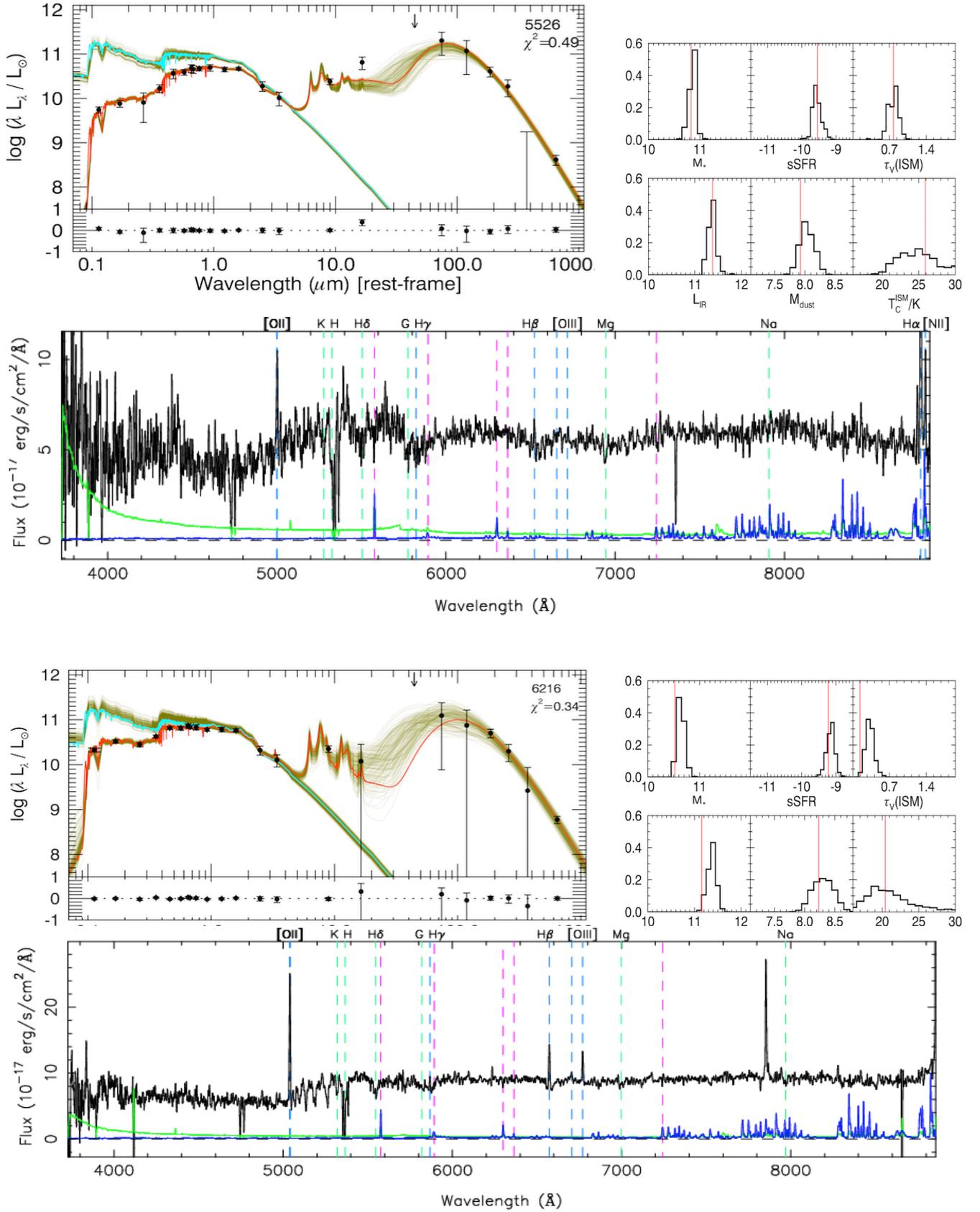

\begin{minipage}{0.62\linewidth}
\vspace{-1cm}
\begin{lpic}[]{5526_sed(0.365,0.34)}
\end{lpic}
\end{minipage}
\begin{minipage}{0.37\linewidth}
\begin{lpic}{5526_pdfs(0.23,0.27)}
\end{lpic}
\end{minipage}
\begin{minipage}{0.98\textwidth}
\vspace{-0.5cm}
\begin{lpic}[]{SDP5526_zspec(0.55,0.53)}
\end{lpic}
\end{minipage}
\vspace{-1.5cm}
\begin{minipage}{0.62\linewidth}
\begin{lpic}[]{6216_sed(0.365,0.34)}
\vspace{0.3cm}
\end{lpic}
\end{minipage}
\begin{minipage}{0.37\linewidth}
\begin{lpic}{6216_pdfs(0.23,0.27)}
\end{lpic}
\end{minipage}
\begin{minipage}{0.98\textwidth}
\begin{lpic}[]{SDP6216_zspec(0.55,0.5)}
\end{lpic}
\end{minipage}
\caption{\label{A5526F} {\sc magphys} best
  fit SED in red with 200 of the next best fits in olive. SDP.5526 optical spectrum shows O[{\sc
      ii}], weak H$\beta$ and strong H$\alpha$ at the red edge of the
  band. SDP.6216 optical spectrum
  showing strong emission in O[{\sc ii}] and H$\beta$ on a blue
  continuum with strong Balmer absorption features.}
\end{figure*}

An isolated, blue and vigorously star forming galaxy with
$\Ms=3.8\times10^{10}\,\msun$ and $f_g=0.33$. The optical GAMA
spectrum (Fig.~\ref{A5526F}) shows strong lines of O[{\sc ii}] and
H$\beta$ and weaker O[{\sc iii}]. In conjunction with the strong
Balmer absorption lines, this indicates a recent burst of
star-formation, which is now declining. The {\sc magphys} SED fit is
shown in Fig.~\ref{A5526F} and is in agreement with the optical
spectrum, favouring a high sSFR with low dust extinction. The optical
{\em gri} image (Fig.~\ref{6216F}) shows a sharp stellar caustic to
the south-east and disturbed clumpy star forming regions to the
North. This asymmetry suggests a possible recent interaction either
with another galaxy (now subsumed), accretion from the cosmic web or
simply disk instability. There are good detections of \CI and CO, with
a relatively narrow line profile (FWHM$\sim 250$\kms), as expected for
this low inclination galaxy. The morphology of the dust and \CI
emission is different in this source. The \CI is centrally
concentrated while the continuum morphology is complex: low level
emission is visible across the northern half of the optical disk in
the tapered image (grey contours in the right panel of Fig.~\ref{6216F}). There is a bright
and compact (but not point-like) dust source in the northern spiral
arm, where the three colour optical image shows clumpy regions of star
formation (upper inset). The dust clump could be interpreted either as
an obscured SF region in the disk or a background SMG. The flux of the
dust clump is 0.43~mJy. Finally, there is extended clumpy dust
emission to the East of the galaxy; it does not coincide with any
optical features but may be tracing cold gas associated with the outer
disk. Three values for the 850\mic continuum flux are given in
Table~\ref{CIdataT}. The first includes the compact clump and the
emission in the disk, the second is a robust minimum flux which
subtracts the compact clump and does not include the Eastern
extension. The third is a robust maximum which includes all three
components. Careful inspection of the optical/NIR images and the
morphology of the compact clump leads us to prefer the explanation
that it is part of the galaxy disk and so we regard the first flux as
the most plausible estimate, and this is the one we use in our
analysis of the gas mass in Section~\ref{gasS}.

\subsubsection*{SDP.6418}
This appears to be an interacting system with the main galaxy,
SDP.6418.a, having $\Ms=6.2\times10^{10}\,\msun$ and $f_g=0.18$. The
GAMA spectrum (Fig.~\ref{A6418F}) shows a blue continuum with very
strong O[{\sc ii}] emission, and also significant O[{\sc iii}] and
H$\beta$ with H$\delta$ in absorption. There is H$\alpha$ just visible
at the red end of the spectrum. The smaller diffuse neighbour of
similar colour to the east, has $z_{ph}=0.42$ \citep{KIDSMLPz}. A
reasonably strong 850\mic continuum source in the primary beam allowed
us to self-calibrate this image. The dust continuum in both low
redshift galaxies is detected, but is weak and extended. This source
is a challenge to interpret as we do not find CO or \CI line emission
at the systemic velocity of the optical galaxy, but there is \CI
emission offset both red-ward and blue-ward. The \CI spectra did have
problematic spectral baselines in the region around $-200\rightarrow
0\,\rm{kms^{-1}}$. We can be sure that there is no strong signal in
that region, but cannot say anything else. Continuum subtraction was
performed with {\sc imcontsub} in order to try to flatten the
baselines but an order 3 polynomial was required which leads to extra
uncertainty in the line fluxes. Careful inspection before and after
subtraction shows that the subtraction has not produced any spectral
features which were not there already in the original cube. There are
three \CI components which are outside the systemic velocity range:
$v1\sim -400$\kms and $v2\sim +250$\kms are shown by the orange and
red contours in Fig.~\ref{6418F}(top left) and $v3\sim+450$\kms is shown in
green, though $v3$ is less convincing in the spectrum. The spectral
profiles are shown individually for each velocity component in
Fig.~\ref{6418F} with the colour of the spectrum matching the colour
of the contours. The profile for the $v1$ component (which is present
in both the main galaxy and the satellite) is shown as a solid orange
line for the main galaxy and a dotted brown line for the
satellite. The complex kinematics for this source, coupled with the
presence of a neighbour at a similar redshift indicates that there is
a disturbance of the gas but further observations would be necessary
to be certain which of these features are real and related to an
interaction.

We perform the analysis on gas mass in Section~\ref{gasS} using two
versions of the \CI fluxes: conservatively using just the $v2$
component, because this has the least offset from the systemic optical
velocity, and secondly using all three components which would be a
maximum. There is no significant CO emission detectable at any of
these velocities, nor at the systemic velocity. The ratio of line
luminosity \lci/\lcoa ($=\rm{R_{CI}}$) in this galaxy is thus higher
than we see in other sources, ranging from $0.54<\rm{R_{CI}}<0.94$,
using the measured CO flux which is only 2$\sigma$ significance. Using
the 3$\sigma$ upper limit on the CO flux gives a lower limit for this
source of $\rm{R_{CI}}> 0.35$. The average value in Milky Way clouds
is $\rm{R_{CI}} \sim 0.15$ \citep{Frerking1989}, while the average value seen in
QSO is $\rm{R_{CI}}\sim 0.29$ \citep{Walter2011}.

\subsubsection*{SDP.6451}
\begin{figure*}
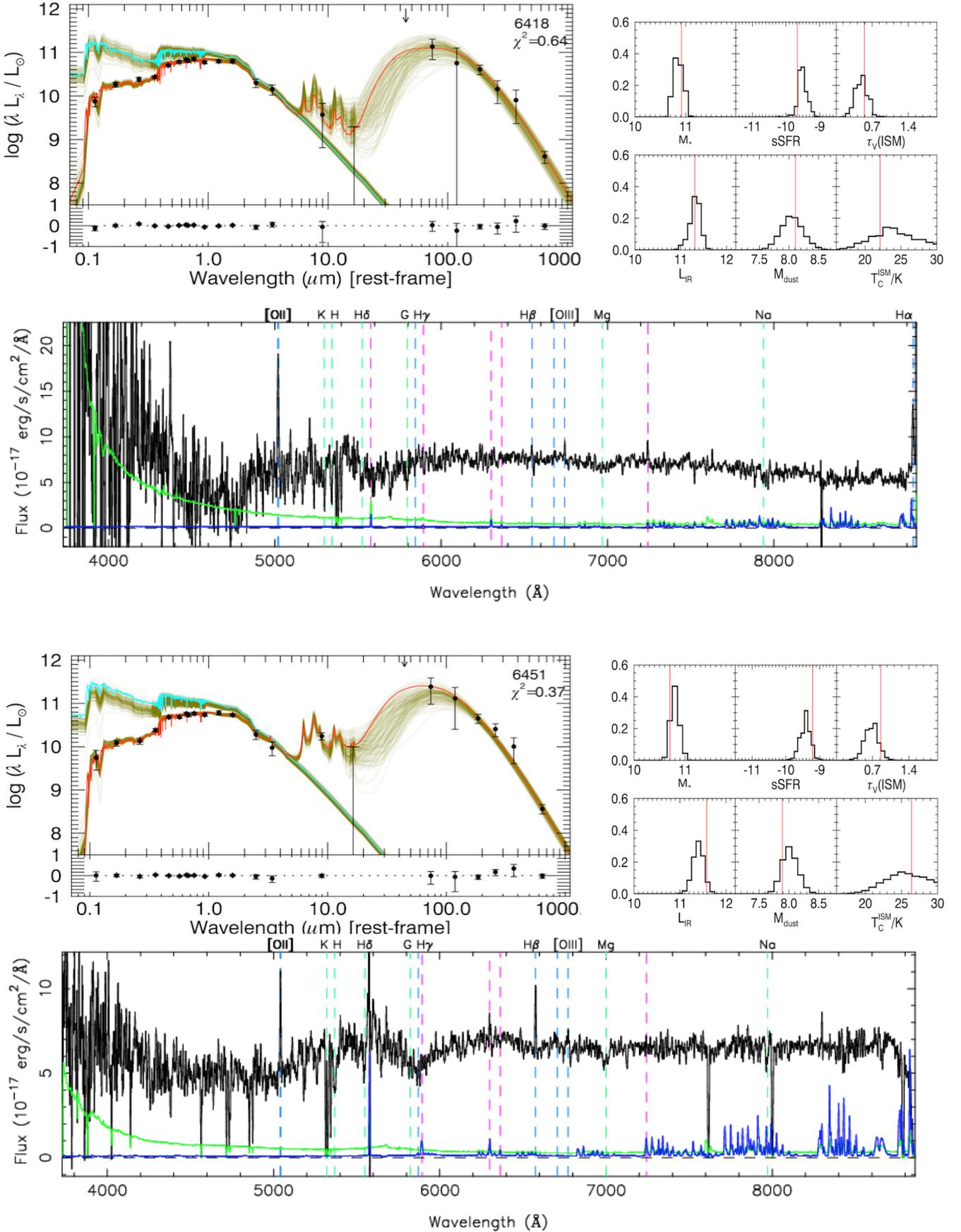

\begin{minipage}{0.62\linewidth}
\vspace{-1.1cm}
\begin{lpic}[]{6418_sed(0.365,0.34)}
\end{lpic}
\end{minipage}
\begin{minipage}{0.37\linewidth}
\vspace{-0.9cm}
\begin{lpic}{6418_pdfs(0.23,0.27)}
\end{lpic}
\end{minipage}
\begin{minipage}{0.98\textwidth}
\vspace{-0.6cm}
\begin{lpic}[]{SDP6418_zspec(0.55,0.53)}
\end{lpic}
\end{minipage}
\vspace{-1cm}
\begin{minipage}{0.62\linewidth}
\begin{lpic}[]{6451_sed_z4cor(0.365,0.34)}
\end{lpic}
\end{minipage}
\begin{minipage}{0.37\linewidth}
\begin{lpic}{6451_pdfs_z4cor(0.23,0.27)}
\end{lpic}
\end{minipage}
\begin{minipage}{0.98\textwidth}
\begin{lpic}[]{SDP6541_zspec(0.55,0.53)}
\end{lpic}
\end{minipage}
\caption{\label{A6418F} {\sc magphys} best fit SED in red with 200 of
  the next best fits in olive. SDP.6418 optical spectrum shows a blue continuum with strong O{\sc ii}, weak
  H$\beta$ and O{\sc iii} and strong H$\alpha$ at the red edge of the
  spectrum. SDP.6451 optical spectrum shows strong O[{\sc ii}] and H$\beta$
  in emission indicating high current levels of SFR. There is only
  weak O[{\sc iii}] indicating that this is not an AGN.}
\end{figure*}

This galaxy has $\Ms=5.6\times10^{10}\,\msun$ and $f_g=0.17$. The GAMA
spectrum (Fig.~\ref{A6418F}) has strong O[{\sc ii}] and H$\beta$ in
emission indicating ongoing star formation. There is only weak O[{\sc
    iii}] indicating that this is not an AGN
\citep{Lamareille2010}. Both dust and \CI line emission are strongly
detected, while CO is much weaker. Figure~\ref{6451F}(top right) shows
the CO contours on tapered \CI showing a roughly 2$\sigma$ detection
of flux at the same location and same velocity range as \CI. There is
a wide blue wing in the \CI line profile which is extended to the east
of the galaxy. At full resolution this appears to have structure,
possibly spiral arms or tidal material. The tapered map suggests that
this may be another source where the dust and \CI may be partially
resolved out. The flux presented in Table~\ref{CIdataT} for \CI is
integrated over the full extent of the emission from $-350\rightarrow
+150$\kms. The CO flux estimate is measured in an aperture over a
similar velocity range. The dust emission appears to be
morphologically different to the \CI, there are two distinct clumps
either side of the optical nucleus and no extension to the east as is
apparent in \CI (Fig.~\ref{6451F}). The
$\rm{R_{CI}}=0.69^{+0.51}_{-0.27}$ for this source is also very high.

\section{Flux boosting by SMG}
\label{boostS}
\normalsize

Flux boosting by the high redshift dusty galaxies in these fields
affects all the {\em Herschel} SPIRE fluxes of our target $z=0.35$
sources, but more so in the 350 and 500\mic bands for two
reasons. Firstly, the {\em Herschel} beam is larger with increasing
wavelength and so a contaminant at distance $r$ will have a higher
weighting in the beam profile. Secondly, the observed sub-mm colour of
the high redshift dust source will be redder (relatively brighter at
500\mic) compared to the target source, as the observed frame samples
closer to the peak of the SED. The percentage contamination at 500\mic
from the high-z SMG is thus larger than that at 250\mic. In order to
produce reasonably accurate physical parameters from SED fitting, we
wanted to estimate the possible contamination from the high redshift
galaxy and investigate its impact on the results of the SED
fitting. More details are given in \citet{Dunne2020smg}.
Briefly, we took the ALMA $S_{850}$ for each high redshift dusty
galaxy and calculated the flux that it should have in each of the {\em
  Herschel} SPIRE bands, using plausible values of $T_d=38$~K,
$\beta=1.8$ and $z=1.5-5$ for SMG
\citep{Chapman2005,daCunha2015,Stach2019}. Any redshift where the
predicted signal should be highly visible in the H-ATLAS maps at the
position of the SMG was ruled out. From the remaining possible
redshifts we calculate the contribution of the high redshift dusty
galaxy to the {\em Herschel} fluxes of the $z=0.35$ galaxy by
weighting its predicted flux at each band by the beam attenuation at
the location of the $z=0.35$ galaxy relative to the position of the
SMG. We then correct the H-ATLAS DR-1 photometry in
Table~\ref{photomT} for the possible contamination and fit the SEDs
again with {\sc magphys}. We choose the redshift at which the
correction produces the lowest overall $\chi^2$ for the SED fit and
the best estimates of the contamination in the {\em Herschel} bands are listed in
Table~\ref{SMGcontamT}.\footnote{The redshift and \td are roughly
  degenerate in this process but we are not attempting to constrain
  either; merely we wish to retrieve FIR colours which are most
  compatible with the PACS and ALMA 850\mic photometry, which we know
  to be uncontaminated}. 

\begin{table}
\caption{\label{SMGcontamT} Estimated boost to the $z=0.35$ H-ATLAS
  photometry due to the high redshift dusty galaxies within the {\em
    Herschel} beam, from Dunne et al. {\em submitted}.}
\begin{tabular}{lccc}
\hline
SMG &   $\rm{S^C_{250}}$ & $\rm{S^C_{350}}$ & $\rm{S^C_{500}}$ \\
   &      (mJy)        &   (mJy)       & (mJy) \\
\hline
1160  & 13.6 & 16.9 & 16.8\\
2173  & 9.2 & 6.7 & 3.3\\
3366  & 12.5 & 7.7 & 3.4\\
4104  &  12.7 & 11.1 & 6.2\\
5347  & 2.0 & 3.4 & 3.0\\
6418  & 1.2 & 3.5 & 4.7\\  
6451  & 1.6 & 4.2 & 5.1\\
\hline
\end{tabular}
\flushleft{\small Estimated contamination to the H-ATLAS SPIRE 
  fluxes in Table~\ref{photomT}.}
\end{table}

\section{{\sc magphys} SED fits and optical spectra for the sample.}

\label{lastpage}
\end{document}